\documentclass[11pt, oneside]{article}   	

\usepackage{jheppub}

\usepackage{amsmath}
\usepackage{amssymb}
\usepackage{amsfonts}
\usepackage{amsthm}
\usepackage{amsbsy}
\usepackage{array}

\usepackage{subcaption}

\usepackage{mathtools}
\usepackage{xcolor}

\usepackage{ytableau}
\usepackage{young}
\usepackage[vcentermath]{youngtab}

\usepackage{tikz}
\usetikzlibrary{decorations.markings}

\usepackage{pepper}

\theoremstyle{definition}

\theoremstyle{theorem}

\theoremstyle{definition}

\tikzset{
  dnode/.style={
	circle,
	draw,
	fill=black,
	inner sep=1pt,
	minimum size=1pt
  },
  spinning/.style={
	thick,
	postaction={
		decorate,
		decoration={
			markings,
			mark=at position 0.5 with {\arrow{>}}
		}
	}
  }
}

\newcommand{\myng}[1]{\,{\tiny\yng #1}\,}
\newcommand{\jyng}{\myng{(2)}\!\cdots\!\myng{(1)}}

\newcommand{\rmel}[5]{
	\Bigg(
	{{#1}\atop{#2}}
	\Bigg\vert
	{#3}
	\Bigg\vert
	{{#4}\atop {#5}}
	\Bigg)
}

\newcommand{\rcgc}[6]{
	\Bigg(
	{{#1}\atop{#2}}
	{{#3}\atop {#4}}
	\Bigg\vert
	{{#5}\atop {#6}}
	\Bigg)	
}

\newcommand{\rcgcConj}[6]{
	\Bigg(
	{{#1}\atop{#2}}
	\Bigg\vert
	{{#3}\atop {#4}}
	{{#5}\atop {#6}}
	\Bigg)	
}

\newcommand{\seq}[1]{{\mathfrak{\MakeUppercase{#1}}}}

\newcommand{\sixj}[6]{
	{
		\left\{
			\begin{matrix}
			 #1 & #2 & #3 \\
			 #4 & #5 & #6
			\end{matrix}
		\right\}
	}
}

\newcommand{\struct}[8]{
	\left[
	\begin{matrix}
		#1 & #2 & #3 & #4 \\
		#5 & #6 & #7 & #8	
	\end{matrix}
	\right]
}

\title{Casimir recursion relations for general conformal blocks}
\author{Petr Kravchuk}
\affiliation{Walter Burke Institute for Theoretical Physics, Caltech, Pasadena, California 91125, USA 
}
\date{}							

\abstract{
We study the structure of series expansions of general spinning conformal blocks. We find that the terms in these expansions are naturally expressed by means of special functions related to matrix elements of $Spin(d)$ representations in Gelfand-Tsetlin basis, of which the Gegenbauer polynomials are a special case. We study the properties of these functions and explain how they can be computed in practice. We show how the Casimir equation in Dolan-Osborn coordinates leads to a simple one-step recursion relation for the coefficients of the series expansion of general spinning conformal block. The form of this recursion relation is determined by $6j$ symbols of $Spin(d-1)$. In particular, it can be written down in closed form in $d=3$, $d=4$, for seed blocks in general dimensions, or in any other situation when the required $6j$ symbols can be computed. We work out several explicit examples and briefly discuss how our recursion relation can be used for efficient numerical computation of general conformal blocks.
}

\preprint{CALT-TH 2017-050}

\begin{document}

\maketitle

\section{Introduction}

Numerical conformal bootstrap is a very general and powerful approach to quantum conformal filed theories (CFTs), based on the idea of analyzing the crossing symmetry~\cite{Ferrara:1973yt,Polyakov:1974gs,Mack:1975jr} of correlation functions in unitary CFTs by methods of semidefinite programming~\cite{Rattazzi:2008pe,Poland:2011ey,Kos:2013tga,Kos:2014bka,Simmons-Duffin:2015qma}. In recent years, this approach has proven to be extremely useful in extracting non-perturbative information about concrete CFTs, such as the critical exponents and structure constants of 3d Ising CFT, $O(N)$ and Gross-Neveu models~\cite{ElShowk:2012ht,El-Showk:2014dwa,Kos:2013tga,Kos:2014bka,Kos:2015mba,Kos:2016ysd,Iliesiu:2015qra,Simmons-Duffin:2016wlq,Iliesiu:2017nrv}, as well as a host of other results~\cite{Rychkov:2009ij,Caracciolo:2009bx,Rattazzi:2010gj,Poland:2010wg,Rattazzi:2010yc,Vichi:2011ux,Rychkov:2011et,Liendo:2012hy,ElShowk:2012hu,Gliozzi:2013ysa,Alday:2013opa,Gaiotto:2013nva,Berkooz:2014yda,Nakayama:2014lva,Nakayama:2014yia,Chester:2014fya,Caracciolo:2014cxa,Nakayama:2014sba,Paulos:2014vya,Bae:2014hia,Beem:2014zpa,Chester:2014gqa,Bobev:2015jxa,Chester:2015qca,Beem:2015aoa,Rejon-Barrera:2015bpa,Poland:2015mta,Lemos:2015awa,Kim:2015oca,Lin:2015wcg,Chester:2015lej,Chester:2016wrc,Behan:2016dtz,Dey:2016zbg,Nakayama:2016knq,El-Showk:2016mxr,Li:2016wdp,Pang:2016xno,Lin:2016gcl,Lemos:2016xke,Beem:2016wfs,Li:2017ddj,Collier:2017shs,Cornagliotto:2017dup,Gliozzi:2017hni,Gopakumar:2016cpb,Rychkov:2017tpc,Nakayama:2017vdd,Chang:2017xmr,Dymarsky:2017xzb}. Crossing symmetry of the four-point functions of such fundamental operators as spin-1 conserved currents or the energy-momentum tensor has also been instrumental in deriving universal constraints valid for general CFTs~\cite{Dymarsky:2017xzb,Dymarsky:2017yzx}.

The practical implementation of numerical conformal bootstrap relies heavily on two technical requirements: the knowledge of conformal blocks and the ability to efficiently solve the semidefinite programs. An efficient semidefinite solver~\texttt{SDPB}, designed specifically for bootstrap applications, was introduced in~\cite{Simmons-Duffin:2015qma}. This solver is able to solve the most general semidefinite programs which typically arise in conformal bootstrap, thus eliminating the technical obstructions related to semidefinite programming. The situation with conformal blocks is different. The simplest conformal blocks -- those with external scalar operators -- are very well studied by now and there exist simple and efficient techniques for their computation~\cite{DO1,DO2,DO3,Kos:2013tga,Hogervorst:2013kva,Hogervorst:2013sma}. Some of these techniques, such as Zamolodchikov-like recursion relations, iterative/analytic solutions of conformal Casimir equations or shadow integrals have been extended to conformal blocks of operators with spins~\cite{Penedones:2015aga,Iliesiu:2015akf,Echeverri:2016dun,Costa:2016xah,Costa:2016hju,Dymarsky:2017xzb,SimmonsDuffin:2012uy,Rejon-Barrera:2015bpa}. Another approach to spinning conformal blocks is to relate them to simpler conformal blocks by means of differential operators~\cite{Costa:2011dw,Iliesiu:2015qra,Echeverri:2015rwa}; recently it was shown that the most general conformal blocks can be reduced in this way to scalar blocks~\cite{Karateev:2017jgd}. While these methods do allow us to calculate any given non-supersymmetric conformal block, all of them currently require a nontrivial amount of case-specific analysis.

In order to facilitate the conformal bootstrap studies with spinning operators it is therefore desirable to have a simple and general algorithm for numerical computation of conformal blocks which can be implemented on a computer, ideally avoiding the need for symbolic algebra. The first step in this direction was undertaken in~\cite{Kravchuk:2016qvl}, where a general classification and construction of conformally-invariant tensor structures was given. In this paper, we take another step towards this goal by formulating a general Casimir recursion relation for the $z$-coordinate series expansion of general spinning conformal blocks in any number of dimensions. For a conformal block exchanging a primary operator $\cO$, the recursion relation takes the form
\be
	(C(\Delta_{p+1},\widetilde\bfm_d)-C(\cO))\,\Lambda^{ba}_{p+1,\widetilde\bfm_d}=\sum_{\bfm_d\in\myng{(1)}\otimes\widetilde\bfm_d}
	(\bar\gamma_{p,\bfm_d,\widetilde\bfm_d}\Lambda_{p,\bfm_d}\gamma_{p,\bfm_d,\widetilde\bfm_d})^{ba}.
\ee
where the matrices $\Lambda_{p,\bfm_d}$ encode the contribution of descendants at level $p$ and in $Spin(d)$ representation $\bfm_d$ in $z$-coordinates, $\Delta_p=\Delta_\cO+p$, $C$ give the conformal Casimir eigenvalues, while $\gamma$ and $\bar\gamma$ are some matrices. Similar recursion relations have been recently considered in~\cite{Costa:2016xah}. Our improvement over these results is in that the structure of our recursion relation is much simpler (in particular, it is one-step, i.e.\ relates levels $p$ and $p+1$, similarly to the scalar recursion relation in~\cite{Hogervorst:2013sma}) and we are able to remain completely general and write the coefficients $\gamma$ and $\bar\gamma$ in terms $6j$ symbols (or Racah coefficients) of $Spin(d-1)$. Thus, in our form, the Casimir recursion relations can be immediately translated into a computer algorithm in all cases when the $6j$ can be computed algorithmically. This includes the general conformal blocks in $3$ and $4$ dimensions as well as seed blocks in general dimensions. Importantly, since we solve all representation-theoretic questions in terms of Clebsch-Gordan coefficients and $6j$ symbols, our analysis is applicable to all spin representations without any caveats, i.e.\ it applies equally well to spinor representations and is free from the redundancies which plague the less abstract approaches in low dimensions.\footnote{Assuming, of course, that Clebsch-Gordan coefficients are known.}

This paper consists of three main parts. The first part is section~\ref{sec:reptheory} in which we review the basics of the representation theory of $Spin(d)$ and give a brief summary of the required facts from the theory of Gelfand-Tsetlin (GT) bases. The advantage of GT bases is that they allows us to work very explicitly with completely general representations in arbitrary $d$, at the same time being perfectly compatible with the conformal frame construction of~\cite{Kravchuk:2016qvl}. Moreover, many explicit formulas for matrix elements and Clebsch-Gordan coefficients are available in these bases. These facts make them our main computational tool in this paper.

In section~\ref{sec:structure} we use these tools to study the contribution of a general $\bR\times Spin(d)$ (dilatations$\times$rotations) multiplet to a given four-point function. In section~\ref{sec:structure:Contribution} we express the answer in terms of an explicit basis of three- and four-point functions (constructed using the Clebsch-Gordan coefficients of $Spin(d-1)$). The functions $P$ which replace the Gegenbauer polynomials (which appear in scalar correlation functions) are some particular matrix elements of $e^{\theta M_{12}}$ in a GT basis. In sections~\ref{sec:structure:ScalarCorrelators}-\ref{sec:structure:SeedConformalBlocks} we consider the $\bR\times Spin(d)$ contributions in some simple special cases. In section~\ref{sec:structure:FolkloreTheorem} we prove the folklore theorem which states that the number of four-point tensor structures is equal to the number of classes of conformal blocks.
In section~\ref{sec:structure:Pfunctions} we study the properties of $P$-functions and explain how they can be efficiently computed in practice by organizing them in so-called ``matroms''~\cite{RepresentationsAndSpecialFunctions} and deriving a recursion relation for these matroms. We also discuss the simplifications in the low-dimensional cases of $d=3$ and $d=4$. In appendix~\ref{app:tensors} we relate the functions $P$ to irreducible projectors studied recently in~\cite{Costa:2016hju} in the case of tensor representations.

In section~\ref{sec:casimir} we study the Casimir recursion relations for general conformal blocks. We start by rederiving the scalar result of~\cite{Hogervorst:2013sma} in section~\ref{sec:casimir:ReviewScalar} using an abstract group-theoretic approach. In section~\ref{sec:casimir:SpinningConformalBlocks} we extend this approach to general representations and derive the formulas~\eqref{eq:gammadefn} and~\eqref{eq:bargammadefn} for $\gamma$ and $\bar\gamma$ in terms of $6j$ symbols of $Spin(d-1)$. In sections~\ref{sec:casimir:General3d}-\ref{sec:casimir:SeedBlocks} we discuss how these $6j$ symbols simplify in the case $d=3$ and for the seed blocks in general $d$.\footnote{We do not discuss the case of general blocks in $d=4$, where these $6j$ symbols are also known, only to keep the size of the paper reasonable -- the application of the general formula is completely mechanical.} For more specific examples we explicitly work out the recursion relations for scalar-fermion seed blocks in $d=3$ and $d=2n$ and compare them to the known results. In section~\ref{sec:implementationremarks} we briefly discuss the problems associated with a practical solution of the Casimir recursion relation and suggest some possible workarounds.

We conclude in section~\ref{sec:conclusions}. The appendices~\ref{app:conformalalgebra} and~\ref{app:formulae} contain some explicit formulas and details on our conventions. The appendix~\ref{app:ScalarFermionChecks} elaborates on comparison to known results. In appendix~\ref{app:tensors} we explain the relation between GT and Cartesian bases for tensor representations.

\section{Representation theory of $Spin(d)$}
\label{sec:reptheory}
We will be studying conformal blocks for the most general representations of $Spin(d)$, which requires a certain amount of mathematical machinery. In this section we review the relevant representation theory and establish important notation.

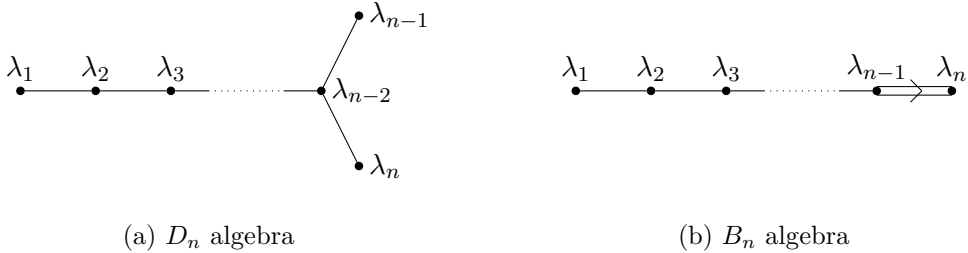
\begin{figure}[h!]
\centering
\begin{subfigure}[b]{0.45\textwidth}
\begin{tikzpicture}
	\path[use as bounding box] (-1,-1.5) rectangle (6,1.5);
	\node (l1) at (0,0) [dnode] {};
	\node (l2) at (1,0) [dnode] {};
	\node (l3) at (2,0) [dnode] {};

	\node (lnm2) at (4,0) [dnode] {};
	\node (lnm1) at (4.5,1) [dnode] {};
	\node (ln) at (4.5,-1) [dnode] {};
	
	\draw (l1) -- (l2) -- (l3) -- (2.5,0);
	\draw [dotted] (2.5,0) -- (3.5,0);
	\draw (3.5,0) -- (lnm2);
	\draw (lnm2) -- (lnm1);
	\draw (lnm2) -- (ln);
	
	\node at (l1) [above] {$\lambda_1$};
	\node at (l2) [above] {$\lambda_2$};
	\node at (l3) [above] {$\lambda_3$};

	\node at (lnm2) [right] {$\lambda_{n-2}$};
	\node at (lnm1) [right] {$\lambda_{n-1}$};
	\node at (ln) [right] {$\lambda_n$};
\end{tikzpicture}
\caption{$D_n$ algebra}\label{fig:dn}
\end{subfigure}
~
\begin{subfigure}[b]{0.45\textwidth}
\begin{tikzpicture}
	\path[use as bounding box] (-1,-1.5) rectangle (6,1.5);
	\node (l1) at (0,0) [dnode] {};
	\node (l2) at (1,0) [dnode] {};
	\node (l3) at (2,0) [dnode] {};
	
	\node (lnm1) at (4,0) [dnode] {};
	\node (ln) at (5,0) [dnode] {};
	
	\draw (l1) -- (l2) -- (l3) -- (2.5,0);
	\draw [dotted] (2.5,0) -- (3.5,0);
	\draw (3.5,0) -- (lnm2);
	\draw (lnm1.north) -- (ln.north);
	\draw (lnm1.south) -- (ln.south);
	\draw (4.6,0) -- (4.45,0.15);
	\draw (4.6,0) -- (4.45,-0.15);
	
	\node at (l1) [above] {$\lambda_1$};
	\node at (l2) [above] {$\lambda_2$};
	\node at (l3) [above] {$\lambda_3$};
	
	\node at (lnm1) [above] {$\lambda_{n-1}$};
	\node at (ln) [above] {$\lambda_{n}$};
\end{tikzpicture}
\caption{$B_n$ algebra}\label{fig:bn}
\end{subfigure}
\caption{Dynkin diagrams of $\mathfrak{so}(d)$ algebras.}
\end{figure}

We will be working exclusively in the Euclidean signature (the results can be easily translated to Lorentz signature by Wick rotation). This means that we work with the compact real form of $Spin(d)$, which double covers $SO(d)$. As is well known, the basic properties of these groups depend on the parity of $d$. If $d=2n$, then the Lie algebra of $Spin(d)$ is the simple\footnote{Semi-simple for $d=4$: $D_2=A_1\oplus A_1$ is equivalent to two copies of $\frak{su}_2$ algebra.} rank-$n$ Lie algebra $D_n$ with Dynkin diagram shown in Fig.~\ref{fig:dn}. If $d=2n+1$ then the relevant algebra is the simple rank-$n$ Lie algebra $B_n$ with Dynkin diagram shown in Fig.~\ref{fig:bn}.

It is standard to specify the irreducible representations\footnote{We are interested in representations over $\bC$, since the physical Hilbert space is complex. However, we often treat the representations which are real (in the sense of being representable by real matrices) as being over $\bR$.} by non-negative integral Dynkin labels $\lambda_i$ associated to the nodes in the Dynkin diagram. The representations in which only one $\lambda_i$ is non-zero and equal to $1$ are called the fundamental representations. The fundamental representation associated with $\lambda_1$ (i.e.\ the one with labels $\lambda_i=\delta_{i1}$) is the fundamental vector representation $\bR^d$.\footnote{Unless $d\leq 4$ when $\lambda_1$ corresponds to one of the spinor representations.} More generally, the fundamental representations associated with $\lambda_i$ with $i<d/2-1$ are the exterior powers of the vector representation, $\wedge^i \bR^d$. The nodes $\lambda_{n-1}=a,\,\lambda_{n}=c$ in $D_n$ case correspond to the two chiral spinor representations. Similarly, the node $\lambda_n=b$ corresponds to the unique spinor representation in $B_n$ case. A general representation can be obtained by tensoring the above ``fundamental'' representations together and taking the irreducible component with the highest weight (i.e.\ by imposing the maximal symmetry and tracelessness conditions on the resulting tensors).

For us it will be more convenient to label the representations by generalized Young diagrams, constructed as follows. To a given set of Dynkin labels of $Spin(d)$ we associate a vector of numbers $\bfm_d$ with components, for $d=2n$,
\begin{align}
m_{d,1}&=\lambda_1+\lambda_2+\ldots+\lambda_{n-2}+\frac{a+c}{2},\\
m_{d,2}&=\lambda_2+\lambda_3+\ldots+\lambda_{n-2}+\frac{a+c}{2},\\
&\qquad\vdots\nn\\
m_{d,n-2}&=\lambda_{n-2}+\frac{a+c}{2},\\
m_{d,n-1}&=\frac{a+c}{2},\\
m_{d,n}&=\frac{a-c}{2},
\end{align}
and for $d=2n+1$,
\be
m_{d,1}&=\lambda_1+\lambda_2+\ldots+\lambda_{n-1}+\frac{b}{2},\\
m_{d,2}&=\lambda_2+\lambda_3+\ldots+\lambda_{n-1}+\frac{b}{2},\\
&\qquad\vdots\nn\\
m_{d,n-1}&=\lambda_{n-1}+\frac{b}{2},\\
m_{d,n}&=\frac{b}{2}.
\ee
This gives all possible sequences satisfying
\be
\begin{array}{ll}
m_{d,1}\geq m_{d,2}\geq \ldots m_{d,n-1}\geq |m_{d,n}|,&\quad\text{for $d=2n$},\\
m_{d,1}\geq m_{d,2}\geq \ldots m_{d,n}\geq 0,&\quad\text{for $d=2n+1$},
\end{array}
\ee
and consisting either entirely of intergers (bosonic representations) or entirely of half-integers (fermionic representations). The dimensions of these irreducible representations are given in appendix~\ref{app:formulae}.

When $\bfm_d$ is bosonic, we can think of $|m_{d,k}|$ as giving the length of $k$-th row in a Young diagram, with the caveat that for $d=2n$ the diagrams of height $n$ can correspond to self-dual tensors ($m_{d,n}>0$) or anti-self-dual tensors ($m_{d,n}<0$). Because of that, we will often represent the vectors $\bfm_d$ by Young diagrams, for example,
\be
	(5,0,0,\ldots)&=\myng{(5)},\\
	(5,3,1,0,\ldots)&=\myng{(5,3,1)},\\
	(0,0,\ldots)&=\bullet.
\ee
Note that we denote the empty diagram corresponding to the trivial representation by $\bullet$. We will also sometimes use the notation
\be
	\bfj&\equiv \jyng\quad (j\text{ boxes}),\\
	(\bfj,\myng{(2,1)})&\equiv \myng{(4,2,1)}\raisebox{7pt}{$\cdots\!\myng{(1)}$}\quad (j\text{ boxes in 1st row}).
\ee
Note, however, that we do \textit{not} restrict our analysis to bosonic representations only. 

For future convenience, we define
\be
	|\bfm_d|=\sum_{k=1}^n|m_{d,k}|,
\ee
which gives the number of boxes when $\bfm_d$ can be represented by a Young diagram. 

\paragraph{Examples}

For example, consider $d=2$. Strictly speaking, this case does not fall under the above discussion, since $Spin(2)$ is not semi-simple. However, the vectors $\bfm_2$ can still be used to label the representations, and this will be important to us in the following. The vectors $\bfm_2$ are one-dimensional, with a single (half-)integral entry $m=m_{2,1}$. The corresponding representation is the one-dimensional representation which associates to rotation $e^{\phi M_{12}}$ the phase factor $e^{-im\phi}$.\footnote{We choose the minus sign for future convenience.} This is $4\pi$-periodic for half-integral $m$, corresponding to the need to consider the double-cover $Spin(2)$ instead of $SO(2)$.

Now consider $d=3$ corresponding to $B_1$ case. In this case the vector $\bfm_3$ consists of a single component equal to $b/2$, where $b$ is the unique Dynkin label. In other words $\bfm_3=(j)$ where $j$ is the usual spin of $Spin(3)$.

The case $d=4$ corresponds to $D_2$. We have two Dynkin labels, which we will denote by $l_L=a/2,l_R=c/2$. For example, the vector representation is given by $(l_L, l_R)=(\thalf,\thalf)$, while the Dirac spinors are $(\thalf,0)\oplus(0,\thalf)$. The vector $\bfm_4$ is two dimensional with the components,
\be
	\bfm_4=(l_L+l_R,l_L-l_R).
\ee
We see that for traceless-symmetric representations with $l_L=l_R$ we recover the one row Young diagram, while for example for the representations $(1,0)$ or $(0,1)$ we recover the diagram $\myng{(1,1)}$ with self- or anti-self-duality condition.

\subsection{Dimensional reduction}
\label{sec:reptheory:DimensionalReduction}
Labeling the representations by the vectors $\bfm_d$ is convenient for describing the rule for dimensional reduction from $Spin(d)$ to $Spin(d-1)$. More precisely, an irreducible representation $\bfm_d$ decomposes into a direct sum of irreducible representations $\bfm_{d-1}$ of $Spin(d-1)$, which we can write as
\be
\bfm_d=\bigoplus_{\bfm_{d-1}\in\,\bfm_d} N^{\bfm_d}_{\bfm_{d-1}}\,\bfm_{d-1},
\ee
where $ N^{\bfm_d}_{\bfm_{d-1}}$ denote the multiplicity with which $\bfm_{d-1}$ appears in the irreducible decomposition of $\bfm_d$. It turns out that all multiplicities are equal to one, 
\be
N^{\bfm_d}_{\bfm_{d-1}}=1,\quad \forall\bfm_{d-1}\in\bfm_d.
\ee
We say that dimensional reduction is multiplicity-free. The representations $\bfm_{d-1}\in\bfm_d$ are described by the following rule~\cite{RepresentationsAndSpecialFunctions}

\paragraph{From $Spin(2n+1)$ to $Spin(2n)$:} For an irreducible representation $\bfm_{d}$ of $Spin(d)$, $d=2n+1$, and an irreducible representation $\bfm_{d-1}$ of $Spin(d-1)$ the relation $\bfm_{d-1}\in\,\bfm_d$ holds iff both representations are of the same statistics (fermionic or bosonic) and satisfy
\be
	m_{d,1}\geq m_{d-1,1}\geq m_{d,2}\geq m_{d-1,2}\geq\ldots\geq m_{d,n}\geq |m_{d-1,n}|\geq 0.
\ee
\paragraph{From $Spin(2n)$ to $Spin(2n-1)$:}
For an irreducible representation $\bfm_{d}$ of $Spin(d)$, $d=2n$, and an irreducible representation $\bfm_{d-1}$ of $Spin(d-1)$ the relation $\bfm_{d-1}\in\,\bfm_d$ holds iff both representations are of the same statistics (fermionic or bosonic) and satisfy
\be
	m_{d,1}\geq m_{d-1,1}\geq m_{d,2}\geq m_{d-1,2}\geq\ldots\geq m_{d-1,n-1}\geq |m_{d,n}|\geq 0.
\ee

\paragraph{Examples} Consider first the reduction from $Spin(4)$ to $Spin(3)$. The constraint is
\be
	m_{4,1}\geq m_{3,1}\geq |m_{4,2}|,
\ee
which in terms of $j,l_L, l_R$ reads
\be
	l_L+l_R\geq j\geq |l_L-l_R|.
\ee
Together with the constraint that the Fermi/Bose statistics is preserved, we find that 
\be
	j=|l_L-l_R|, |l_L-l_R|+1,\ldots, l_L+l_R.
\ee
This is the same as saying that $j\in l_L\otimes l_R$, where $l_L$ and $l_R$ are interpreted as $Spin(3)$ spins, which coincides with the familiar reduction rule.

Consider now the reduction from $Spin(3)$ to $Spin(2)$. For a given $\bfm_{3}=(j)$ we have the following constraint on $\bfm_2=(m)$,
\be
	j\geq |m|\geq 0,
\ee
and $m$ should be (half-)integral simultaneously with $j$. In other words, $m=-j,-j+1,\ldots j$. It is no accident that the relation between $j$ and $m$ is the same as in the basis elements $|j,m\>$, because the $Spin(2)$ irreps are one-dimensional. This in fact is a very powerful observation which generalizes to higher dimensions, as we now discuss.

\subsection{Gelfand-Tsetlin basis}
\label{sec:reptheory:GelfandTsetlin}
The fact that the dimensional reduction is multiplicity-free allows one to define a convenient basis for the irreducible representations of $Spin(d)$. To construct it, one first fixes a sequence of subgroups
\be\label{eq:groupsequence}
	Spin(d)\supset Spin(d-1)\supset Spin(d-2)\supset\ldots\supset Spin(2).
\ee
In practice, we pick an orthonormal basis $e_1,\ldots e_d$ in $\bR^d$, and the $Spin(d-k)$ subgroup in the above sequence is defined as the one preserving the basis elements $e_1,\ldots, e_k$.
Then, given a representation $\bfm_{d}$, we can consider an irreducible component $\bfm_{d-1}\in\,\bfm_{d}$ with respect to $Spin(d-1)$. Since the dimensional reduction is multiplicity-free, by specifying the numbers $\bfm_{d-1}$ we uniquely select an $Spin(d-1)$-irreducible subspace inside the representation space $V_{\bfm_d}$ of the representation $\bfm_d$. We can then continue to build a sequence
\be\label{eq:admissiblesequences}
	\bfm_d\ni\bfm_{d-1}\ni\bfm_{d-2}\ni\ldots\ni\bfm_2,
\ee
which uniquely selects a $Spin(2)$-irreducible subspace inside $V_{\bfm_{d}}$. Since $Spin(2)$ is abelian, all such subspaces are one-dimensional. Therefore, if we in addition make a choice of phases, the above sequence specifies a unit vector in $V_{\bfm_{d}}$.

Let us now denote a sequence of $\bfm_k$, $k=d,d-1,\ldots,2$ by $\seq{m}_d$. Call a sequence $\seq{m}_d$ admissible if \eqref{eq:admissiblesequences} is satisfied. The above construction associates to each admissible sequence a vector $|\seq{m}_d\rangle$ in $V_{\bfm_d}$. It is an easy exercise to show that the set of  $|\seq{m}_d\rangle$ over all admissible sequences (with $\bfm_d$ fixed) forms an orthonormal basis in $V_{\bfm_d}$. This is the Gelfand-Tsetlin (GT) basis \cite{GTbasis}, and the sequences $\seq{m}_d$ are known as Gelfand-Tsetlin patterns.

Analogously to the well-known formulas for the matrix elements of $Spin(3)$ generators between the $|j,m\rangle$ states, Gelfand and Tsetlin have derived formulas for the matrix elements of $Spin(d)$ generators in Gelfand-Tsetlin basis for arbitrary representations \cite{GTbasis,RepresentationsAndSpecialFunctions,ReductionFactors,GouldFirst}. We provide these formulas for reference in section~\ref{sec:reptheory:ClebshGordan} and appendix~\ref{app:formulae}. Availability of such general formulas is one of the reasons why Gelfand-Tsetlin bases are useful. For our purposes the more important reason is that these bases play nicely with the inclusions $\eqref{eq:groupsequence}$, which appear naturally in construction of conformally invariant tensor structures \cite{Kravchuk:2016qvl}.

\paragraph{Choice of phases} Before proceeding further, let us make a general comment about the choice of phases for vectors $|\seq{m}_d\>$. This choice is not going to be important in the discussion that follows -- it only influences the explicit expressions for $Spin(d)$ matrix elements, Clebsch-Gordan coefficients, etc. Therefore, we should only worry about it when we compute these quantities, and we can make a choice which is the most convenient for our purposes. For example the formulas given in appendix~\ref{app:formulae} correspond to some particular choice of phases. We have made this choice so that it is compatible with the explicit constructions in the the examples below, unless explicitly stated otherwise.

\paragraph{Notation}

As we mentioned above, for us the utility of GT bases comes from their compatibility with the nested sequence~\eqref{eq:groupsequence}, which plays an important role in classification of conformally-invariant tensor structures~\cite{Kravchuk:2016qvl}. Unfortunately, this means that we will have to dive into the structure of the sequences $\seq{m}_d$ quite often. Because of that, it is important to establish a well-defined notation.

Firstly, we will always explicitly write the space dimension $d$ to which a weight $\bfm_d$ corresponds as a subscript. Secondly, the GT patterns in representation with highest weight $\bfm_d$ will be denoted by the capital Fraktur letter $\seq{m}_d$. Distinct patterns in the same $\bfm_d$ will be distinguished by primes, i.e.\ $\seq{m}'_d$. The subscript on the pattern indicates the dimension $d$ corresponding to the first weight in the pattern. This weight is kept fixed and equal to $\bfm_d$ when we write summation as
\be
	\sum_{\seq{m}_d}.
\ee
In all summations it is assumed implicitly that only admissible sequences are included.

Furthermore, $\bfm_k$ for $k\leq d'$ is always used to denote the components of the GT pattern $\seq{m}_{d'}$. In particular, this means that the pattern $\seq{m}_{d-1}$ is the tail\footnote{This is slightly in tension with our convention on primes. We will understand that $\seq{m}'_{d-1}$ is the tail of $\seq{m}'_{d}$, i.e.\ $\bfm'_{d-1}$ is not necessarily the same as $\bfm_{d-1}$ (which would be the case if we gave the priority to the prime notation rule and understood $\seq{m}'_{d-1}$ as another pattern in $\bfm_{d-1}$). Note also that by this convention $\bfm_{d-1}'\in\bfm_d$, etc.} of the pattern $\seq{m}_d$ and we have, for example,
\be
	\sum_{\seq{m}_d}\equiv\sum_{\bfm_{d-1}}\sum_{\seq{m}_{d-1}}.
\ee
We also occasionally write $\seq{m}_d=\bfm_d\,\seq{m}_{d-1}$, etc, arranging the right hand side either vertically or horizontally, whichever way leads to more compact expressions. We also sometimes write out the GT patterns explicitly as
\be
	\seq{m}_d\equiv \bfm_d,\bfm_{d-1},\ldots,\bfm_{2}.
\ee
If we have $\bfm_k=\bullet$, then necessarily $\bfm_i=\bullet$ for $i\leq k$. We therefore often write the patterns out only to the first trivial representation, replacing the rest by dots. For example,
\be
	\seq{m}_d=\myng{(5,2)},\myng{(3)},\bullet,\ldots
\ee
has $\bfm_k=\bullet$ for all $k\leq d-2$.

Different representations and patterns are distinguished either by different letters (i.e.\ $\bfu_d$ and $\seq{u}_d$ vs $\bfm_d$ and $\seq{m}_d$), accents other than primes (i.e.\ $\widetilde\bfm_d$ and $\widetilde{\seq{m}}_d$ vs $\bfm_d$ and $\seq{m}_d$), or upper indices (i.e.\ $\bfm^1_d$ and $\seq{m}_d^1$ vs $\bfm_d$ and $\seq{m}_d$). To reiterate, the lower index only ``addresses'' inside one pattern.

Our final comment concerns the use of GT patterns as indices. We will assume that the upper GT indices, such as
\be
	\cO^{\seq{m}_d},
\ee
behave as ket states $|\seq{m}_d\>$, while the lower indices behave as the dual bra states $\<\seq{m}_d|$, i.e.\
\be
	[M_{\mu\nu},\cO^{\seq{m}_d}(0)]&=\sum_{\seq{m}'_d} \<\seq{m}'_d|M_{\mu\nu}|\seq{m}_d\>\cO^{\seq{m}'_d}(0),\\
	[\cO_{\seq{m}_d}(0),M_{\mu\nu}]&=\sum_{\seq{m}'_d} \<\seq{m}_d|M_{\mu\nu}|\seq{m}'_d\>\cO_{\seq{m}'_d}(0).
\ee

\subsubsection{Bilinear parings}

The most basic invariants of $Spin(d)$ are the bilinear parings, such as the paring between a representation and its dual, or the invariant inner product in real representations. A bilinear pairing between irreducible representations $\bfm_d$ and $\bfu_d$ is a singlet in the tensor product
\be
	\bfm_d\otimes \bfu_d.
\ee
Schur's lemma implies that there is at most one such singlet, which exists iff $\bfm_d=\overline{\bfu_d}$, i.e.\ when the representations are mutually dual (equivalently, complex conjugate). The duality acts on the $Spin(d)$ irreps as follows. For odd $d$ all irreps are self-dual, $\bfm_{2n+1}=\overline{\bfm_{2n+1}}$, as well as for $d$ divisible by $4$, $\bfm_{4k}=\overline{\bfm_{4k}}$. For $d=4k+2$ the duality acts non-trivially by exchanging the spinor nodes on $D_{2k+1}$ Dynkin diagram, resulting in
\be
	m_{4k+2,i}&=\overline{m_{4k+2}}{}_{,i}, \quad i < n=2k+1,\\
	m_{4k+2,2k+1}&=-\overline{m_{4k+2}}{}_{,2k+1}.
\ee

It is quite easy to write down the formula for the singlet in $\bfm_d\otimes \overline{\bfm_d}$ in GT basis. Indeed, it has to be singlet under all groups in~\ref{eq:groupsequence} and thus the above discussion implies that it must be of the form
\be\label{eq:singletansatz}
	\sum_{\seq{m}_d}\zeta_{\seq{m}_d}|\seq{m}_d\>\otimes|\overline{\seq{m}_d}\>,
\ee
where $\overline{\seq{m}_d}$ is obtained from GT pattern $\seq{m}_d$ by replacing all representations with their duals, and the coefficients $\zeta_{\seq{m}_d}$ are yet to be determined. Let us define
\be
	&(-1)^{\bfm_{2n+1}}=1,\\
	&(-1)^{\bfm_{4k}}=1,\\
	&(-1)^{\bfm_{4k+2}}=(-1)^{m_{4k+2,2k+1}},\\
	&(-1)^{\seq{m}_d}=\prod_{k=2}^d(-1)^{\bfm_k}.
\ee
With the choice of phases as in appendix~\ref{app:formulae}, the coefficients $\zeta_{\seq{m}_d}$ are proportional to $(-1)^{\seq{m}_d}$.\footnote{We have not proven this statement, but we have checked it on a large sample of representations in various dimensions.} In what follows, we will use the notation
\be
	\<\seq{m}_d,\overline{\seq{m}'_d}|0\>\equiv \zeta_{\seq{m}_d}\delta_{\seq{m}_d,{\seq{m}'_d}},
\ee
so that the singlet~\eqref{eq:singletansatz} can be written as
\be\label{eq:2jsymbolusage}
	\sum_{\seq{m}_d,{\seq{m'}_d}} \<\seq{m}_d,\overline{\seq{m}'_d}|0\>\,\,|\seq{m}_d\>\otimes|\overline{\seq{m}'_d}\>.
\ee
Note that this is a special case of Clebsch-Gordan coefficients, which suggests the normalization condition 
\be
	\sum_{\seq{u}_d,\seq{m}_d}\<0|\seq{m}_d,{\seq{u}_d}\>\<{\seq{m}_d},\seq{u}_d|0\>\equiv	\sum_{\seq{u}_d,\seq{m}_d}(\<\seq{m}_d,{\seq{u}_d}|0\>)^*\<{\seq{m}_d},\seq{u}_d|0\> =1.
\ee
It corresponds to the requirement that~\eqref{eq:2jsymbolusage} has unit norm. This implies
\be\label{eq:2jfinal}
	\<\seq{m}_d,\overline{\seq{m}'_d}|0\>\equiv \frac{(-1)^{\seq{m}_d}}{\sqrt{\dim \bfm_d}}\delta_{\seq{m}_d,{\seq{m}'_d}}.
\ee
Whenever $\bfm_d=\overline{\bfm_d}$ these coefficients have a definite symmetry under permutation of the two tensor factors. For bosonic representations they are always symmetric, while for fermionic they are symmetric if $d=0,1,7 \mod 8$ and anti-symmetric for $d=3,4,5\mod 8$, as can be easily verified by using the explicit formula above.\footnote{If these coefficients are symmetric, then the self-dual $\bfm_d$ is real and otherwise it is pseudo-real (quaternionic). This statement is specific to Euclidean signature (in Lorentzian dual and complex conjugate representations are not the same), but the symmetry properties are signature-independent.} Fermionic representations are never self-dual for $d=2,6\mod 8$.

\subsubsection{Vector representation}
\label{sec:reptheory:GelfandTsetlin:vector}
To gain some familiarity with GT bases, it is perhaps a good idea to start with the vector representation of $Spin(d)$. The vector representation is also going to play an extremely important role in section~\ref{sec:casimir}.

First of all, for $d>3$, under dimensional reduction the $d$-dimensional vector representation splits into two irreducible components -- a scalar and a $(d-1)$-dimensional vector. For $d=3$ we obtain three representations, the $+1,\bullet,-1$ representations of $Spin(2)$. This means that the GT basis for vector representation consists of the following elements,
\be
&|\myng{(1)},\bullet,\bullet,\ldots,\bullet,\bullet\>,\label{eq:vece1}\\
&|\myng{(1)},\myng{(1)},\bullet,\ldots,\bullet,\bullet\>,\label{eq:vece2}\\
&|\myng{(1)},\myng{(1)},\myng{(1)},\ldots,\bullet,\bullet\>,\\
&\vdots\nn\\
&|\myng{(1)},\myng{(1)},\myng{(1)},\ldots,\myng{(1)},\bullet\>,\\
&|\myng{(1)},\myng{(1)},\myng{(1)},\ldots,\myng{(1)},+1\>,\\
&|\myng{(1)},\myng{(1)},\myng{(1)},\ldots,\myng{(1)},-1\>.
\ee
Given that each sequence contains the $d-1$ irreps~\eqref{eq:admissiblesequences}, it is easy to see that the above gives exactly $d$ basis vectors.

Let us consider the element~\eqref{eq:vece1}. By definition, it lives in the trivial representation of $Spin(d-k)$ for $k\geq 1$ and thus has to be proportional to $e_1$. Similarly,~\eqref{eq:vece2} is invariant for $k\geq 2$ and thus has to be a linear combination of $e_1$ and $e_2$. Since it also has to be orthogonal to~\eqref{eq:vece1}, it can only be proportional to $e_2$. Repeating this argument, and making a choice of phases, we find 
\be
&|\myng{(1)},\bullet,\bullet,\ldots,\bullet,\bullet\>= (-1)^d e_1,\label{eq:vece1phase}\\
&|\myng{(1)},\myng{(1)},\bullet,\ldots,\bullet,\bullet\>= (-1)^{d-1} e_2,\label{eq:vece2phase}\\
&|\myng{(1)},\myng{(1)},\myng{(1)},\ldots,\bullet,\bullet\>= (-1)^{d-2} e_3,\\
&\vdots\nn\\
&|\myng{(1)},\myng{(1)},\myng{(1)},\ldots,\myng{(1)},\bullet\>= (-1)^{3} e_{d-2},\\
&|\myng{(1)},\myng{(1)},\myng{(1)},\ldots,\myng{(1)},+1\>=(-1)^2 \frac{e_{d-1}+i e_d}{\sqrt{2}},\label{eq:veceplus}\\
&|\myng{(1)},\myng{(1)},\myng{(1)},\ldots,\myng{(1)},-1\>=(-1)^1 \frac{e_{d-1}-i e_d}{\sqrt 2}.\label{eq:veceminus}
\ee
In the above expressions the phases are chosen to be consistent with the formulas for the matrix elements in appendix~\ref{app:formulae} and the interpretation that $M_{ij}$ ``rotates from $i$ to $j$'',
\be
M_{ij}e_i=e_j.
\ee
Note that according to our conventions for $Spin(2)$ representations described earlier, we have
\be
M_{d-1,d}|\myng{(1)},\myng{(1)},\myng{(1)},\ldots,\myng{(1)},\pm 1\>=\mp i |\myng{(1)},\myng{(1)},\myng{(1)},\ldots,\myng{(1)},\pm 1\>.
\ee 
This approach generalizes to other representations. In appendix~\ref{app:tensors} we consider the relation between GT and Cartesian bases in tensor representations of $Spin(d)$.

Let us now look at the inner product between vectors. Note that $\bfm_{4k+2,2k+1}$ is only non-zero in the GT patterns~\eqref{eq:veceplus} and~\eqref{eq:veceminus} and for $k=0$. Thus $(-1)^{\seq{m}_d}$ is $-1$ for these two patterns and $1$ otherwise. Finally, these two patterns are mutually dual, while all other patterns are self-dual, so that according to~\eqref{eq:2jsymbolusage} and~\eqref{eq:2jfinal} we get the following pairing, up to normalization,
\be
	&|\myng{(1)},\bullet,\ldots\>\otimes |\myng{(1)},\bullet,\ldots\>+|\myng{(1)},\myng{(1)},\bullet,\ldots\>\otimes |\myng{(1)},\myng{(1)},\bullet,\ldots\>+\ldots\nn\\
	&- |\myng{(1)},\ldots,\myng{(1)},+1\>\otimes |\myng{(1)},\ldots,\myng{(1)},-1\> - |\myng{(1)},\ldots,\myng{(1)},-1\>\otimes |\myng{(1)},\ldots,\myng{(1)},+1\>. 
\ee
From~\eqref{eq:vece1phase}-\eqref{eq:veceminus} we see that this is equal to
\be
	\sum_{i=1}^d e_i\otimes e_i,
\ee
which is the usual pairing between vectors.

\subsubsection{General representations in 3 dimensions}
\label{sec:reptheory:GelfandTsetlin:3d}
We now consider the case of general representations in $d=3$ ($n=1$). As before, the representations $\bfm_3$ are labeled by a (half-)integer $j\equiv m_{3,1}\geq 0$, which is the usual spin, and the representations $\bfm_{2}$ are labeled by a (half-)integer $m\equiv m_{2,1}$. The representations $\bfm_{2}\in\bfm_{3}$ are given by $m=-j,-j+1,\ldots,j$. The GT basis vectors are then
\be
|\seq{m}_3\rangle\equiv|\bfm_3,\bfm_2\rangle\equiv   |j,m\rangle.
\ee
We can choose conventions such that this coincides with the basis of $Spin(3)$ representations familiar from the theory of angular momenta. Indeed, let us first define the anti-Hermitian generators
\be
	I_\mu=\half\epsilon_{\mu\nu\lambda}M_{\nu\lambda},
\ee
which are then subject to the commutation relation (see appendix~\ref{app:conformalalgebra}),
\be
	[I_\mu,I_\nu]=\epsilon_{\mu\nu\lambda}I_\lambda.
\ee
Their Hermitian analogues $J_{\mu}=iI_\mu$ satisfy the familiar $Spin(3)$ commutation relations
\be
	[J_\mu,J_\nu]=i\epsilon_{\mu\nu\lambda}J_\lambda.
\ee
If we now define
\be\label{eq:indexrelabeling}
	\hat 1\equiv2,\quad \hat 2\equiv3,\quad \hat 3\equiv 1,
\ee
then the operators $J_{\hat \mu}$ satisfy the same commutation relations. By definition, we have
\be
	J_{\hat 3}|j,m\>=i I_{1}|j,m\>=i M_{23}|j,m\>=i(-im)|j,m\>=m|j,m\>.
\ee
We have performed the index relabeling~\eqref{eq:indexrelabeling} precisely so that $|j,m\>$ are eigenstates of $J_{\hat 3}$, making contact with standard angular momentum conventions. In particular, the standard~\cite{edmonds1996angular} formulas for action of $J_{\hat\mu}$ coincide with $d=3$ case of formulas in appendix~\ref{app:formulae}.

\subsubsection{General representations in 4 dimensions}
In $d=4$ ($n=2$), we have 
\be
	\bfm_4&=(\ell_1,\ell_2)=(l_L+l_R,l_L-l_R),\\
	\bfm_3&=j=|l_L-l_R|,|l_L-l_R|+1,\ldots,l_L+l_R,\quad\Leftrightarrow\quad j\in l_L\otimes l_R,\\
	\bfm_2&=m=-j,-j+1,\ldots, j,
\ee
and thus we can write
\be\label{eq:4dGTbasis}
	|\seq{m}_4\>\equiv |l_L,l_R; j,m\>.
\ee
It will be convenient to connect this to the basis which arises from the exceptional isomorphism $Spin(4)\simeq SU(2)\times SU(2)$. To define this latter basis, we write
\be
	Q_\mu\equiv M_{1\mu},\quad I_{\mu}\equiv \frac{1}{2}\epsilon_{\mu\nu\lambda}M_{\nu\lambda}, \quad\mu,\nu,\lambda\in \{2,3,4\},
\ee
where $\epsilon_{234}=1$.
Then the Hermitian operators 
\be
	J^L_\mu\equiv i I^L_\mu\equiv\frac{i}{2}(I_\mu+Q_\mu),\quad J^R_\mu\equiv iI^R_\mu\equiv\frac{i}{2}(I_\mu-Q_\mu)
\ee
obey the commutation relations
\be
	[J^L_\mu,J^L_\nu]&=i\epsilon_{\mu\nu\lambda} J^L_\lambda,\\
	[J^R_\mu,J^R_\nu]&=i\epsilon_{\mu\nu\lambda} J^R_\lambda,\\
	[J^L_\mu,J^R_\nu]&=0.
\ee
We can then define, similarly to 3 dimensions,
\be
	\hat 1\equiv 3,\quad \hat 2\equiv 4,\quad	\hat 3\equiv 2,
\ee
and construct the conventional basis states for the algebras $J^L_{\hat\mu},J^R_{\hat\mu}$,
\be\label{eq:4dchiralbasis}
	|l_L,m_L;l_R,m_R\>
\ee
subject to the usual condition
\be
	J^L_{\hat 3}|l_L,m_L;l_R,m_R\>&=m_L|l_L,m_L;l_R,m_R\>,\\
	J^R_{\hat 3}|l_L,m_L;l_R,m_R\>&=m_R|l_L,m_L;l_R,m_R\>.
\ee

Let us now relate the bases~\eqref{eq:4dGTbasis} and~\eqref{eq:4dchiralbasis}. First, note that the generators $J_{\hat\mu}\equiv iI_{\hat \mu}$ of the $Spin(3)$ which preserves the first axis are given by
\be
	J_{\hat\mu}=J_{\hat\mu}^L+J_{\hat\mu}^R,
\ee
and thus under this $Spin(3)$ the state~\eqref{eq:4dchiralbasis} transforms as a tensor product state in $l_L\otimes l_R$. We can therefore simply set
\be\label{eq:4dGTchiralrelation}
	|l_L,l_R;j,m\>\equiv\sum_{m_L+m_R=m} \<l_L,m_L;l_R,m_R|j,m\>|l_L,m_L;l_R,m_R\>,
\ee
where
\be
	\<l_L,m_L;l_R,m_R|j,m\>
\ee
are the Clebsch-Gordan coefficients of $Spin(3)$. It is easy to check that this definition is consistent with the definition of GT basis. Note that~\eqref{eq:4dGTchiralrelation} essentially fixes our choice of phases through the phases of $Spin(3)$ CG coefficients. The resulting phase conventions are consistent with appendix~\ref{app:formulae} if one uses CG coefficients $\<j_1,m_1;j_2;m_2|j,m\>$ which differ from~\cite{edmonds1996angular} by a factor of $i^{j-j_1-j_2}$.\footnote{These CG coefficients will still differ from the vector CG coefficients of~\ref{app:formulae} by a factor of $-i$ when $j=j_1$ and $j_2=1$, but the matrix elements in 4d will be consistent.}

For future reference, let us give the expression for $M_{12}=Q_2$. We have
\be\label{eq:4dchiralM12}
	M_{12}=Q_2=-iJ^L_2+iJ^R_2=-iJ^L_{\hat 3}+iJ^R_{\hat 3}.
\ee 

\subsection{Clebsch-Gordan coefficients and matrix elements}
\label{sec:reptheory:ClebshGordan}
In the next sections we will find that a lot of calculations (for example, three-point tensor structures and Casimir recursion relations) involve manipulations with Clebsch-Gordan coefficients (CG coefficients). In this section we therefore discuss the structure of these coefficients in GT bases.

CG coefficients essentially establish an equivalence between a tensor product and its decomposition into irreducible representations,
\be
	V_{\bfm_d^1}\otimes V_{\bfm_d^2}\simeq \bigoplus_{\bfm_d\in\bfm_d^1\otimes \bfm_d^2} V_{\bfm_d}.
\ee
More specifically, we have the relation between basis vectors
\be
	|\seq{m}^1_{d}\seq{m}^2_{d}\>=\sum_{\bfm_d\in\bfm_d^1\otimes\bfm_d^2}\sum_{\seq{m}_d}\<\seq{m}_d|\seq{m}^1_{d}\seq{m}^2_{d}\>\,\,|\seq{m}_d\>,
\ee
where $\<\seq{m}_d|\seq{m}^1_{d}\seq{m}^2_{d}\>$ are the CG coefficients. This equation has to be modified somewhat if there are multiplicities in the tensor product,
\be\label{eq:CGcoeffs}
	|\seq{m}^1_{d}\seq{m}^2_{d}\>=\sum_{(\bfm_d,t)\in\bfm_d^1\otimes\bfm_d^2}\sum_{\seq{m}_d}\<\seq{m}_d,t|\seq{m}^1_{d}\seq{m}^2_{d}\>\,\,|\seq{m}_d,t\>.
\ee
Here $t$ counts the possible degeneracy. Inverse transformation is given by
\be
	|\seq{m}_d,t\>=\sum_{\seq{m}^1_{d}\seq{m}^2_{d}}\<\seq{m}^1_{d}\seq{m}^2_{d}|\seq{m}_d,t\>\,\,|\seq{m}^1_{d}\seq{m}^2_{d}\>,
\ee
where $\<\seq{m}^1_{d}\seq{m}^2_{d}|\seq{m}_d,t\>=\<\seq{m}_d,t|\seq{m}^1_{d}\seq{m}^2_{d}\>^*$.
Note that there is an ambiguity in the definition of CG coefficients. Indeed, the decomposition
\be
	|\seq{m}^1_{d}\seq{m}^2_{d}\>=\sum_{(\bfm_d,t)\in\bfm_d^1\otimes\bfm_d^2}\sum_{\seq{m}_d,t'}U_{tt'}\<\seq{m}_d,t'|\seq{m}^1_{d}\seq{m}^2_{d}\>\,\,|\seq{m}_d,t\>,
\ee
where $U$ is a unitary matrix, is also perfectly fine from the point of view of $Spin(d)$ invariance. One thus has to fix this freedom for every choice of $\bfm_d^1$ and $\bfm_d^2$. We will not try to fix the general conventions here, and work on a case-by-case basis in the examples.

GT bases exhibit a set of relations between the CG coefficients of the nested groups~\eqref{eq:groupsequence}. Indeed, let us write the GT patters in CG coefficients~\eqref{eq:CGcoeffs} in the form $\seq{m}_d=\bfm_d\,\seq{m}_{d-1}$,
\be
	\<\seq{m}_d,t|\seq{m}^1_{d}\seq{m}^2_{d}\>\equiv\<\bfm_d\,\seq{m}_{d-1},t|\bfm_d^1\,\seq{m}^1_{d-1};\bfm_d^2\,\seq{m}^2_{d-1}\>.
\ee
Thinking about $Spin(d-1)$-invariance, we see that must necessarily have
\be\label{eq:isoscalar}
	\<\bfm_d\,\seq{m}_{d-1}\,t|\bfm_d^1,\seq{m}^2_{d-1};\bfm_d^2\,\seq{m}^2_{d-1}\>=
	\sum_{t'}\left(
	{\bfm_d\atop\bfm_{d-1}}
	\Bigg\vert
	{\bfm_d^1\atop\bfm_{d-1}^1}
	{\bfm_d^2\atop\bfm_{d-1}^2}
	\right)_{tt'}\<\seq{m}_{d-1},t'|\seq{m}_{d-1}^1\seq{m}_{d-1}^2\>
\ee
where the constants
\be
	\left(
	{\bfm_d\atop\bfm_{d-1}}
	\Bigg\vert
	{\bfm_d^1\atop\bfm_{d-1}^1}
	{\bfm_d^2\atop\bfm_{d-1}^2}
	\right)_{tt'}
\ee
are the so-called $Spin(d):Spin(d-1)$ isoscalar factors,\footnote{Also known as reduced CG, reduced Wigner coefficients, or reduction factors.} while $\<\seq{m}_{d-1},t'|\seq{m}_{d-1}^1\seq{m}_{d-1}^2\>$ are the CG coefficients of $Spin(d-1)$. This can be iterated, and since the CG coefficients of $Spin(2)$ are extremely simple,
\be
	\<m|m^1m^2\>=\delta_{m,m^1+m^2},
\ee
it follows that the knowledge of CG coefficients of $Spin$ groups is equivalent to the knowledge of the isoscalar factors.

For example, the $Spin(3):Spin(2)$ isoscalar factors are essentially the $Spin(3)$ CG coefficients, due to the aforementioned triviality of $Spin(2)$ CG coefficients. One can show that the $Spin(4):Spin(3)$ isoscalar factors are essentially equivalent to $Spin(3)$ $9j$ symbols~\cite{ninej}.

For our applications we in principle need the most general CG coefficients of $Spin(d-1)$ groups -- simply the knowledge of all possible conformally-invariant three-point tensor structures already implies the knowledge of all possible $Spin(d-1)$ CG coefficients (see section~\ref{sec:structure:Contribution}). We are not aware of a general formula for $Spin(d-1)$ CG coefficients valid for general $d$.\footnote{See~\cite{GouldI,GouldII} for partial progress in this direction.} For the most physically relevant cases $d=4,3$ one can use the well-known CG coefficients of $Spin(3)\simeq SU(2)$ or the trivial CG coefficients of $Spin(2)\simeq U(1)$. Due to the exceptional isomorphism $Spin(4)\simeq SU(2)\times SU(2)$, we also know the general CG coefficients of $Spin(d-1)$ for $d=5$. Let us note that the case $d\geq 6$ is qualitatively different since tensor products in $Spin(5)$ and larger groups are not multiplicity-free. Luckily, for each particular choice of a four-point function there is only a finite number of relevant three-point tensor structures and thus also of $Spin(d-1)$ CG coefficients. For any given tensor product, the problem of finding CG coefficients is a finite-dimensional linear algebra problem and can in principle be solved on a computer, although phase conventions and resolution of multiplicities need to be carefully addressed. See~\cite{Caprio:2010tj} for an approach to $Spin(5)$ CG coefficients.

For the applications to Casimir recursion relations, we will need a special infinite class of CG coefficients of $Spin(d)$ -- the CG coefficients involving a vector representation. The good news are that these CG coefficients are known for general $d$ in closed form.

\paragraph{$Spin(d)$ matrix elements and Clebsch-Gordan coefficients with vector representation}

It turns out that Clebsch-Gordan coefficients for vector representation are closely related to the matrix elements of $Spin(d)$ generators. Indeed, let us consider the matrix elements of $M_{1\mu}$,
\be
	M_{1\mu}|\seq{m}_d\>=\sum_{\seq{m}'_{d}}\<\seq{m}'_{d}|M_{1\mu}|\seq{m}_d\> \, |\seq{m}'_{d}\>,\qquad \bfm'_d=\bfm_d.
\ee
The piece $M_{1\mu}|\seq{m}_d\>$ transforms under $Spin(d-1)$ in the representation $\myng{(1)}\otimes \bfm_{d-1}$. The vectors on the right, on the other hand, transform in irreducible representations of $Spin(d-1)$. For fixed $\bfm_d,\bfm_{d-1}$ this therefore has precisely the form required of a CG decomposition, so that we have
\be
	\<\seq{m}'_{d}|M_{1\mu}|\seq{m}_d\>=\Bigg(
		{\bfm_d\atop \bfm'_{d-1}}
		\Bigg\vert
		M^{\myng{(1)}}
		\Bigg\vert
		{\bfm_d\atop \bfm_{d-1}}
	\Bigg)
	\<\seq{m}'_{d-1}|\seq{m}_{d-1},\mu\>
\ee
for some constants
\be\label{eq:reducedmatelement}
\Bigg(
{\bfm'_d\atop \bfm'_{d-1}}
\Bigg\vert
M^{\myng{(1)}}
\Bigg\vert
{\bfm_d\atop \bfm_{d-1}}
\Bigg)
\ee
known as reduced matrix elements. This is essentially a version of Wigner-Eckart theorem. Note that the tensor product with vector representation is always multiplicity free and thus we don't need any extra labels. This follows from Brauer's formula~\cite{Brauer} and the fact that all weights in the vector representation have multiplicity $1$. The $\myng{(1)}$ label for $M$ is supposed to indicate that we are looking at $M_{1\mu}$ which is a vector under $Spin(d-1)$. 

Let us consider an example by setting $\mu=2$ which is equivalent to $\mu=[\myng{(1)},\bullet,\ldots]$ in terms of GT patterns. We then find
\be\label{eq:M12reduced}
	\<\seq{m}'_{d}|M_{12}|\seq{m}_{d}\>=&
	(-1)^{d-1}\Bigg(
	{\bfm_d\atop \bfm'_{d-1}}
	\Bigg\vert
	M^{\myng{(1)}}
	\Bigg\vert
	{\bfm_d\atop \bfm_{d-1}}
	\Bigg)
	\<\seq{m}'_{d-1}|\seq{m}_{d-1};\myng{(1)},\bullet,\ldots\>\nn\\
	=&
	(-1)^{d-1}\Bigg(
	{\bfm_d\atop \bfm'_{d-1}}
	\Bigg\vert
	M^{\myng{(1)}}
	\Bigg\vert
	{\bfm_d\atop \bfm_{d-1}}
	\Bigg)
	\left(
	{\bfm'_{d-1}\atop\bfm'_{d-2}}
	\Bigg\vert
	{\bfm_{d-1}\atop\bfm_{d-2}}
	{\myng{(1)}\atop\bullet}
	\right)
	\<\seq{m}'_{d-2}|\seq{m}_{d-2};\bullet,\ldots\>\nn\\
	=&	(-1)^{d-1}\Bigg(
	{\bfm_d\atop \bfm'_{d-1}}
	\Bigg\vert
	M^{\myng{(1)}}
	\Bigg\vert
	{\bfm_d\atop \bfm_{d-1}}
	\Bigg)
	\left(
	{\bfm'_{d-1}\atop\bfm'_{d-2}}
	\Bigg\vert
	{\bfm_{d-1}\atop\bfm_{d-2}}
	{\myng{(1)}\atop\bullet}
	\right)\delta_{\seq{m}_{d-2},\seq{m}'_{d-2}}.
\ee
Here we used the definition of the isoscalar factor~\eqref{eq:isoscalar} and the triviality of CG coefficients when one of the factors is the trivial representation. We also made use of the relation~\eqref{eq:vece2phase}. Note that this implies the constraint $\bfm'_{d-1}\in\myng{(1)}\otimes \bfm_{d-1}$. Due to the structure of the nested sequence~\eqref{eq:groupsequence} the matrix elements of $M_{k,k+1}$ for all $1\leq k\leq d-1$ follow from the matrix elements of $M_{12}$ for $Spin(d-k+1)$. It is an easy exercise to show that $M_{k,k+1}$ generate the whole Lie algebra of $Spin(d)$.

We therefore find that the reduced matrix elements~\eqref{eq:reducedmatelement} and the simplest vector isoscalar factors
\be\label{eq:isoscalarelementary}
	\left(
	{\bfm_{d}\atop\bfm_{d-1}}
	{\myng{(1)}\atop\bullet}
	\Bigg\vert
	{\bfm'_{d}\atop\bfm_{d-1}}
	\right)
\ee
allow the computation of the most general $Spin(d)$ matrix elements. There exist relatively simple closed-form expressions for these quantities~\cite{GouldFirst,ReductionFactors}, which we provide in appendix~\ref{app:formulae} for the ease of reference.\footnote{Note that our phase conventions differ from those in~\cite{GouldFirst,ReductionFactors}.} 

These quantities in fact also completely determine the vector CG coefficients. Indeed, given the isoscalar factor~\eqref{eq:isoscalarelementary}, it only remains to find the second isoscalar factor\footnote{For $d=3$ we can have $(\pm 1)$ instead of lower $\myng{(1)}$ in~\eqref{eq:isocalarcomplete}. The corresponding isoscalar factors can be obtained completely analogously. See appendix~\ref{app:formulae:3dcomment}.}
\be\label{eq:isocalarcomplete}
	\left(
{\bfm_{d-1}\atop\bfm_{d-2}}
{\myng{(1)}\atop\myng{(1)}}
\Bigg\vert
{\bfm'_{d-1}\atop\bfm'_{d-2}}
\right).
\ee
It can be easily computed by considering the expression
\be
	\<\seq{m}_{d};\myng{(1)},\bullet,\ldots|M_{12}|\seq{m}'_{d}\>
\ee
and evaluating it via isoscalar factors and reduced matrix elements in two different ways (acting with $M$ on the left and on the right). Action on the left produces, among other terms, the term
\be
	\<\seq{m}_{d};\myng{(1)},\myng{(1)},\bullet,\ldots|\seq{m}'_{d}\>,
\ee
which is proportional to the sought for isoscalar factor. See appendix~\ref{app:formulae:Isoscalar} for details.

\section{Structure of spinning correlation functions and conformal blocks}
\label{sec:structure}

In this section we apply the formalism of GT bases to study the general structure of radially-quantized correlators or conformal blocks. At this stage, no distinction is made between correlation functions and individual conformal blocks, so we use these two terms interchangeably.

\subsection{Contribution of a $\bR\times Spin(d)$-multiplet}
\label{sec:structure:Contribution}

Consider a 4-point correlation function, radially quantized so that the points $1$ and $2$ lie inside the unit sphere, whereas the points $3$ and $4$ lie outside (or on) the unit sphere. One can then insert a complete basis of states on the unit sphere, organized in representations of $\bR\times Spin(d)$ (dilatations $\times$ rotations), and ask what is the contribution of a single representation. This question was answered in \cite{Hogervorst:2013sma} for four-point functions with external scalar operators, exchanging traceless-symmetric tensors on the unit sphere (the only representations allowed in this this case). The case of four-point functions of tensor operators was addressed in~\cite{Costa:2016xah}. Unfortunately, as mentioned in the introduction, the approach of~\cite{Costa:2016xah} requires a non-trivial amount of case-by-case analysis and the knowledge of irreducible projectors. The goal of this section is to give a more general alternative treatment.

For concreteness, we will work in the radial kinematics of \cite{Hogervorst:2013sma}.\footnote{The same approach also works in other kinematics. For examples, we will switch to Dolan-Osborn~\cite{DO1,DO2} kinematics in section~\ref{sec:casimir}. The analysis in that case is only slightly different due to the presence of an operator at infinity.} Namely, we chose an orthonormal basis in $\bR^d$, labeling the axes by integers from $1$ to $d$, and we introduce a complex coordinate $w$ in plane 1-2 as
\be
	w=x_1+ix_2.
\ee
We then place all four operators in this plane, setting their coordinates to 
\be
	w_1=-\rho,\quad w_2=\rho\quad w_3=1,\quad w_4=-1,
\ee
for some $\rho\in \bC$. Any non-coincident configuration of four points can be brought to a configuration of the above form by a conformal transformation, with $\rho$ being related to the familiar cross-ratios $u$ and $v$. We assume $|\rho|<1$.

We also fix the sequence of groups \eqref{eq:groupsequence}, defining $Spin(d-k)$ to be the subgroup of $Spin(d)$ which fixes the first $k$ axes. This defines for us Gelfand-Tsetlin bases for the representations of $Spin(d)$. We will accordingly denote the primary operators by
\be
	\cO_i^{\seq{m}^i_d}(w_i),
\ee
where the sequences $\seq{m}^i_d$ label the Gelfand-Tsetlin basis vectors as in section~\ref{sec:reptheory:GelfandTsetlin}, and we use the upper index $i$ to label the operators in order to avoid confusion with the dimension label, $\seq{m}^i_d=\bfm^i_d,\bfm^i_{d-1},\ldots,\bfm^i_2$.

We are interested in the radially-quantized four-point function 
\be\label{eq:fourptstart}
	\langle 0|\cO_4^{\seq{m}^4_{d}}(-1) \cO_3^{\seq{m}^3_{d}}(1)\cO_2^{\seq{m}^2_{d}}(\rho)\cO_1^{\seq{m}^1_{d}}(-\rho)|0\rangle.
\ee
It turns out that it is more convenient to work with
\be\label{eq:fourptconvenient}
	\langle 0|\cO_4^{\seq{m}^4_{d}}(-1) \cO_3^{\seq{m}^3_{d}}(1)r^{D}e^{\theta M_{12}}\cO_2^{\seq{m}^2_{d}}(1)\cO_1^{\seq{m}^1_{d}}(-1)|0\rangle,
\ee
where $\rho=r e^{i\theta}$, $D$ is the dilatation operator and $M_{\mu\nu}$ is the anti-hermitian rotation generator in the plane $\mu$-$\nu$.\footnote{See appendix~\ref{app:conformalalgebra} for our conventions on conformal algebra. Our definition of $M_{\mu\nu}$ differs by a sign from e.g.~\cite{Simmons-Duffin:2016gjk}.} The relation between~\eqref{eq:fourptstart} and~\eqref{eq:fourptconvenient} is given by
\begin{multline}
	\langle 0|\cO_4^{\seq{m}^4_{d}}(-1) \cO_3^{\seq{m}^3_{d}}(1)r^{D}e^{\theta M_{12}}\cO_2^{\seq{m}^2_{d}}(1)\cO_1^{\seq{m}^1_{d}}(-1)|0\rangle=\\=
	r^{\Delta_1+\Delta_2}\sum_{\seq{m}^{\prime 1}_d,\seq{m}^{\prime 2}_d}R^{\seq{m}^1_{d}}_{\seq{m}^{\prime 1}_d}(\theta)R^{\seq{m}^2_{d}}_{\seq{m}^{\prime 2}_d}(\theta)\langle 0|\cO_4^{\seq{m}^4_{d}}(-1) \cO_3^{\seq{m}^3_{d}}(1)\cO_2^{\seq{m}^{\prime 2}_d}(\rho)\cO_1^{\seq{m}^{\prime 1}_d}(-\rho)|0\rangle,
\end{multline}
where $R$ are the matrix elements of the rotations in the plane 1-2 in Gelfand-Tsetlin basis,
\be
	R^{\seq{m}^i_{d}}_{\seq{m}^{\prime i}_d}(\theta)=\<\seq{m}^{\prime i}_d|e^{\theta M_{12}}|\seq{m}^i_{d}\>.
\ee
Recall that according to our conventions the primed patterns belong to the same representations as unprimed ones. Clearly, the two forms can be used interchangeably. The reader may recognize the factor $r^{-\Delta_1-\Delta_2}$, which appears in many formulas for scalar four-point functions, and is often stripped off as in here by multiplying by $r^{+\Delta_1+\Delta_2}$. The matrices $R$ play a similar role for the spinning degrees of freedom.\footnote{Importantly, the action of $R$ here is only on the labels of the external operators. Because it commutes with the stabilizer group $Spin(d-2)$ of four points, it can be though of as a change of the basis of four-point tensor structures. We study the matrix elements such as $R$ further in sections~\ref{sec:structure:Contribution:matrixelements} and~\ref{sec:structure:Pfunctions}.}

Consider now a contribution from a $\bR\times Spin(d)$ multiplet with scaling dimension $\Delta$ and in representation $\bfm_d$ of $Spin(d)$,
\begin{multline}\label{eq:bigscaryformula}
	\sum_{\seq{m}_{d}}\langle 0|\cO_4^{\seq{m}^4_{d}}(-1) \cO_3^{\seq{m}^3_{d}}(1)|\Delta,\seq{m}_{d}\rangle\langle\Delta,\seq{m}_{d}|r^{D}e^{\theta M_{12}}\cO_2^{\seq{m}^2_{d}}(1)\cO_1^{\seq{m}^1_{d}}(-1)|0\rangle=\\=
	\sum_{\seq{m}_{d},\seq{m}'_{d}} r^\Delta\langle 0|\cO_4^{\seq{m}^4_{d}}(-1) \cO_3^{\seq{m}^3_{d}}(1)|\Delta,\seq{m}_{d}\rangle\langle\seq{m}_{d}|e^{\theta M_{12}}|\seq{m}'_{d}\rangle\langle\Delta,\seq{m}'_{d}|\cO_2^{\seq{m}^2_{d}}(1)\cO_1^{\seq{m}^1_{d}}(-1)|0\rangle.
\end{multline}
Here $\bfm'_d=\bfm_d$. This expression consists of three main ingredients: the two three-point functions
\be\label{eq:threeptfunctions}
\langle 0|\cO_4^{\seq{m}^4_{d}}(-1) \cO_3^{\seq{m}^3_{d}}(1)|\Delta,\seq{m}_{d}\rangle\quad\text{and}\quad\langle\Delta,\seq{m}'_{d}|\cO_2^{\seq{m}^2_{d}}(1)\cO_1^{\seq{m}^1_{d}}(-1)|0\rangle,
\ee
and the matrix elements
\be\label{eq:matrixelements}
\langle\seq{m}_{d}|e^{\theta M_{12}}|\seq{m}'_{d}\rangle.
\ee
In order to proceed further, we need to understand the structure of these objects.

\subsubsection{Three-point functions} The three-point functions \eqref{eq:threeptfunctions} are some tensors in the Gelfand-Tsetlin indices, whose values are constrained by the requirement of conformal invariance. To be precise, for three-point functions involving $\bR\times Spin(d)$ multiplets, the only intrinsic restrictions come from $\bR\times Spin(d)$ invariance.\footnote{The extrinsic restrictions, relating the contribution of the descendant multiplets to the primary, are discussed in section~\ref{sec:casimir}.} Of these, only the $Spin(d-1)$ subgroup which fixes the first axis imposes the restriction directly on \eqref{eq:threeptfunctions}, while the other generators in $\bR\times Spin(d)$ can be used to determine the values of these three-point functions for different positions of $\cO_i$ (we have essentially done this above). Even in the case when the $\bR\times Spin(d)$ multiplet in question is a conformal primary, $Spin(d-1)$-invariance is the only restriction on the tensors \eqref{eq:threeptfunctions} \cite{Kravchuk:2016qvl}.

In particular, the allowed tensor structures for, e.g. 
\be
	\langle\Delta,\seq{m}'_{d}|\cO_2^{\seq{m}^2_{d}}(1)\cO_1^{\seq{m}^1_{d}}(-1)|0\rangle
\ee
are in one-to-one correspondence with the $Spin(d-1)$ invariant subspace
\be
	\left(\overline{\bfm_d}\otimes \bfm^1_d\otimes\bfm^2_d\right)^{Spin(d-1)},
\ee
where the bar indicates taking the dual\footnote{Equivalently complex-conjugate, since all representations of compact $Spin(d)$ are unitary.} representation. Because dimensional reduction is multiplicity-free, such singlets are in one-to-one correspondence with singlets in
\be
	\overline{\bfm_{d-1}'}\otimes \bfm^1_{d-1}\otimes\bfm^2_{d-1}
\ee
over all $\bfm'_{d-1}\in\,\bfm_d$, $\bfm^i_{d-1}\in\,\bfm^i_d$. Such a singlet exists whenever $\bfm'_{d-1}$ appears in $\bfm^1_{d-1}\otimes\bfm^2_{d-2}$, in which case we write
\be\label{eq:tensormultiplicity}
	(\bfm'_{d-1},t')\in \bfm^1_{d-1}\otimes\bfm^2_{d-1},
\ee
where the extra label $t'$ is needed if $\bfm'_{d-1}$ appears in the tensor product with multiplicity.\footnote{If $d\leq 5$, then tensor products in $Spin(d-1)$ are multiplicity-free and the sum over $t'$ can be dropped.} If~\eqref{eq:tensormultiplicity} holds, we can build an invariant using $Spin(d-1)$ Clebsch-Gordan coefficients. More explicitly, we have
\be\label{eq:right3pt}
	\langle\Delta,\seq{m}'_{d}|\cO_2^{\seq{m}^2_{d}}(1)\cO_1^{\seq{m}^1_{d}}(-1)|0\rangle=\sum_{t'}\lambda^{\bfm^1_{d-1},\bfm^2_{d-1}}_{\bfm'_{d-1},t'}\<\seq{m}'_{d-1},t'|\seq{m}^1_{d-1},\seq{m}^2_{d-1}\>,
\ee
where $\lambda$'s are the three-point coefficients unconstrained by symmetry, and we recall that $\seq{m}_{d-1}$ is defined as
\be
	\seq{m}_{d}=\bfm_d,\bfm_{d-1},\ldots,\bfm_2\Longrightarrow\seq{m}_{d-1}\equiv\bfm_{d-1},\bfm_{d-2},\ldots,\bfm_2.
\ee
It is understood that if $\bfm'_{d-1} \notin \bfm^1_{d-1}\otimes\bfm^2_{d-1}$, then the Clebsch-Gordan coefficient vanishes and the corresponding $\lambda$ is undefined. 

Analogously, for the second three-point function we have\footnote{The coefficients $\bar\lambda$ are in general not complex conjugates of $\lambda$.}
\be\label{eq:left3pt}
\langle 0|\cO_4^{\seq{m}^4_{d}}(-1) \cO_3^{\seq{m}^3_{d}}(1)|\Delta,\seq{m}_{d}\rangle=\sum_t \bar\lambda^{\bfm_{d-1}^3,\bfm_{d-1}^4}_{\bfm_{d-1},t}\<0|\seq{m}^3_{d-1},\seq{m}^4_{d-1},\seq{m}_{d-1},t\>,
\ee
where we now have a $3j$ symbol instead of Clebsch-Gordan coefficients (the distinction is of course rather formal).

Note that~\eqref{eq:right3pt} and~\eqref{eq:left3pt} give a somewhat unusual way of writing the three-point function, since the spin indices of the operators directly select which three-point coefficients $\lambda$ appear in the right hand side. A perhaps more intuitive equivalent form of~\eqref{eq:right3pt} is
\be
	\sum_{\widetilde\bfm^i_{d-1}}\sum_{\widetilde{\bfm}'_{d-1},t'}\lambda^{\widetilde\bfm^1_{d-1},\widetilde\bfm^2_{d-1}}_{\widetilde{\bfm}'_{d-1},t'}\left\{\delta_{\bfm^1_{d-1},\widetilde\bfm^1_{d-1}}\delta_{\bfm^2_{d-1},\widetilde\bfm^2_{d-1}}\delta_{\bfm'_{d-1},\widetilde{\bfm}'_{d-1}}\<\seq{m}'_{d-1},t'|\seq{m}^1_{d-1},\seq{m}^2_{d-1}\>\right\},
\ee
where the object in the curly braces is the three point tensor structure, and it is made explicit that the three-point coefficients are labeled by two $Spin(d-1)$ representations $\widetilde\bfm^1_{d-1}$ and $\widetilde\bfm^2_{d-1}$ and a pair $(\widetilde{\bfm}_{d-1}',t')\in\widetilde\bfm^1_{d-1}\otimes\widetilde\bfm^2_{d-1}$. We will sometimes use a shorthand notation to denote such composite labels. Namely, for the right three point function we use the label
\be\label{eq:3ptrightshorthand}
	a&=(\widetilde\bfm_{d-1}^1,\widetilde\bfm_{d-1}^2,\widetilde{\bfm}'_{d-1},t'),\qquad 
	(\widetilde{\bfm}'_{d-1},t')\in \widetilde\bfm_{d-1}^1\otimes\widetilde\bfm_{d-1}^2.
\ee	
Similarly, for the left three-point function we use
\be\label{eq:3ptleftshorthand}
	b&=(\widetilde\bfm_{d-1}^3,\widetilde\bfm_{d-1}^4,\widetilde{\bfm}_{d-1},t),
	\qquad (\overline{\widetilde{\bfm}_{d-1}},t)\in \widetilde\bfm_{d-1}^3\otimes\widetilde\bfm_{d-1}^4.
\ee

It is instructive to consider the case of 3 dimensions. In this case, we are considering the three-point functions
\be
	\< \Delta, j',m'|\cO_2^{j_2,m_2}(1)\cO_1^{j_1,m_1}(-1)|0\>.
\ee
The $Spin(2)$ invariance basically tells us that the spin projection has to be conserved, $m'=m_1+m_2$, and the Spin(2) Clebsch-Gordan coefficients are
\be
	\<m'|m_1,m_2\>=\delta_{m',m_1+m_2}.
\ee
We can therefore write
\be\label{eq:threeptright3d}
	\< \Delta, j',m'|\cO_2^{j_2,m_2}(1)\cO_1^{j_1,m_1}(-1)|0\>=\delta_{m',m_1+m_2}\lambda^{m_1,m_2}_{m'}.
\ee
Analogously, for the other three-point function we have
\be\label{eq:threeptleft3d}
	\<0|\cO_4^{j_4,m_4}(-1)\cO_3^{j_3,m_3}(-1)|\Delta,j,m\>=\bar\lambda^{m_3,m_4}_{m}\delta_{0,m_3+m_4+m}.
\ee
We discuss the 3d case further in section~\ref{sec:structure:General3dCorrelators}.

In order to study the most general four-point functions, we need to know the most general three-point functions~\eqref{eq:right3pt} and~\eqref{eq:left3pt} and thus the most general $Spin(d-1)$ CG coefficients. Unfortunately, as discussed in section~\ref{sec:reptheory:ClebshGordan}, to the best of our knowledge there is no general closed-form expression for such CG coefficients valid for general $d$ available in the literature, but there are important special cases when such expressions are available. 

Besides the cases considered in section~\ref{sec:reptheory:ClebshGordan}, an important scenario is when, say, $\bfm^1_d=\bfm^4_d=\bullet$, in which case the required CG coefficients are trivial in any $d$. This happens, for example, in a certain choice of four-point functions for the so-called seed blocks. These are the simplest conformal blocks which exchange a given intermediate $Spin(d)$ representation $\bfm_d$. We discuss this case further in section~\ref{sec:structure:SeedConformalBlocks}.

\subsubsection{Matrix elements}
\label{sec:structure:Contribution:matrixelements}
Consider now the matrix elements~\eqref{eq:matrixelements}. An important feature is that the $Spin(d)$ element $e^{\theta M_{12}}$ commutes with the standard $Spin(d-2)$ subgroup which fixes the axes 1 and 2. On the other hand, the $Spin(d)$ representation $\bfm_d$ decomposes into irreducibles under $Spin(d-2)$, and by Schur's lemma this implies that $e^{\theta M_{12}}$ acts by identity times a constant inside of these irreducible components. More precisely, we have
\be\label{eq:Pdefinition}
\<\seq{m}_{d}|e^{\theta M_{12}}|\seq{m}'_{d}\>=P^{\bfm_d,\bfm_{d-2}}_{\bfm_{d-1},\bfm'_{d-1}}(\theta)\delta_{\seq{m}_{d-2},\seq{m}'_{d-2}}.
\ee
One can arrive at the same conclusion by examining~\eqref{eq:M12reduced}. The functions $P^{\bfm_d,\bfm_{d-2}}_{\bfm_{d-1},\bfm'_{d-1}}(\theta)$ will play the role of Gegenbauer polynomials for the spinning conformal blocks. We will describe their structure, basic properties, and how to compute them in section~\ref{sec:structure:Pfunctions}. For now, note that they are labeled by an $Spin(d)$ representation $\bfm_d$, two $Spin(d-1)$ representations $\bfm_{d-1},\bfm'_{d-1}\in\,\bfm_d$, and one $Spin(d-2)$ representation $\bfm_{d-2}\in\,\bfm_{d-1},\bfm'_{d-1}$.

It is again useful to look at the case of three dimensions. Here, $Spin(d-2)=Spin(1)$ is trivial, and according to \eqref{eq:Pdefinition} we have (recall that $\bfm_3\equiv j$ and $\bfm_2\equiv m$)
\be\label{eq:P3d}
	P^{j}_{m,m'}(\theta)=\<j,m|e^{\theta M_{12}}|j,m'\>=\<j,m|e^{-i\theta J_{\hat 2}}|j,m'\>=d^j_{m,m'}(-\theta),
\ee
where $d^j_{m,m'}(\theta)$ is the small Wigner $d$-matrix familiar from the representation theory of $Spin(3)$. For other examples see section~\ref{sec:structure:Pfunctions} and appendix~\ref{app:tensors}.

\subsubsection{Putting everything together} We can now combine~\eqref{eq:right3pt}, \eqref{eq:left3pt} and \eqref{eq:Pdefinition} to rewrite~\eqref{eq:bigscaryformula} in the following terrifying form,
\be\label{eq:biggerscaryformula}
\sum_{\seq{m}_{d}}\langle 0|\cO_4^{\seq{m}^4_{d}}(-1) \cO_3^{\seq{m}^3_{d}}(1)|&\Delta,\seq{m}_{d}\rangle\langle\Delta,\seq{m}_{d}|r^{D}e^{\theta M_{12}}\cO_2^{\seq{m}^2_{d}}(1)\cO_1^{\seq{m}^1_{d}}(-1)|0\rangle=\nn\\
	=&
	\sum_{\widetilde{\bfm}^i_{d-1}}\sum_{\bfm_{d-1},t\atop \bfm'_{d-1},t'}\sum_{\bfm_{d-2}}
	\lambda^{\widetilde\bfm^1_{d-1},\widetilde\bfm^2_{d-1}}_{\bfm'_{d-1},t'}
	\bar\lambda^{\widetilde\bfm_{d-1}^3,\widetilde\bfm_{d-1}^4}_{\bfm_{d-1},t}
	r^\Delta P^{\bfm_d,\bfm_{d-2}}_{\bfm_{d-1},\bfm'_{d-1}}(\theta)\times\nn\\
	&\times\left[
	{\seq{m}^3_{d}\atop\seq{m}^4_{d}}\Big\vert
		{\widetilde{\bfm}^3_{d-1}\atop\widetilde{\bfm}^4_{d-1}}
		\,\bfm_{d-1},t
		\Big\vert
		\bfm_{d-2}
		\Big\vert
		\bfm'_{d-1},t'\,{\widetilde{\bfm}^1_{d-1}\atop\widetilde{\bfm}^2_{d-1}}\Big\vert
	{\seq{m}^1_{d}\atop\seq{m}^2_{d}}
	\right],
\ee
where following selection rules on the summation variables hold,
\be\label{eq:selectionrules}
	\widetilde\bfm_{d-1}^i&\in \bfm_d^i,\nn\\
	({\bfm_{d-1}'},t')&\in\widetilde\bfm_{d-1}^1\otimes \widetilde\bfm_{d-1}^2,\nn\\
	(\overline{\bfm_{d-1}},t)&\in \widetilde\bfm_{d-1}^3\otimes \widetilde\bfm_{d-1}^4,\nn\\
	\bfm_{d-2}&\in \bfm_{d-1},\bfm'_{d-1}\in \bfm_d.
\ee
Using the shorthand notation~\eqref{eq:3ptrightshorthand} and~\eqref{eq:3ptleftshorthand} for the three-point tensor structures, we can rewrite~\eqref{eq:biggerscaryformula} as
\be\label{eq:smallunicornformula}
=\sum_{a,b}\sum_{\bfm_{d-2}}
\lambda^{a}
\bar\lambda^{b}
r^\Delta P^{\bfm_d,\bfm_{d-2}}_{\bfm_{d-1},\bfm'_{d-1}}(\theta)\times\left[
{\seq{m}^3_{d}\atop\seq{m}^4_{d}}\Big\vert b
\Big\vert
\bfm_{d-2}
\Big\vert
a\Big\vert
{\seq{m}^1_{d}\atop\seq{m}^2_{d}}
\right].	
\ee
We have also introduced a four-point tensor structure 
\be\label{eq:fourptdemo}
\left[
{\seq{m}^3_{d}\atop\seq{m}^4_{d}}\Big\vert
{\widetilde{\bfm}^3_{d-1}\atop\widetilde{\bfm}^4_{d-1}}
\,\bfm_{d-1},t
\Big\vert
\bfm_{d-2}
\Big\vert
\bfm'_{d-1},t'\,{\widetilde{\bfm}^1_{d-1}\atop\widetilde{\bfm}^2_{d-1}}\Big\vert
{\seq{m}^1_{d}\atop\seq{m}^2_{d}}
\right]
\ee
which we will define momentarily. Before doing that, let us comment briefly on the structure of~\eqref{eq:biggerscaryformula} and~\eqref{eq:smallunicornformula}. 

There are two complications compared to the case of external scalar operators. First, there are many possible three-point tensor structures, and we have to sum over the contributions from different pairs of three-point structures. This is done in the first two sums in \eqref{eq:biggerscaryformula} or equivalently the first sum in~\eqref{eq:smallunicornformula}. Indeed, according to the discussion around \eqref{eq:right3pt}, the set $a=(\widetilde\bfm^1_{d-1},\widetilde\bfm^2_{d-1},\bfm'_{d-1},t)$ such that $\bfm'_{d-1},t$ selects an irreducible component in $\widetilde\bfm^1_{d-1}\otimes\widetilde\bfm^2_{d-1}$ uniquely determines a three-point tensor structure for the operators $1$ and $2$, and an analogous statement holds for $b$ and the operators $3$ and $4$. Second, there are many four-point structures, and a single pair of three-point structures can contribute to many four-point structures. This is the last sum in~\eqref{eq:biggerscaryformula} and~\eqref{eq:smallunicornformula}. As we discuss below, the role of $\bfm_{d-2}$ representation is to specify a way of gluing the two three-point structures into a four-point structure. Note that the three-point structures do not depend on $\bfm_{d-2}$, but the angular functions $P$ and the four-point tensor structures do. We stress that the structures \eqref{eq:fourptdemo} form a basis of all four-point tensor structures, as we now explain.

The definition of \eqref{eq:fourptdemo} follows straightforwardly from the construction,
\be\label{eq:fourptdefn}
&\left[
{\seq{m}^3_{d}\atop\seq{m}^4_{d}}\Big\vert
{\widetilde{\bfm}^3_{d-1}\atop\widetilde{\bfm}^4_{d-1}}
\,\bfm_{d-1},t
\Big\vert
\bfm_{d-2}
\Big\vert
\bfm'_{d-1},t'\,{\widetilde{\bfm}^1_{d-1}\atop\widetilde{\bfm}^2_{d-1}}\Big\vert
{\seq{m}^1_{d}\atop\seq{m}^2_{d}}
\right]=\nn\\
&=\sum_{\seq{m}_{d-2},\seq{m}'_{d-2}}\<0|\seq{m}^3_{d-1},\seq{m}^4_{d-1},\seq{m}_{d-1},t\>\delta_{\seq{m}_{d-2},\seq{m}'_{d-2}}\<\seq{m}'_{d-1},t'|\seq{m}^1_{d-1},\seq{m}^2_{d-1}\>\times\nn\\
&\quad\times\delta_{\bfm^1_{d-1},\widetilde\bfm^1_{d-1}}\delta_{\bfm^2_{d-1},\widetilde\bfm^2_{d-1}}\delta_{\bfm^3_{d-1},\widetilde\bfm^3_{d-1}}\delta_{\bfm^4_{d-1},\widetilde\bfm^4_{d-1}}.
\ee
Here $\bfm'_{d-2}=\bfm_{d-2}$. Note that for every choice of $\widetilde\bfm^i_{d-1},\bfm_{d-1},\bfm_{d-1}',\bfm_{d-2},t,t'$, this is a function of $\seq{m}^i_d$,
i.e.\ an element of
\be
	\bfm^1_d\otimes\bfm^2_d\otimes\bfm^3_d\otimes\bfm^4_d.
\ee
Furthermore, it is clear from the definition that it is $Spin(d-2)$ invariant. This means that it is an element of
\be\label{eq:fourptspace}
	\left(\bfm^1_d\otimes\bfm^2_d\otimes\bfm^3_d\otimes\bfm^4_d\right)^{Spin(d-2)},
\ee
which is the space of four-point tensor structures~\cite{Kravchuk:2016qvl,Schomerus:2016epl}. 

The set of structures~\eqref{eq:fourptdefn} with the parameters restricted by~\eqref{eq:selectionrules} spans \eqref{eq:fourptspace}. Indeed, we have
\be
\left(\bfm^1_d\otimes\bfm^2_d\otimes\bfm^3_d\otimes\bfm^4_d\right)^{Spin(d-2)}=\bigoplus_{\bfm^{12}_{d-1}\in \bfm^1_d\otimes\bfm^2_d\atop\overline{\bfm^{34}_{d-1}}\in \bfm^3_d\otimes\bfm^4_d}(\bfm^{12}_{d-1}\otimes \overline{\bfm^{34}_{d-1}})^{Spin(d-2)},
\ee
where the sum is taken with multiplicities. Because the dimensional reduction is multiplicity-free, we have that $Spin(d-2)$ singlets in $\bfm^{12}_{d-1}\otimes \overline{\bfm^{34}_{d-1}}$ are in one-to-one correspondence with $\bfm_{d-2}^{1234}\in \bfm^{12}_{d-1}, \bfm^{34}_{d-1}$.

This enumeration is implemented by~\eqref{eq:selectionrules} as follows. By specifying $\widetilde\bfm^{1}_{d-1},\widetilde\bfm^{2}_{d-1},\bfm'_{d-1},t'$ we first select a general $Spin(d-1)$ representation $\bfm_{d-1}^{12}\simeq \bfm_{d-1}'$ in $\bfm^1_d\otimes\bfm^2_d$. Similarly, $\widetilde\bfm^{3}_{d-1},\widetilde\bfm^{4}_{d-1},\overline{\bfm_{d-1}},t$ select a general $Spin(d-1)$ irrep $\overline{\bfm_{d-1}^{34}}\simeq \overline{\bfm_{d-1}}$ in $\bfm^3_d\otimes\bfm^4_d$. The ``gluing'' representation $\bfm_{d-2}^{1234}$ is then identified with $\bfm_{d-2}$.

\subsection{Example: Scalar correlators}
\label{sec:structure:ScalarCorrelators}
Let us see how we can recover the Genegenbauer expansion for scalar four-point functions. For scalars we have $\bfm_d^i=(0,\ldots 0)=\bullet$, and the only Gelfand-Tsetlin patterns are $[\mathbf{\bullet}]\equiv(\bullet,\ldots\bullet)$. Similarly, $\widetilde \bfm_{d-1}^i=\bullet$.
In \eqref{eq:biggerscaryformula} we only need to sum over $\bfm'_{d-1}\in \widetilde\bfm^1_{d-1}\otimes \widetilde\bfm^2_{d-1}$, thus only $\bfm'_{d-1}=\bullet$ is allowed and there is no need in $t'$ label. Similarly, $\bfm_{d-1}=\bullet$. The sum over $\bfm_{d-2}$ is restricted to $\bfm_{d-2}\in\,\bfm_{d-1},\bfm'_{d-1}$, and thus we only have $\bfm_{d-2}=\bullet$. The unique component of the unique four-point structure is
\be
\left[
{[\bullet]\atop[\bullet]}
\Big\vert
{\bullet\atop\bullet}
\,\bullet
\Big\vert
\bullet
\Big\vert
\bullet\,
{\bullet\atop\bullet}
\Big\vert
{[\bullet]\atop[\bullet]}
\right]=1.
\ee
Equation \eqref{eq:biggerscaryformula} collapses then to
\begin{multline}
\label{eq:scalarcorrelatorcontribution}
\sum_{\seq{m}_d}\langle 0|\cO_4^{[\bullet]}(-1) \cO_3^{[\bullet]}(1)|\Delta,\seq{m}_{d}\rangle\langle\Delta,\seq{m}_{d}|r^{D}e^{\theta M_{12}}\cO_2^{[\bullet]}(1)\cO_1^{[\bullet]}(-1)|0\rangle=
\lambda^{\bullet,\bullet}_{\bullet}
\bar\lambda^{\bullet,\bullet}_{\bullet}
r^\Delta P^{\bfm_d,\bullet}_{\bullet,\bullet}(\theta).
\end{multline}
We need $\bfm_{d-1},\bfm'_{d-1}\in\,\bfm_d$, and thus for scalars we get the condition $\bfm_d\ni\bullet$, which is only satisfied if $\bfm_d$ is traceless-symmetric, $\bfm_d=\bfj=(j,0,\ldots,0)$. Finally, as we show in~\eqref{eq:gegenbauer} later in this section, $P^{\bfj,\bullet}_{\bullet,\bullet}(\theta)$ is proportional to a Gegenbauer polynomial. Taking~\eqref{eq:gegenbauer} into account, we reproduce the result of~\cite{Hogervorst:2013sma}
\be
\sum_{\seq{m}_d}\langle 0|\cO_4^{[\bullet]}(-1) \cO_3^{[\bullet]}(1)|\Delta,\seq{m}_{d}\rangle\langle\Delta,\seq{m}_{d}|r^{D}e^{\theta M_{12}}\cO_2^{[\bullet]}(1)\cO_1^{[\bullet]}(-1)|0\rangle=
\lambda^{\bullet,\bullet}_{\bullet}
\bar\lambda^{\bullet,\bullet}_{\bullet}
r^\Delta \frac{C_j^{(\nu)}(\theta)}{C_j^{(\nu)}(1)}.
\ee

\subsection{Example: General 3d correlators}
\label{sec:structure:General3dCorrelators}

Consider now the case $d=3$. Let us first write the four-point tensor structure~\eqref{eq:fourptdefn}. Since $d=3$, the sums in~\eqref{eq:fourptdefn} are trivial, as well as $\bfm_{d-2}$ is. Furthermore, $Spin(d-1)=Spin(2)$ tensor products are multiplicity-free, so the labels $t$ and $t'$ are also trivial. We then find, using~\eqref{eq:threeptleft3d} and~\eqref{eq:threeptright3d},
\be
\left[
{j_3,m_3\atop j_4,m_4}\Big\vert{\widetilde m_3\atop \widetilde m_4}
\,m
\Big\vert
\Big\vert
m'\,{\widetilde m_1\atop \widetilde m_2}\Big\vert
{j_1,m_1\atop j_2,m_2}
\right]=\delta_{m',m_1+m_2}\delta_{0,m+m_3+m_4}\delta_{m_1,\widetilde m_1}\delta_{m_2,\widetilde m_2}\delta_{m_3,\widetilde m_3}\delta_{m_4,\widetilde m_4}.
\ee
Since the tensor product of $Spin(2)$ representations $\widetilde m_1,\widetilde m_2$ contains only one representation, $\widetilde m_1+\widetilde m_2$, we do not need to specify $m'$ separately. The same holds for $m$. We can thus simplify this tensor structure as
\be\label{eq:fourpt3d}
\left[
{j_3,m_3\atop j_4,m_4}\Big\vert{\widetilde m_3\atop \widetilde m_4}
{\widetilde m_1\atop \widetilde m_2}\Big\vert
{j_1,m_1\atop j_2,m_2}
\right]\equiv\delta_{m_1,\widetilde m_1}\delta_{m_2,\widetilde m_2}\delta_{m_3,\widetilde m_3}\delta_{m_4,\widetilde m_4}.
\ee
Before moving further, let us understand the meaning of this expression. It is a four-point tensor structure in the sense that by fixing $\widetilde m_i$ we have a tensor with indices $m_i$, i.e.\ an element of 
\be
	j_1\otimes j_2\otimes j_3\otimes j_4.
\ee
Note that these structures form a complete basis for such tensors, which is consistent with the fact that $Spin(d-2)=Spin(1)$ is trivial and so there is no invariance constraint on conformal frame four-point structures~\cite{Kravchuk:2016qvl}.\footnote{One can be more pedantic by taking $Spin(1)=\bZ_2$, in which case there is a constraint which simply says that $\widetilde m_1+\widetilde m_2+\widetilde m_3+\widetilde m_4$ (equivalently, $j_1+j_2+j_3+j_4$) must be an integer, i.e.\ the correlator should contain an even number of fermions.}

As noted above, we can essentially drop $\bfm_{d-1},\bfm_{d-1}',\bfm_{d-2},t,t'$ in~\eqref{eq:biggerscaryformula}. Using~\eqref{eq:P3d} and~\eqref{eq:fourpt3d} we can rewrite~\eqref{eq:biggerscaryformula} as
\be\label{eq:3dcontribution}
	&\sum_m\<0|\cO_4^{j_4,m_4}(-1)\cO_3^{j_3,m_3}(1)|\Delta,j,m\>\<\Delta,j,m|r^De^{\theta M_{12}}\cO_2^{j_2,m_2}(1)\cO_1^{j_1,m_1}(-1)|0\>=\nn\\
	&\quad=\sum_{\widetilde m^i}\lambda^{\widetilde m_1,\widetilde m_2}\bar\lambda^{\widetilde m_3,\widetilde m_4}r^\Delta d^j_{-\widetilde m_3-\widetilde m_4,\widetilde m_1+\widetilde m_2}(-\theta)\left[
	{j_3,m_3\atop j_4,m_4}\Big\vert{\widetilde m_3\atop \widetilde m_4}
	{\widetilde m_1\atop \widetilde m_2}\Big\vert
	{j_1,m_1\atop j_2,m_2}
	\right],
\ee
where summation is over
\be
	\widetilde m_i=-j_i,-j_i+1,\ldots j_i,
\ee
and the last line of~\eqref{eq:selectionrules} also restricts
\be
	|\widetilde m_1+\widetilde m_2|,\,|\widetilde m_3+\widetilde m_4|\leq j
\ee
as well as that $\widetilde m_1+\widetilde m_2$ and $\widetilde m_3+\widetilde m_4$ are integral or half-integral simultaneously with $j$, so that small Wigner $d$-matrix is well-defined.

\subsection{Example: General 4d correlators}
\label{sec:structure:General4dCorrelators}

We now consider the case of the general correlation functions in $d=4$. The usefulness of this example comes from the fact that while being not very different from the most general case, it can still be formulated using only the familiar ingredients from representation theory of $Spin(d-1)=Spin(3)\simeq SU(2)$. 

First, we need to construct the three-point tensor structures. Consider for example the right tensor structure~\eqref{eq:right3pt} parametrized by the data~\eqref{eq:3ptrightshorthand}. We can write in 4d
\be
	a=(\widetilde j_1,\widetilde j_2,\widetilde j'),
\ee
where $\widetilde j_i\in l^i_L\otimes l^i_R$ and $\widetilde j'\in l_L\otimes l_R$ where $(l_L,l_R)$ is the representation of the exchanged operator. The constraint in~\eqref{eq:3ptrightshorthand} then takes form $\widetilde j'\in \widetilde j_1\otimes \widetilde j_2$. In particular, we do not need a multiplicity label because the tensor products in $Spin(3)$ are multiplicity-free. The three-point functions take the form
\be
	\<\Delta,l_L,l_R;j',m'|\cO_2^{j_2,m_2}(1)\cO_1^{j_1,m_1}(-1)|0\>
	=&\lambda_{(j_1,j_2,j')}\<j',m'|j_1,m_1;j_2,m_2\>\nn\\
	=&\sum_{a=(\widetilde j_1,\widetilde j_2,\widetilde j')} \lambda_{(\widetilde j_1,\widetilde j_2,\widetilde j')}\{\delta_{j_1\widetilde j_1}\delta_{j_2\widetilde j_2}\delta_{j'\widetilde j'}\<j',m'|j_1,m_1;j_2,m_2\>\}.
\ee
Here, for notational simplicity, we have omitted the $\bfm_4^i$ part of the GT pattern for the primary operators $\cO_i$. The second line of this equation gives the more traditional form of the three-point functions as a sum over tensor structures labeled by $a$. Finally, $\<j',m'|j_1,m_1;j_2,m_2\>$ is the $SU(2)$ Clebsch-Gordan coefficient. Similarly, for the left three-point function we have
\be
	\<0|\cO_4^{j_4,m_4}(-1)\cO_3^{j_3,m_3}(1)|\Delta,l_L,l_R;j,m\>
	=&\bar\lambda_{(j_3,j_4,j)}\<0|j_4,m_4;j_3,m_3;j,m\>\nn\\
	=&\sum_{b=(\widetilde j_3,\widetilde j_4, \widetilde j)}\lambda_{(\widetilde j_3,\widetilde j_4,\widetilde j)}\{\delta_{j_3,\widetilde j_3}\delta_{j_4,\widetilde j_4}\delta_{j,\widetilde j}\<0|j_4,m_4;j_3,m_3;j,m\>\},
\ee
and the constraint from~\eqref{eq:3ptleftshorthand} is simply $\widetilde j\in \widetilde j_3\otimes \widetilde j_4$ since all $SU(2)$ irreps are self-conjugate. Here $\<0|j_4,m_4;j_3,m_3;j,m\>$ is essentially the $SU(2)$ $3j$ symbol. Note that this parametrization of three-point structures is essentially the same as the one mentioned in~\cite{Cuomo:2017wme}.

The four-point tensor structures~\eqref{eq:fourptdefn} can also be computed as 
\be\label{eq:4d4pt}
&\left[
{j_3,m_3 \atop j_4, m_4}\Big\vert
{\widetilde{j}_3\atop\widetilde{j}_4}
\,j
\Big\vert
m
\Big\vert
j'\,{\widetilde{j}_1\atop\widetilde{j}_2}\Big\vert
{j_1,m_1\atop j_2,m_2}
\right]=\<0|j_4,m_4;j_3,m_3;j,m\>\<j',m|j_1,m_1;j_2,m_2\>\delta_{j_1,\widetilde j_1}\delta_{j_2,\widetilde j_2}\delta_{j_3,\widetilde j_3}\delta_{j_4,\widetilde j_4}.
\ee
Recall that the labels $m_i$ parametrize the representations of the $Spin(2)$ which rotates in the plane 3-4. This plane is orthogonal to the plane 1-2 in which we place our operators, and thus this $Spin(2)$ it the stabilizer group of the four points and, as usual, the four-point tensor structures have to be invariant under it. Using the constraints $m_4+m_3+m=0$ and $m=m_1+m_2$ coming from the CG coefficients, we find $m_4+m_3+m_2+m_1=0$ which is precisely the required invariance condition. Of course, this comes as no surprise since it was guaranteed by construction. Note that this basis of four-point tensor structures is different from the one in~\cite{Cuomo:2017wme}, since it is not an eigenbasis for rotations in plane 1-2.

The final formula~\eqref{eq:biggerscaryformula} takes the following form in $4d$,
\begin{multline}
	\sum_{j,m}\<0|\cO_4^{j_4,m_4}(-1)\cO_3^{j_3,m_3}(1)|\Delta,l_L,l_R;j,m\>\<\Delta,l_L,l_R;j,m|e^{\theta M_{12}}r^D \cO_2^{j_2,m_2}(1)\cO_1^{j_1,m_1}(-1)|0\>=\\
	=\sum_{a,b}\sum_{m}\lambda^a\bar\lambda^b \left[
	{j_3,m_3 \atop j_4, m_4}\Big\vert
	b
	\Big\vert
	m
	\Big\vert
	a\Big\vert
	{j_1,m_1\atop j_2,m_2}
	\right]P^{l_L,l_R;m}_{\widetilde j,\widetilde j'}(\theta),
\end{multline}
where the four-point tensor structure and the three-point labels $a,b$ are described above, while the $P$-function is given below in section~\ref{sec:structure:Pfunctions} by equation~\eqref{eq:4dP}. The range of summation over $m$ is restricted to be $-\mathrm{min}(\widetilde j,\widetilde j'),-\mathrm{min}(\widetilde j,\widetilde j')+1,\ldots, \mathrm{min}(\widetilde j,\widetilde j')$.

\subsection{Example: Seed conformal blocks in general dimensions}
\label{sec:structure:SeedConformalBlocks}

Our last example concerns an especially simple case which occurs for every $d$. The simplification is based on the fact that the CG coefficients are trivial when one of the representations is trivial. Choosing two of the four operators operators to be scalars, we can ensure that the CG coefficients for both the right and the left three-point function simplify, with the correlator itself still being sufficiently general. If fact, as will be clear from the construction, the so-called seed blocks for arbitrary intermediate representations can be chosen to be of this form. 

Let us choose the operators $\cO_1$ and $\cO_3$ to be scalars. Then the general result~\eqref{eq:biggerscaryformula} simplifies as
\be\label{eq:seedlead}
\sum_{\seq{m}_{d}}\langle 0|\cO_4^{\seq{m}^4_{d}}(-1) \cO_3(1)|&\Delta,\seq{m}_{d}\rangle\langle\Delta,\seq{m}_{d}|r^{D}e^{\theta M_{12}}\cO_2^{\seq{m}^2_{d}}(1)\cO_1(-1)|0\rangle=\nn\\
=&
\sum_{\widetilde{\bfm}^i_{d-1}}\sum_{\bfm_{d-2}}
\lambda^{\bullet,\widetilde\bfm^2_{d-1}}_{\widetilde\bfm^2_{d-1}}
\bar\lambda^{\bullet,\widetilde\bfm_{d-1}^4}_{\overline{\widetilde\bfm_{d-1}^4}}
r^\Delta P^{\bfm_d,\bfm_{d-2}}_{\widetilde\bfm^2_{d-1},\overline{\widetilde\bfm_{d-1}^4}}(\theta)\times\nn\\
&\times\left[
{\bullet\atop\seq{m}^4_{d}}\Big\vert
{\bullet\atop\widetilde{\bfm}^4_{d-1}}
\,\overline{\widetilde\bfm_{d-1}^4}
\Big\vert
\bfm_{d-2}
\Big\vert
\widetilde\bfm^2_{d-1}\,{\bullet\atop\widetilde{\bfm}^2_{d-1}}\Big\vert
{\bullet\atop\seq{m}^2_{d}}
\right],
\ee
with the four point-structures given by the specialization of~\eqref{eq:fourptdefn},
\be\label{eq:seed4pt}
&\left[
{\bullet\atop\seq{m}^4_{d}}\Big\vert
{\bullet\atop\widetilde{\bfm}^4_{d-1}}
\,\overline{\widetilde\bfm_{d-1}^4}
\Big\vert
\bfm_{d-2}
\Big\vert
\widetilde\bfm^2_{d-1}
\,{\bullet\atop\widetilde{\bfm}^2_{d-1}}\Big\vert
{\bullet\atop\seq{m}^2_{d}}
\right]=\nn\\
&=\sum_{\seq{m}_{d-2},\seq{m}'_{d-2}}\<0|\seq{m}^4_{d-1},\seq{m}_{d-1}\>\delta_{\seq{m}_{d-2},\seq{m}'_{d-2}}\<\seq{m}'_{d-1}|\seq{m}^2_{d-1}\>\delta_{\bfm^2_{d-1},\widetilde\bfm^2_{d-1}}\delta_{\bfm^4_{d-1},\widetilde\bfm^4_{d-1}}=\nn\\
&=\sum_{\seq{m}_{d-2},\seq{m}'_{d-2}}\frac{(-1)^{\seq{m}_{d-1}^4}}{\sqrt{\dim \widetilde\bfm_{d-1}^4}}\delta_{\seq{m}_{d-1}^4,\overline{\seq{m}_{d-1}}}\delta_{\seq{m}_{d-2},\seq{m}'_{d-2}}\delta_{\seq{m}'_{d-1},\seq{m}^2_{d-1}}\delta_{\bfm^2_{d-1},\widetilde\bfm^2_{d-1}}\delta_{\bfm^4_{d-1},\widetilde\bfm^4_{d-1}}=\nn\\
&=\frac{(-1)^{\seq{m}_{d-1}^4}}{\sqrt{\dim \widetilde\bfm_{d-1}^4}}\delta_{\bfm^2_{d-1},\widetilde\bfm^2_{d-1}}\delta_{\bfm^4_{d-1},\widetilde\bfm^4_{d-1}}\delta_{\bfm^4_{d-2},\overline{\bfm_{d-2}}}\delta_{\bfm^2_{d-2},\bfm_{d-2}}\delta_{\seq{m}^4_{d-2},\overline{\seq{m}^2_{d-2}}},
\ee
where we made use of~\eqref{eq:2jfinal}. The constraints~\eqref{eq:selectionrules} reduce in this case to 
\be
	\widetilde\bfm_{d-1}^i\in &\,\bfm_d^i,\quad i=2,4,\\
	\bfm_{d-2}\in &\,\widetilde\bfm_{d-1}^2\in \bfm_d,\\
	\bfm_{d-2}\in &\,\overline{\widetilde\bfm_{d-1}^4}\in \bfm_d.
\ee

Note that for any $\bfm_d$ there exists a choice of $\bfm_d^i$ such that these constraints can be satisfied, and thus arbitrary intermediate representations can be exchanged in this simplified setup. In fact, for a given $\bfm_d$, in even $d$, we can always choose $\bfm_d^i$ so that there is a unique choice available for $\widetilde\bfm_{d-1}^i$ (and thus a unique three-point function on either side). For this, set, for example\footnote{This is choice is different from the one used in $d=4$ in~\cite{Cuomo:2017wme}. In fact, in even $d$ it doesn't matter what we choose $m^2_{d,n}$ to be, and the choice in~\cite{Cuomo:2017wme} corresponds to $m^2_{d,n}=|m_{d,n}|$. Our choice~\eqref{eq:seedsetup} has the advantage that is also works in odd dimensions, see below. Also, note that there is some freedom in choosing $\bfm_d^4$ independently of $\bfm_d^2$.}
\be\label{eq:seedsetup}
	m^2_{d,k}&=m_{d,k+1},\quad 1\leq k<n,\nn\\
	m^2_{d,n}&=0\text{ or }\thalf,\nn\\
	\bfm^4_{d}&=\overline{\bfm^2_{d}},
\ee
where the choice in the second equality is determined by the statistics of $\bfm_d$.
In odd $d$, this only reduces down to two choices for each of $\widetilde\bfm_{d-1}^i$ if the representations are fermionic (but still one choice for bosonic representations). This is because in the case of odd $d$ the outer automorphism of $Spin(d-1)$ (given by reflection) necessarily acts non-trivially on fermionic representations of $Spin(d-1)$, but trivially on the representations of $Spin(d)$. Therefore, the number of three-point tensor structures involving fermionic representations is always even, and we simply cannot have less than 2 non-trivial structures.

If we think about the state $|\Delta,\seq{m}_d\>$ as being a conformal primary, then the choices of external representations described above give us a valid choice for the so-called seed blocks for exchange of primary $\bfm_d$ -- they lead to the minimum number of three-point tensor structures on both sides of the four-point function. The equations~\eqref{eq:seedlead} and~\eqref{eq:seed4pt} then give the leading contribution to the OPE limit of such seed conformal blocks. 

As a concrete example, consider the scalar-fermion blocks in even dimensions. Specifically, we take
\be
	\bfm_{d}^2&=(\thalf,\ldots,\thalf,+\thalf),\\
	\bfm_{d}^4&=(\thalf,\ldots,\thalf,-\thalf).
\ee
This is slightly different from the prescription~\eqref{eq:seedsetup} unless $d=4k+2$, but it is more convenient to have a uniform choice of representations for all even $d$. Under dimensional reduction both $\bfm_{d}^2$ and $\bfm_{d}^4$ restrict to a single representation, and thus necessarily
\be
	\widetilde\bfm^2_{d-1}=\widetilde\bfm^4_{d-1}=(\thalf,\ldots,\thalf).
\ee
These representations further restrict to a direct sum of $(\thalf,\ldots,+\thalf)$ and $(\thalf,\ldots,-\thalf)$ in $d-2$ dimensions, so that there are two four-point tensor structures
\be\label{eq:ferm4pt}
t_\pm\equiv \left[
{\bullet\atop\seq{m}^4_{d}}\Big\vert
{\bullet\atop(\thalf,\ldots,\thalf)}
\,(\thalf,\ldots,\thalf)
\Big\vert
(\thalf,\ldots,\pm\thalf)
\Big\vert
(\thalf,\ldots,\thalf)
\,{\bullet\atop(\thalf,\ldots,\thalf)}\Big\vert
{\bullet\atop\seq{m}^2_{d}}
\right].
\ee
Correspondingly, there are two types of $\bfm_d$ that can be exchanged, each with a single three-point tensor structure on either side,
\be
	\bfm_d^\pm=(j,\thalf,\ldots,\pm\thalf).
\ee
From~\eqref{eq:biggerscaryformula} we find that the contribution of the representation $\bfm^\pm_d$ to the four-point function~\eqref{eq:fourptconvenient} is given by
\be
	\sum_{\pm} \Lambda_\pm r^\Delta\left( P^{(j,\thalf,\ldots,\pm\thalf),(\thalf,\ldots,+\thalf)}_{(\thalf,\ldots,\thalf)(\thalf,\ldots,\thalf)}(\theta)t_+
	+
	P^{(j,\thalf,\ldots,\pm\thalf),(\thalf,\ldots,-\thalf)}_{(\thalf,\ldots,\thalf)(\thalf,\ldots,\thalf)}(\theta)t_-\right),
\ee
where 
\be
	\Lambda_\pm = \lambda^{\bullet,\widetilde\bfm^2_{d-1}}_{\widetilde\bfm^2_{d-1},\bfm_d^\pm}
	\bar\lambda^{\bullet,\widetilde\bfm^4_{d-1}}_{\widetilde\bfm^4_{d-1},\bfm_d^\pm}.
\ee
Here $\bfm^\pm_d$ index of OPE coefficients labels the exchanged representation. We find explicit expressions for the above $P$-functions in section~\ref{sec:structure:Pfunctions:computation}, with the result given in~\eqref{eq:fermionmatrom}. 

\subsection{Example: Conformal block/Four-point tensor structure correspondence}
\label{sec:structure:FolkloreTheorem}

As another simple application of the above formalism, let us discuss the folklore theorem which states that the number of classes of conformal blocks which contribute to a given four-point function is equal to the number of four-point tensor structures~\cite{Dymarsky:2013wla,Elkhidir:2014woa}. We will consider the simplest case where the only relevant symmetry is the connected conformal group (i.e.\ no space parity or permutation symmetries for identical operators). In our formalism this theorem becomes essentially a tautology. Because of that, this section basically reiterates what was already said, with a slightly different focus.

First, let us explain what is meant by classes of conformal blocks. Each conformal block contributing to a four-point function is parametrized by the dimension $\Delta$ and the $Spin(d)$ representation $\bfm_d$ of the exchanged primary operator, as well as by a pair of three-point functions $a$ and $b$. From the previous discussion, we can parametrize the three-point functions as follows,
\be\label{eq:folklore3pt}
a&=(\widetilde\bfm_{d-1}^1,\widetilde\bfm_{d-1}^2,\bfm'_{d-1},t'),\nn\\
b&=(\widetilde\bfm_{d-1}^3,\widetilde\bfm_{d-1}^4,\bfm_{d-1},t),
\ee
subject to~\eqref{eq:selectionrules}. In particular, the constraint 
\be\label{eq:folkloreconstraint0}
\widetilde\bfm_{d-1}^i\in\bfm_d^i
\ee
gives us finitely many choices for $\widetilde\bfm_{d-1}^i$ for fixed $\bfm_{d}^i$ and the constraints
\be\label{eq:folkloreconstraints}
(\bfm'_{d-1},t')&\in \widetilde\bfm_{d-1}^1\otimes\widetilde\bfm_{d-1}^2,\nn\\
(\overline{\bfm_{d-1}},t)&\in \widetilde\bfm_{d-1}^3\otimes\widetilde\bfm_{d-1}^4,
\ee
thus give us finitely many choices of $(\bfm'_{d-1},t')$ and $(\bfm_{d-1},t)$. The intermediate representation is then constrained by $\bfm_{d-1},\bfm'_{d-1}\in \bfm_{d}$. This leaves \textit{infinitely} many choices of $\bfm_d$ for a given four-point function. However, the allowed $\bfm_d$ organize into natural families. Indeed, let us denote $\bfm_d=(j,\widetilde \bfm_{d-2})$, i.e.\ $j$ is the length of the first row of the generalized Young diagram of $j$ and $\widetilde \bfm_{d-2}$ encodes the remaining rows.\footnote{Note that indeed $\widetilde\bfm_{d-2}$ is always a dominant weight for $Spin(d-2)$.} The following two statements are then equivalent,
\be
\bfm_{d-1},\bfm'_{d-1}\in \bfm_{d}=(j,\widetilde\bfm_{d-2})	\quad\Longleftrightarrow\quad \widetilde\bfm_{d-2}\in\bfm_{d-1},\bfm'_{d-1}\text{ and }j\geq m_{d-1,1},m_{d-1,1}'.
\ee
This leaves only a finite number of choices for $\widetilde\bfm_{d-2}$.

The infiniteness of the number of conformal blocks is therefore only due to the generic parameters $\Delta$ and $j$. If we consider any two conformal blocks differing by only these two parameters to belong to the same \textit{class}, we obtain a finite set of classes parametrized by a pair of three-point structures~\eqref{eq:folklore3pt} subject to~\eqref{eq:folkloreconstraint0}-\eqref{eq:folkloreconstraints} and a $\widetilde\bfm_{d-2}$ subject to
\be
\widetilde\bfm_{d-2}\in\bfm_{d-1},\bfm'_{d-1}.
\ee
The statement of the theorem is that the number of such classes is equal to the number of four-point tensor structures. Indeed, we already saw that the four-point tensor structures~\eqref{eq:fourptdefn} are parameterized by exactly the same data.

For conformal blocks this statement is, strictly speaking, only a counting statement and thus it would be interesting to get a more physical understanding of this. Note however that the matroms $\bfP_{\bfm_{d-1},\bfm'_{d-1}}$, as discussed in section~\ref{sec:structure:Pfunctions:computation} link together, in some sense, the spaces of $\bR\times Spin(d)$ blocks and four-point tensor structures.

\subsection{$P$-functions}
\label{sec:structure:Pfunctions}
In this section we discuss general properties of the GT matrix elements $P$, as well as their explicit calculation in various situations. This section is rather technical and mostly independent from the sections to follow, and thus can be skipped on the first reading.

\subsubsection{Basic properties}
First, recall the definition~\eqref{eq:Pdefinition}
\be
\<\seq{m}_{d}|e^{\theta M_{12}}|\seq{m}'_{d}\>=P^{\bfm_d,\bfm_{d-2}}_{\bfm_{d-1},\bfm'_{d-1}}(\theta)\delta_{\seq{m}_{d-2},\seq{m}'_{d-2}}.
\ee
There are a lot of properties of $P$ which follow immediately from this definition as a matrix element. For example, the simplest property of $P$ is given by substituting $\theta=0$,
\be
P^{\bfm_d,\bfm_{d-2}}_{\bfm_{d-1},\bfm'_{d-1}}(0)=\delta_{\bfm_{d-1},\bfm'_{d-1}}.
\ee
Furthermore, $P$ is $2\pi$-periodic for bosonic representations and $2\pi$-antiperiodic for fermionic representations. More generally, we know from the standard representation theory arguments that the spectrum of $iM_{12}$ consists of (half-)integers ranging from $-m_{d,1}$ to $m_{d,1}$, and thus all $P$-functions have the form
\be
P^{\bfm_d,\bfm_{d-2}}_{\bfm_{d-1},\bfm'_{d-1}}(\theta)=\sum_{m=-m_{d,1}}^{m_{d,1}} c_m e^{im\theta},
\ee
where $c_m$ are coefficients which depend on the indices of $P$, some of which may vanish.

Reality properties can be obtained by applying Hermitian conjugation to the definition above and noting that $M_{\mu\nu}$ are anti-Hermitian, resulting in, for real $\theta$,
\be\label{eq:hermitianconj}
\left(P^{\bfm_d,\bfm_{d-2}}_{\bfm_{d-1},\bfm'_{d-1}}(\theta)\right)^*=P^{\bfm_d,\bfm_{d-2}}_{\bfm'_{d-1},\bfm_{d-1}}(-\theta).
\ee
Note that we also have
\be
\sum_{\seq{m}'_d}\<0|\seq{m}_d\overline{\seq{m}'}_d\>\<\overline{\seq{m}'}_d|e^{\theta M_{12}}|\overline{\seq{m}''}_d\>=	\sum_{\seq{m}'_d}\<0|\seq{m}'_d\overline{\seq{m}''}_d\>\<\seq{m}'_d|e^{-\theta M_{12}}|\seq{m}_d\>
\ee
due to the invariance of $\<0|\seq{m}_d\overline{\seq{m}'}_d\>$. Contracting with $\<\seq{m}'''_d\overline{\seq{m}''}_d|0\>$ on both sides we find
\be
\dim\bfm_{d}\sum_{\seq{m}'_d,\seq{m}''_d}\<0|\seq{m}_d\overline{\seq{m}'}_d\>\<\overline{\seq{m}'}_d|e^{\theta M_{12}}|\overline{\seq{m}''}_d\>\<\seq{m}'''_d\overline{\seq{m}''}_d|0\>=\<\seq{m}'''_d|e^{-\theta M_{12}}|\seq{m}_d\>.
\ee
This implies, in terms of $P$-functions,
\be
(-1)^{\bfm_{d-1}-\bfm'_{d-1}}P^{\overline{\bfm_d},\overline{\bfm_{d-2}}}_{\overline{\bfm_{d-1}},\overline{\bfm'_{d-1}}}(\theta)=P^{\bfm_d,\bfm_{d-2}}_{\bfm'_{d-1},\bfm_{d-1}}(-\theta)=\left(P^{\bfm_d,\bfm_{d-2}}_{\bfm_{d-1},\bfm'_{d-1}}(\theta)\right)^*.
\ee

The group composition property for the matrix elements
\be
\sum_{\seq{m}_d'}\<\seq{m}_{d}|e^{\theta_1 M_{12}}|\seq{m}'_{d}\>\<\seq{m}'_{d}|e^{\theta_2 M_{12}}|\seq{m}''_{d}\>=
\<\seq{m}_{d}|e^{(\theta_1+\theta_2) M_{12}}|\seq{m}''_{d}\>
\ee
gives the sum rule
\be
\sum_{\bfm_{d-1}'}P^{\bfm_d,\bfm_{d-2}}_{\bfm_{d-1},\bfm'_{d-1}}(\theta_1)P^{\bfm_d,\bfm_{d-2}}_{\bfm_{d-1}',\bfm''_{d-1}}(\theta_2)=P^{\bfm_d,\bfm_{d-2}}_{\bfm_{d-1},\bfm''_{d-1}}(\theta_1+\theta_2).
\ee
In particular, substituting $\theta_2=-\theta_1$, $\bfm''_{d-1}=\bfm_{d-1}$, we find, for real $\theta$,
\be
\sum_{\bfm'_{d-1}}|P^{\bfm_d,\bfm_{d-2}}_{\bfm_{d-1},\bfm'_{d-1}}(\theta_1)|^2=1,
\ee
and thus
\be
|P^{\bfm_d,\bfm_{d-2}}_{\bfm_{d-1},\bfm'_{d-1}}(\theta)|\leq 1.
\ee

\subsubsection{Orthogonality relations}

The matrix elements of group representations obey Schur orthogonality relations which read as
\be\label{eq:schurortho1}
\int_{Spin(d)}\<\seq{m}_d|R|\seq{m}'_d\>(\<\widetilde{\seq{m}}_d|R|\widetilde{\seq{m}}'_d\>)^* dR = \frac{1}{\dim\bfm_d}\delta_{\seq{m}_d,\widetilde{\seq{m}}_d}\delta_{\seq{m}'_d,\widetilde{\seq{m}}'_d}.
\ee
Here the $\delta$-symbols also compare $\bfm_d$ with $\widetilde\bfm_d$. The group integral in the left hand side is understood to be over Haar measure normalized as
\be
\int_{Spin(d)}dR=1.
\ee
Let us set $\seq{m}_{d-1}=\widetilde{\seq{m}}_{d-1}$ and $\seq{m}'_{d-1}=\widetilde{\seq{m}}'_{d-1}$ in~\eqref{eq:schurortho1} and do $Spin(d-1)$ sums. Equation~\eqref{eq:schurortho1} then becomes
\be
&\sum_{\seq{m}_{d-1},\seq{m}'_{d-1}}\int_{Spin(d)}\<\bfm_d\seq{m}_{d-1}|R|\bfm_d\seq{m}'_{d-1}\>\<\widetilde{\bfm}_d{\seq{m}}'_{d-1}|R^{-1}|\widetilde{\bfm}_d{\seq{m}}_{d-1}\>dR=\nn\\
&=\frac{\dim \bfm_{d-1}\dim\bfm'_{d-1}}{\dim\bfm_d}\delta _{\bfm_d,\widetilde\bfm_d}.
\ee
We then write $R$ as $R=KAK'$, where $A=e^{\theta M_{12}}$ for some $\theta$ and $K,K'\in Spin(d-1)$.\footnote{This follows from a standard choice of coordinates on $Spin(d)$, which follows from $Spin(d)/Spin(d-1)=S^{d-1}$ : an element on the sphere can be obtained from a fixed point by $KA$ and $K'$ comes from $Spin(d-1)$ equivalence class.} In the left hand side $K$ and $K'$ cancel out due to $Spin(d-1)$ invariance of the contractions, resulting in
\be
&\sum_{\seq{m}_{d-1},\seq{m}'_{d-1}}\int_{Spin(d)}\<\bfm_d\seq{m}_{d-1}|e^{\theta(R) M_{12}}|\bfm_d\seq{m}'_{d-1}\>\<\widetilde{\bfm}_d{\seq{m}}'_{d-1}|e^{-\theta(R) M_{12}}|\widetilde{\bfm}_d{\seq{m}}_{d-1}\>dR=\nn\\
&=\sum_{\bfm_{d-2}}\dim\bfm_{d-2}\, \int_{Spin(d)}P^{\bfm_d,\bfm_{d-2}}_{\bfm_{d-1},\bfm'_{d-1}}(\theta(R))\left(P^{\widetilde\bfm_d,\bfm_{d-2}}_{\bfm_{d-1},\bfm'_{d-1}}(\theta(R))\right)^* dR.
\ee
By using explicit coordinates on $Spin(d)$ one can show that, for $d>2$
\be
\int_{Spin(d)}f(\theta(R))dR=\frac{\Gamma(\frac{d}{2})}{\sqrt\pi \Gamma(\frac{d-1}{2})}\int_0^\pi \sin^{d-2}\theta f(\theta)d\theta.
\ee
Putting everything together, we obtain the following orthogonality relation
\begin{multline}
\sum_{\bfm_{d-2}}\dim\bfm_{d-2}\, \int_{0}^\pi P^{\bfm_d,\bfm_{d-2}}_{\bfm_{d-1},\bfm'_{d-1}}(\theta)\left(P^{\widetilde\bfm_d,\bfm_{d-2}}_{\bfm_{d-1},\bfm'_{d-1}}(\theta)\right)^* \sin^{d-2}\theta d\theta=\\=\frac{\sqrt\pi \Gamma(\frac{d-1}{2})}{\Gamma(\frac{d}{2})}\frac{\dim \bfm_{d-1}\dim\bfm'_{d-1}}{\dim\bfm_d}\delta _{\bfm_d,\widetilde\bfm_d}.
\end{multline}

\subsubsection{Computational techniques}
\label{sec:structure:Pfunctions:computation}

In the remainder of this section we discuss how $P$-functions can be computed in practice, first in general and then in specific examples.

The conceptually simplest computational scheme follows immediately from the definition~\eqref{eq:Pdefinition} as a matrix element of $e^{\theta M_{12}}$. Indeed, since we know the matrix elements of $M_{12}$ (see section~\ref{sec:reptheory:ClebshGordan} and appendix~\ref{app:formulae}), we can find the matrix corresponding to $M_{12}$ in any given representation and then exponentiate it by the standard methods. When doing this, one can reduce the amount of calculation by taking note of the structure of the right hand side of~\eqref{eq:Pdefinition}. Following this strategy, we simultaneously produce
\be
P^{\bfm_d,\bfm_{d-2}}_{\bfm_{d-1},\bfm'_{d-1}}(\theta)
\ee
with fixed $\bfm_d$ and $\bfm_{d-2}$ for all choices of $\bfm_{d-1}$ and $\bfm'_{d-1}$. 

This strategy is therefore somewhat of an overkill for our purposes, since in a four-point function the possible choices of representations $\bfm_{d-1}$ and $\bfm'_{d-1}$ are prescribed by the spins of external representations, while $\bfm_d$ and $\bfm_{d-2}$ take on all the values allowed by each pair of $\bfm_{d-1}$ and $\bfm'_{d-1}$.\footnote{Also, the size of the matrix which one needs to exponentiate grows with the spin $m_{d,1}$, which makes this approach computationally more intensive.} Fortunately, there exist techniques which compute $P^{\bfm_d,\bfm_{d-2}}_{\bfm_{d-1},\bfm'_{d-1}}(\theta)$ for fixed $\bfm_{d-1}$ and $\bfm'_{d-1}$.

Let us fix $\bfm_{d-1}$ and $\bfm'_{d-1}$. Furthermore, write $\bfm_d=(j,\widetilde\bfm_{d-2})$, i.e.\ define $j\equiv m_{d,1}$ and think of the rest of $\bfm_d$ as a $(d-2)$-dimensional weight $\widetilde\bfm_{d-2}$. Note that $\bfm_{d-1},\bfm'_{d-1}\in\bfm_d$ requires $j\geq \max(m_{d-1,1},m'_{d-1,1})$. Assuming that this holds, it is easy to check that the following two statements are equivalent,
\be
\bfm_d\ni \bfm_{d-1},\bfm'_{d-1}\Longleftrightarrow \widetilde\bfm_{d-2}\in \bfm_{d-1},\bfm'_{d-1}.
\ee
In other words, $\widetilde{\bfm}_{d-2}$ satisfies the same requirements as $\bfm_{d-2}$. This means that we can arrange $P^{(j,\widetilde\bfm_{d-2}),\bfm_{d-2}}_{\bfm_{d-1},\bfm'_{d-1}}(\theta)$ into a square matrix $\bfP^j_{\bfm_{d-1},\bfm'_{d-1}}(\theta)$ with rows and columns labeled by $\widetilde{\bfm}_{d-2}$ and $\bfm_{d-2}$ respectively. Such matrices are discussed, for example, in~\cite{RepresentationsAndSpecialFunctions} (and references therein), where they are shown to satisfy certain second-order matrix differential equations, and methods for solving these equations were developed. Following the terminology of~\cite{RepresentationsAndSpecialFunctions}, we will refer to these matrices as ``matroms''. Note that the size of the matrom is independent of $j$ and is only determined by $\bfm_{d-1}$ and $\bfm'_{d-1}$. Furthermore, all (if any) components of a given matrom appear in a given four-point function.

Potentially, the results described in~\cite{RepresentationsAndSpecialFunctions} may allow one to find analytic in $j$ expressions for the matroms $\bfP^j_{\bfm_{d-1},\bfm'_{d-1}}$ in terms of known special functions. Unfortunately, we were not able to devise a complete computational algorithm based on these results.\footnote{It is an interesting problem to complete the results described in~\cite{RepresentationsAndSpecialFunctions} to find a general algorithm for constructing analytic expressions for generic matroms.} However, since in numerical applications one requires $\bfP^j_{\bfm_{d-1},\bfm'_{d-1}}$ for all $j$ up to a certain cutoff, it is convenient to use a recursion relation in $j$ as described below. Expressions analytic in $j$ can still be obtained in a number of cases, as we discuss in the next subsections.

The basic idea is to consider the product
\be
\<\seq{m}_{d}|e^{\theta M_{12}}|\seq{m}'_{d}\>\<\myng{(1)},\bullet,\ldots|e^{\theta M_{12}}|\myng{(1)},\bullet,\ldots\>=\<\seq{m}_{d}|e^{\theta M_{12}}|\seq{m}'_{d}\>\cos\theta.
\ee
The left hand side is a matrix element in $\bfm_{d}\otimes \myng{(1)}$ and thus can be decomposed as a sum of matrix elements in various irreducible representations,
\be
&\<\seq{m}_{d}|e^{\theta M_{12}}|\seq{m}'_{d}\>\<\myng{(1)},\bullet,\ldots|e^{\theta M_{12}}|\myng{(1)},\bullet,\ldots\>=\nn\\
&\quad=\sum_{\widetilde\bfm_d\in\bfm_d\otimes\myng{(1)}}\<\widetilde\bfm_d\seq{m}_{d-1}|e^{\theta M_{12}}|\widetilde\bfm_d\seq{m}'_{d-1}\>
\rcgc{\bfm_d}{\bfm_{d-1}}{\myng{(1)}}{\bullet}{\widetilde\bfm_d}{\bfm_{d-1}}
\rcgc{\bfm_d}{\bfm'_{d-1}}{\myng{(1)}}{\bullet}{\widetilde\bfm_d}{\bfm'_{d-1}}^*.
\ee
One can easily see that in terms of matroms this leads to the following recursion relation,
\be\label{eq:matromrecursion}
\bfA^+_j\bfP^{j+1}+\bfA^-_j\bfP^{j-1}+\bfB_j\bfP^{j}=\cos\theta\,\bfP^j,
\ee
where $\bfA^\pm_j$, $\bfB^j$ are some matrices,\footnote{The matrices $\bfA$ are, importantly, diagonal, which makes it easy to invert $\bfA^+_j$.} and we have suppressed the dependence of everything on $\bfm_{d-1},\bfm'_{d-1}$ for simplicity of notation. Starting from the smallest possible $j$ (for which we can compute $\bfP^j$ by, say, exponentiation), one can use this relation to find $\bfP^j$ for higher $j$.

As an example, consider the matroms in $d=2n$ with $\bfm_{d-1}=\bfm'_{d-1}=(\thalf,\ldots,\thalf)$, which will be useful in the example of section~\ref{sec:structure:SeedConformalBlocks}. There are two representations in the dimensional reduction of $\bfm_{d-1}=\bfm'_{d-1}$, $\bfm_{d-2}=(\thalf,\ldots,\pm\thalf)$, i.e.\ the two fermionic representations in $d-2$ dimensions. We therefore have a $2\times2$ matrom 
\be
\bfP^j = \begin{pmatrix}
	P_{(\thalf,\ldots,\thalf),(\thalf,\ldots,\thalf)}^{(j,\thalf,\ldots,+\thalf),(\thalf,\ldots,+\thalf)} (\theta)& P_{(\thalf,\ldots,\thalf),(\thalf,\ldots,\thalf)}^{(j,\thalf,\ldots,+\thalf),(\thalf,\ldots,-\thalf)} (\theta)\\
	P_{(\thalf,\ldots,\thalf),(\thalf,\ldots,\thalf)}^{(j,\thalf,\ldots,-\thalf),(\thalf,\ldots,+\thalf)} (\theta)& P_{(\thalf,\ldots,\thalf),(\thalf,\ldots,\thalf)}^{(j,\thalf,\ldots,-\thalf),(\thalf,\ldots,-\thalf)} (\theta)
\end{pmatrix}.
\ee
For example, one can easily check that for any $d$
\be
\bfP^{\half} = \begin{pmatrix}
	e^{-i \theta/2} & e^{+i \theta/2} \\
	e^{+i \theta/2} & e^{-i \theta/2}
\end{pmatrix}.
\ee
By using the explicit formulas for the isoscalar factors from appendix~\ref{app:formulae:Isoscalar}, one can show that the recursion relation~\eqref{eq:matromrecursion} reduces in this case to
\be
\frac{j+2n-\tfrac{3}{2}}{j+n-\thalf}\bfP^{j+1}+
\frac{j-\tfrac{1}{2}}{j+n-\tfrac{3}{2}}\bfP^{j-1}+
\frac{n-1}{(j+n-\thalf)(j+n-\tfrac{3}{2})}
\begin{pmatrix}
	0 & 1 \\ 1 & 0
\end{pmatrix}\bfP^j=2\cos\theta\,\bfP^j,
\ee
where $n=d/2$. For instance, applying this relation twice, we find
\be
\bfP^{2+\thalf} = \frac{n-1}{2n-1}
\begin{pmatrix}
	\half \frac{n+1}{n-1} e^{-\frac{5}{2}i\theta}
	+e^{-\frac{1}{2}i\theta}
	+\half e^{+\frac{3}{2}i\theta} 
	&
	\half \frac{n+1}{n-1} e^{+\frac{5}{2}i\theta}
	+e^{+\frac{1}{2}i\theta}
	+\half e^{-\frac{3}{2}i\theta} \\
	\half \frac{n+1}{n-1} e^{+\frac{5}{2}i\theta}
	+e^{+\frac{1}{2}i\theta}
	+\half e^{-\frac{3}{2}i\theta} &
	\half \frac{n+1}{n-1} e^{-\frac{5}{2}i\theta}
	+e^{-\frac{1}{2}i\theta}
	+\half e^{+\frac{3}{2}i\theta} 
\end{pmatrix},
\ee
valid for any $d=2n$. The general solution can be expressed in terms of Jacobi polynomials as\footnote{To find this solution, we first diagonalized the recursion relation and then matched it to the recursion relation for Jacobi polynomials. The Jacobi polynomials entering this expression can in principle be expressed in terms of linear combinations of Gegenbauer polynomials.}
\be\label{eq:fermionmatrom}
\bfP^j=\frac{(j-\thalf)!}{(n-\thalf)_{j-\half}}\left[\cos\tfrac{\theta}{2} P^{(n-\frac{3}{2},n-\frac{1}{2})}_{j-\half}(\cos\theta)\begin{pmatrix}
	1 & 1 \\ 1 & 1
\end{pmatrix}+
\sin\tfrac{\theta}{2}P^{(n-\frac{1}{2},n-\frac{3}{2})}_{j-\half}(\cos\theta)\begin{pmatrix}
	-i & i \\ i & -i
\end{pmatrix}\right].
\ee

\subsubsection{Contribution of $\bR\times Spin(d)$ multiplets in terms of matroms}
\label{sec:structure:Pfunctions:matromForm}

Having introduced the matroms $\bfP$ in the previous subsection, it makes sense to reanalyse~\eqref{eq:smallunicornformula} in terms of them. For fixed $a$ and $b$ as in~\eqref{eq:3ptrightshorthand} and~\eqref{eq:3ptleftshorthand} denote by
\be
\bfT^{ba}
\ee
the column vector built out of four-point tensor structures
\be
\left[
{\seq{m}^3_{d}\atop\seq{m}^4_{d}}\Big\vert b
\Big\vert
\bfm_{d-2}
\Big\vert
a\Big\vert
{\seq{m}^1_{d}\atop\seq{m}^2_{d}}
\right]
\ee
with $\bfm_{d-2}$ running through all allowed values. Also, denote $\bfP^j_{ba}\equiv\bfP^j_{\bfm_{d-1},\bfm'_{d-1}}$. Finally, let 
\be
\mathbf{\Lambda}^{ba}_j
\ee
be the row vector built out of 
\be
\bar\lambda^b_{\Delta,\bfm_d}\lambda^a_{\Delta,\bfm_d}
\ee
corresponding to all $\bfm_d=(j,\widetilde\bfm_{d-2})$ which can contribute to the given pair $a,b$ according to~\eqref{eq:selectionrules}, summed over degenerate multiplets. If we are considering the contribution of a single $\bR\times Spin(d)$ multiplet, then this vector contains a single non-zero element, but at this point it is convenient to also allow several contributions. We then have
\be\label{eq:matromForm}
&\sum_{\bfm_d,m_{d,1}=j}\sum_{a,b}\sum_{\bfm_{d-2}}
\bar\lambda^b_{\Delta,\bfm_d}\lambda^a_{\Delta,\bfm_d}
r^\Delta P^{\bfm_d,\bfm_{d-2}}_{\bfm_{d-1},\bfm'_{d-1}}(\theta)\times\left[
{\seq{m}^3_{d}\atop\seq{m}^4_{d}}\Big\vert b
\Big\vert
\bfm_{d-2}
\Big\vert
a\Big\vert
{\seq{m}^1_{d}\atop\seq{m}^2_{d}}
\right]=\nn\\
&=r^\Delta\,\sum_{a,b} \mathbf{\Lambda}_j^{ba}\cdot \bfP^j_{ba}(\theta)\cdot \bfT^{ba}.
\ee
As we discuss in section~\ref{sec:structure:FolkloreTheorem}, in this equation $\mathbf\Lambda^{ba}_j$ correspond, roughly speaking, to the space of conformal blocks, while $\bfT^{ba}$ correspond to the space of four-point tensor structures. The matroms link these two spaces together, giving a realization of the folklore theorem~\cite{Dymarsky:2013wla,Elkhidir:2014woa} (see section~\ref{sec:structure:FolkloreTheorem}).

In the rest of this section we consider some more explicit examples. First, we recover the Gegenbauer polynomials relevant to the scalar correlation functions and then we consider the low-dimensional cases $d=3$ and $d=4$.

\subsubsection{Scalar matrom}
\label{sec:structure:Pfunctions:scalarmatrom}

Let us consider the simplest $P$-function $P_{\bullet,\bullet}^{\bfj,\bullet}(\theta)$, which is the only component of the simplest scalar matrom $\bfP^j_{\bullet,\bullet}(\theta)$. Analogously to the example considered above, we could write down the recursion relation~\eqref{eq:matromrecursion} for this matrom and recognize that, together with the initial condition $P_{\bullet,\bullet}^{\bullet,\bullet}(\theta)\equiv 1$, it is solved by
\be\label{eq:gegenbauer}
P_{\bullet,\bullet}^{\bfj,\bullet}(\theta)=\frac{C^{(\nu)}_j(\cos\theta)}{C^{(\nu)}_j(1)},
\ee
where $\nu=(d-2)/2$. However, it is instructive to take another approach to arrive at this result. Consider the tensor given by
\be
e_1^{\mu_1}\cdots e_1^{\mu_j}-\text{traces}.
\ee
Obviously, this tensor is an element of $\bfj$ of $Spin(d)$. On the other hand, it transforms under the trivial representation of $Spin(d-1)$. Therefore, we have
\be
e_1^{\mu_1}\cdots e_1^{\mu_j}-\text{traces}\propto |\bfj,\bullet,\ldots\>.
\ee
Acting with $e^{\theta M_{12}}$, we find that
\be
e^{\theta M_{12}}|\bfj,\bullet,\ldots\>\propto e_1^{\mu_1}(\theta)\cdots e_1^{\mu_j}(\theta)-\text{traces},
\ee
where $e_1(\theta)=\cos\theta e_1+\sin\theta e_2$. This implies
\be
P_{\bullet,\bullet}^{\bfj,\bullet}(\theta)=\<\bfj,\bullet,\ldots|e^{\theta M_{12}}|\bfj,\bullet,\ldots\>\propto
(e_{1,\mu_1}\cdots e_{1,\mu_j}-\text{traces})(e_1^{\mu_1}(\theta)\cdots e_1^{\mu_j}(\theta)-\text{traces}).
\ee
The right hand side of this equation is known to be proportional to the Gegenbauer polynomial $C^{(\nu)}_j(e_1\cdot e_1(\theta))=C^{(\nu)}_j(\cos\theta)$. Combining this with the normalization condition $P_{\bullet,\bullet}^{\bfj,\bullet}(0)=1$, we recover~\eqref{eq:gegenbauer}. 

This strategy generalizes to other tensor representations and also allows one to relate $P$-functions to the irreducible projectors studied recently in~\cite{Costa:2016hju}. We discuss this further in appendix~\ref{app:tensors}.

\subsubsection{3 dimensions}

We now consider the case $d=3$. As discussed in section~\ref{sec:structure:Contribution}, the 3-dimensional GT matrix elements $P^j_{m,m'}(\theta)$ are given by~\eqref{eq:P3d},\footnote{We use the convention consistent with Mathematica's \texttt{WignerD[$\{j,m,m'\}$,$\theta$]}.}
\be
P^{j}_{m,m'}(\theta)=\<j,m|e^{\theta M_{12}}|j,m'\>=\<j,m|e^{-i\theta J_{\hat 2}}|j,m'\>=d^j_{m,m'}(-\theta).
\ee
Note that in $3d$ all matroms are $1\times 1$ and coincide with the above functions. There is not much to add here, except for the explicit formula for the small Wigner $d$-matrix $d^j_{m,m'}(\theta)$,
\be
	d^j_{m,m'}(\theta)=(-1)^{m-m'}\sqrt{\frac{(j+m')!(j-m')!}{(j+m)!(j-m)!}}
	\left(\sin\frac{\beta}{2}\right)^{m'-m}\left(\cos\frac{\beta}{2}\right)^{m'+m}
	P^{(m'-m,m'+m)}_{j-m'}(\cos\theta),
\ee
where in this expression $P^{(a,b)}_n$ are the Jacobi polynomials.

\subsubsection{4 dimensions}
In 4d we have the following definition of GT matrix elements $P^{l_L,l_R;m}_{j,j'}(\theta)$,
\be
P^{l_L,l_R;m}_{j,j'}(\theta)=\<l_L,l_R;j,m|e^{\theta M_{12}}|l_L,l_R;j',m\>.
\ee
We can compute them by going to the $SU(2)\times SU(2)$ basis,
\be
P^{l_L,l_R;m}_{j,j'}(\theta)&=\<l_L,l_R;j,m|e^{\theta M_{12}}|l_L,l_R;j',m\>\nn\\
&=\sum_{m_L+m_R=m}\sum_{m'_L+m'_R=m}\<l_L,m_L;l_R,m_R|e^{\theta M_{12}}|l_L,m'_L;l_R,m'_R\>\times\nn\\
&\qquad \times\<j,m|l_L,m_L;l_R,m_R\>\<l_L,m'_L;l_R,m'_R|j',m\>.
\ee
Using~\eqref{eq:4dchiralM12}, we find
\be
\<l_L,m_L;l_R,m_R|e^{\theta M_{12}}|l_L,m'_L;l_R,m'_R\>&=\<l_L,m_L;l_R,m_R|e^{-i\theta J^L_{\hat 3}+i\theta J^R_{\hat 3}}|l_L,m'_L;l_R,m'_R\>\nn\\
&=e^{-i(m_L-m_R)}\delta_{m_Lm_L'}\delta_{m_Rm_R'},
\ee
and thus
\be\label{eq:4dP}
P^{l_L,l_R;m}_{j,j'}(\theta)=\sum_{k=-l_L-l_R}^{l_L+l_R}\Big\<j,m\Big|l_L,\frac{m+k}{2};l_R,\frac{m-k}{2}\Big\>\Big\<l_L,\frac{m+k}{2};l_R,\frac{m-k}{2}\Big|j',m\Big\>e^{-ik\theta}.
\ee
Note that in this formula the summation is over (half-)integral values of $k$ for (half-)integral values of $\ell_1=l_L+l_R$, and whenever the Clebsch-Gordan coefficient is undefined, we assume that it is equal to zero. Thus the range of summation is effectively restricted to
\be
\{-2l_L-m,\ldots,2l_L-m\}\cap\{-2l_R+m,\ldots,2l_R+m\}.
\ee
For example, if $m=l_L+l_R$, then only $k=l_L-l_R$ enters the sum. (Also necessarily $j=j'=l_L+l_R$.)

\section{Casimir equation}
\label{sec:casimir}

In this section we derive Casimir recursion relation for the series expansion of spinning conformal blocks. We first rederive the results of~\cite{Hogervorst:2013sma} for scalar conformal blocks in a more streamlined way and then extend these results to arbitrary spinning conformal blocks. As an example, we explicitly work out the recursion relations for general 3d conformal blocks and for general seed blocks in arbitrary $d$.

In this section we will work in coordinates different from those in section~\ref{sec:structure}. In particular, we set
\be\label{eq:DOcoords}
	w_1=0,\quad w_2=z,\quad w_3=1,\quad w_4=+\infty.
\ee
We use the following definition of $\cO_4(+\infty)$,
\be
	\cO_4(+\infty)\equiv\lim_{L\to+\infty}L^{2\Delta_4}\cO_4(Le_1).
\ee
Note that we do not act in any way on the spin indices of $\cO_4$ when taking this limit.\footnote{When $\cO_4$ is tensor, one often acts on its indices with reflection along $e_1$ when taking this limit. This is done because $\cO_4(\infty)$ defined our way effectively transforms in the representation reflected to $\bfm_d^4$. When $\bfm_d^4$ is tensor, its reflection is equivalent to $\bfm_d^4$ and thus one may find it convenient to act on $\cO_4$ with the map which furnishes this equivalence. More generally, the reflected representation can be different from $\bfm_d^4$ and thus there is no benefit in acting on spin indices of $\cO_4$ within our general treatment.} The results of section~\ref{sec:structure} translate to this case without essential modification (except for changing the insertion point of the operators in all formulas).

We use~\eqref{eq:DOcoords} because the Casimir recursion relations take the simplest form in these coordinates, analogously to the case of scalar blocks~\cite{Hogervorst:2013sma}. The recursion relations in $\rho$-coordinates, unfortunately, take a much more complicated form~\cite{Hogervorst:2013kva,Costa:2016xah}.

\subsection{Review of scalar conformal blocks}
\label{sec:casimir:ReviewScalar}

Consider the scalar conformal block for exchange of a primary operator $\cO$
\be\label{eq:scalarblock}
	G_\cO(s,\phi)\equiv \<0|\phi_4(\infty)\phi_3(1) |\cO| s^{D}e^{\theta M_{12}}\phi_2(1)\phi_1(0)|0\>,
\ee
where $z=se^{i\theta}$, we have used the convention~\eqref{eq:fourptconvenient} for writing the four-point functions,\footnote{In the scalar case~\eqref{eq:fourptconvenient} differs from~\eqref{eq:fourptstart} only by the factor $s^{\Delta_1+\Delta_2}$.} and $|\cO|$ is the projection operator on the conformal family of $\cO$,
\be\label{eq:conformalprojector}
	|\cO|=\sum_{p\geq 0,\bfm_d,\seq{m}_{d},q}|\Delta_p,\seq{m}_{d},q\>\<\Delta_p,\seq{m}_{d},q|,
\ee
where the sum is over an orthonormal basis of descendants of $\cO$. Here $\Delta_p=\Delta_\cO+p$ is the scaling dimension of a level-$p$ descendant, $\bfm_d$ is the $Spin(d)$ representation of the descendant, and $q$ labels the possible degeneracies which arise when there are several descendants in representation $\bfm_d$ at level $p$. 

The results of section~\ref{sec:structure} and in particular~\ref{sec:structure:ScalarCorrelators} tell us what is the most general contribution of a single term of~\eqref{eq:conformalprojector} to~\eqref{eq:scalarblock}. We therefore have
\be\label{eq:scalargengenbauer}
	G_\cO(s,\phi)=\sum_{p=0}^\infty\sum_{j=0}^\infty\sum_q \lambda_{\bullet,p,j,q}^{\bullet,\bullet}\overline{\lambda}_{\bullet,p,j,q}^{\bullet,\bullet}s^{\Delta_p}P_{\bullet,\bullet}^{\bfj,\bullet}(\theta)=
	\sum_{p=0}^\infty \sum_{j=0}^\infty \Lambda_{p,j} s^{\Delta_\cO+p}\frac{C_j^{(\nu)}(\cos\theta)}{C_j^{(\nu)}(1)}.
\ee
We have defined
\be
	\Lambda_{p,j}\equiv \sum_q \lambda_{\bullet,p,j,q}^{\bullet,\bullet}\overline{\lambda}_{\bullet,p,j,q}^{\bullet,\bullet}.
\ee
The range of $j$ is in fact restricted by the spectrum of descendants at each level $p$ so that $|j-j_\cO|\leq p$, but we will ignore this by assuming that $\Lambda_{p,j}=0$ for $p,j$ outside this range. While this expansion respects $\bR\times Spin(d)$ symmetry, it doesn't tell us what the coefficients $\Lambda_{p,j}$ are.

These coefficients are constrained by consistency of expansion~\ref{eq:conformalprojector} with the full conformal symmetry. It was noticed in~\cite{DO2} that it suffices to ensure consistency with the action of the quadratic conformal Casimir operator. Usually this is condition is formulated in a form of differential equation~\cite{DO2,DO3}. When applied to~\eqref{eq:scalargengenbauer}, this equation immediately yields a one-step recursion relation for the coefficients $\Lambda_{p,j}$~\cite{Hogervorst:2013sma},
\be\label{eq:scalarrecursion}
	(C_{p,j}-C_{0,j_\cO})\Lambda_{p,j}=\Gamma^+_{p-1,j-1}\Lambda_{p-1,j-1}+\Gamma^-_{p-1,j+1}\Lambda_{p-1,j+1},
\ee
where coefficients $\Gamma^\pm_{p,j}$ are given by\footnote{In~\cite{Hogervorst:2013sma} these coefficients are given with $\Delta_{12}=\Delta_{34}=0$, but it is trivial to generalize their argument.}
\be\label{eq:gammacoeffs}
	\Gamma^+_{p,j}&=\frac{(\Delta_p+j-\Delta_{12})(\Delta_p+j+\Delta_{34})(j+d-2)}{2j+d-2},\nn\\
	\Gamma^-_{p,j}&=\frac{(\Delta_p-j-d+2-\Delta_{12})(\Delta_p-j-d+2+\Delta_{34})j}{2j+d-2},
\ee
with $\Delta_{ij}=\Delta_i-\Delta_j$, while the Casimir eigenvalues are given by
\be
	C_{p,j}=\Delta_p(\Delta_p-d)+j(j+d-2).
\ee

This result is remarkably simple, much simpler than the intermediate steps in the derivation of~\cite{Hogervorst:2013sma} would suggest. In fact, it is not a priori obvious from that derivation that the recursion relation should take such a simple form. For example, when repeated in $\rho$-coordinates, essentially the same derivation leads to a much more complicated recursion relation. We are therefore motivated to look for a more conceptual derivation of~\eqref{eq:scalarrecursion}, which manifests this simple structure. 

Let us start from the definition of the conformal Casimir operator,
\be
	\cC=D(D-d)+\cC_{Spin(d)}-P\cdot K,
\ee
where $\cC_{Spin(d)}$ is the $Spin(d)$ quadratic Casimir defined as
\be
	\cC_{Spin(d)}=-\frac{1}{2}M_{\mu\nu}M^{\mu\nu}.
\ee
The key property of $\cC$ is that it commutes with all conformal generators and thus acts on all the descendants of $\cO$ by the same eigenvalue as on $\cO$. That eigenvalue can be computed by 
\be
	\cC|\cO\>&= \left(D(D-d)+\cC_{Spin(d)}-P\cdot K\right)|\cO\>=C(\cO)|\cO\>,\\
	C(\cO)&=\Delta_\cO(\Delta_\cO-d)+C_{Spin(d)}(\bfm_d^\cO),
\ee
where we used $K_\mu|\cO\>=0$, and $C_{Spin(d)}(\bfm_d)$ is the $Spin(d)$ quadratic Casimir eigenvalue corresponding to the $Spin(d)$ representation $\bfm_d$. It is given by
\be\label{eq:spindcasimir}
	C_{Spin(d)}(\bfm_d)=\sum_{k=1}^{\floor{d/2}} m_{d,k}(m_{d,k}+d-2k).
\ee
For future convenience, let us define for any (not necessarily primary) $\bR\times Spin(d)$ multiplet the number
\be\label{eq:confcasimir}
	C(\Delta,\bfm_d)\equiv \Delta(\Delta-d)+C_{Spin(d)}(\bfm_d).
\ee
It is the eigenvalue of the operator
\be
	\widetilde{\cC}\equiv {\cC}+P\cdot K=D(D-d)+\cC_{Spin(d)}.
\ee
Note that $P\cdot K=K^\dagger\cdot K\succeq 0$ for $\Delta$ above unitarity bound and thus we always have in such cases
\be\label{eq:CasimirInequality}
	\widetilde\cC\succeq \cC.
\ee

Since $\cC$ takes the same eigenvalue on all states in a conformal multiplet, we have
\be
	|\cO|\cC=|\cO| C(\cO).
\ee
This implies the following operator version of the Casimir equation,
\be
	\<0|\phi_4\phi_3 |\cO|\cC s^{D}e^{\theta M_{12}}\phi_2\phi_1|0\>=
	C(\cO)\<0|\phi_4\phi_3 |\cO| s^{D}e^{\theta M_{12}}\phi_2\phi_1|0\>.
\ee
For notational simplicity, we have omitted the positions of the operators, which are the same as in~\eqref{eq:scalarblock}. The standard Casimir differential equation can be obtained by acting with $\cC$ on the right in the left hand side of this equation and expressing this action in terms of derivatives in $\theta$ and $s$. We will take another approach, rewriting the left hand side instead as
\be
	\<0|\phi_4\phi_3 |\cO|\cC s^{D}e^{\theta M_{12}}\phi_2\phi_1|0\>=&
	\<0|\phi_4\phi_3 |\cO|(\widetilde\cC-P^\mu K_\mu) s^{D}e^{\theta M_{12}}\phi_2\phi_1|0\>\nn\\
	=&\<0|\phi_4\phi_3 |\cO|\widetilde\cC s^{D}e^{\theta M_{12}}\phi_2\phi_1|0\>-\<0|\phi_4\phi_3 |\cO| P^\mu K_\mu s^{D}e^{\theta M_{12}}\phi_2\phi_1|0\>\nn\\
	=&\<0|\phi_4\phi_3 |\cO|\widetilde\cC s^{D}e^{\theta M_{12}}\phi_2\phi_1|0\>-\<0|\phi_4\phi_3 P^\mu |\cO| K_\mu s^{D}e^{\theta M_{12}}\phi_2\phi_1|0\>,
\ee
where in the last line we have used the conformal invariance of the projector $|\cO|$, i.e.\ that it commutes with all conformal generators. Rearranging, we find
\be\label{eq:operatorCasimirEquation}
	\<0|\phi_4\phi_3 |\cO|(\widetilde\cC-\cC) s^{D}e^{\theta M_{12}}\phi_2\phi_1|0\>=\<0|\phi_4\phi_3 P^\mu |\cO| K_\mu s^{D}e^{\theta M_{12}}\phi_2\phi_1|0\>.
\ee
We will now derive the recursion relation~\eqref{eq:scalarrecursion} by evaluating both sides of this equation with the help of~\eqref{eq:conformalprojector}.

\subsubsection{Left hand side}
To warm up, let us consider the left hand side of this equation first. Using~\eqref{eq:conformalprojector}, we find
\be\label{eq:casimirlhsfinal}
	\<0|\phi_4\phi_3& |\cO|(\widetilde\cC-\cC) s^{D}e^{\theta M_{12}}\phi_2\phi_1|0\>\nn\\
	=&\sum_{p,\bfm_d,\seq{m}_{d},q}\<0|\phi_4\phi_3 |\Delta_p,\seq{m}_{d},q\>\<\Delta_p,\seq{m}_{d},q|(\widetilde\cC-\cC) s^{D}e^{\theta M_{12}}\phi_2\phi_1|0\>\nn\\
	=&\sum_{p,\bfm_d,\seq{m}_{d},q}\left(C(\Delta_p,\bfm_d)-C(\Delta_\cO,\bfj_\cO)\right)\<0|\phi_4\phi_3 |\Delta_p,\seq{m}_{d},q\>\<\Delta_p,\seq{m}_{d},q|s^{D}e^{\theta M_{12}}\phi_2\phi_1|0\>\nn\\
	=&\sum_{p=0}^{\infty}\sum_{j=0}^\infty (C_{p,j}-C_{0,j_\cO})\Lambda_{p,j} s^{\Delta_\cO+p}\frac{C_j^{(\nu)}(\cos\phi)}{C_j^{(\nu)}(1)},
\ee
where the last line follows similarly to~\eqref{eq:scalargengenbauer}, and we also made use of the fact that we arranged the descendants into $\bR\times Spin(d)$ multiplets. We can already see that we are on the right track -- the coefficients in this expansion exactly reproduce the left hand side of the recursion relation~\eqref{eq:scalarrecursion}.

\subsubsection{Right hand side}
Let us now analyze the less trivial right hand side of~\eqref{eq:operatorCasimirEquation}. We first look at the contribution of a single term of~\eqref{eq:conformalprojector}. For simplicity of notation, we will omit the degeneracy index $q$ for now and restore it later. We thus consider
\be\label{eq:casimirrhs}
	\sum_{\seq{m}_{d}}\<0|\phi_4\phi_3 P^\mu |\Delta_p,\seq{m}_{d}\>\<\Delta_p,\seq{m}_{d}| K_\mu s^{D}e^{\theta M_{12}}\phi_2\phi_1|0\>.
\ee
\paragraph{Left three-point structure} We will first evaluate the left three-point function by commuting $P$ on the left. We have (see appendix~\ref{app:conformalalgebra} for our conventions on conformal algebra)
\be\label{eq:left3ptstep1}
	\<0|\phi_4(\infty)\phi_3(1) P_\mu |\Delta_p,\seq{m}_{d}\>=-\<0|\phi_4(\infty)\ptl_\mu\phi_3(1)|\Delta_p,\seq{m}_{d}\>.
\ee
The crucial point is that the knowledge of $\<0|\phi_4(\infty)\phi_3(1)|\Delta_p,\seq{m}_{d}\>$ and $\bR\times Spin(d)$ invariance allow us to evaluate 
\be
	\<0|\phi_4(\infty)\phi_3(x)|\Delta_p,\seq{m}_{d}\>
\ee
for any $x\in \bR^d$. In particular, we can compute the right hand side of~\eqref{eq:left3ptstep1}. For example, note that 
\be\label{eq:PscalarAction}
	\<0|\phi_4(\infty)\ptl_1\phi_3(1)|\Delta_p,\seq{m}_{d}\>=&-\<0|\phi_4(\infty)\phi_3(1)(D+\Delta_3-\Delta_4)|\Delta_p,\seq{m}_{d}\>\nn\\
	=&-(\Delta_p+\Delta_3-\Delta_4)\<0|\phi_4(\infty)\phi_3(1)|\Delta_p,\seq{m}_{d}\>.
\ee
Here the first equality follows from action of $D$ on the left while the second equality follows from action on the right. The minus sign in front of $\Delta_4$ is due to the fact that we placed $\cO_4$ at infinity. Analogously, for $\mu\neq 1$,
\be\label{eq:PvectorActionPre}
	\<0|\phi_4(\infty)\ptl_\mu\phi_3(1)|\Delta_p,\seq{m}_{d}\>=-\<0|\phi_4(\infty)\phi_3(1)M_{1 \mu}|\Delta_p,\seq{m}_{d}\>.
\ee
Here we can act with $M_{1\mu}$ on the right by using the representation $\bfm_d$ for $M_{1\mu}$. As we discussed in section~\ref{sec:reptheory:ClebshGordan}, such actions can be described by means of a reduced matrix element,
\be
	\<\seq{m}'_{d}|M^{1\, \seq{u}_{d-1}}|\seq{m}_{d}\>=
	\Bigg(
		{\bfm_{d}\atop\bfm'_{d-1}}
	\Bigg\vert
		M^{\myng{(1)}}
	\Bigg\vert
		{\bfm_{d}\atop\bfm_{d-1}}
	\Bigg)\<\seq{m}'_{d-1}|\seq{m}_{d-1}\seq{u}_{d-1}\>.
\ee
We conclude
\be\label{eq:PvectorAction}
	\<0|\phi_4(\infty)\phi_3(1)P^{\seq{u}_d}|\Delta_p,\seq{m}_{d}\>=\sum_{\seq{m}'_{d}}&
	\Bigg(
	{\bfm_{d}\atop\bfm'_{d-1}}
	\Bigg\vert
	M^{\myng{(1)}}
	\Bigg\vert
	{\bfm_{d}\atop\bfm_{d-1}}
	\Bigg)\<\seq{m}'_{d-1}|\seq{m}_{d-1}\seq{u}_{d-1}\>\times\nn\\
	&\quad\times\<0|\phi_4(\infty)\phi_3(1)|\Delta_p,\seq{m}'_{d}\>,
\ee
where $\bfu_d=\bfu_{d-1}=\myng{(1)}$.

Note that the states $P^{\seq{u}_d}|\Delta_p,\seq{m}_{d}\>$ are just some other descendants of $\cO$. It is convenient to decompose them into the irreducible representations of $Spin(d)$ by defining the states
\be\label{eq:leftirrdescendants}
	|P,\Delta_p,\bfm_d;\widetilde{\seq{m}}_{d}\>\equiv\sum_{\seq{m}_{d},\seq{u}_d}\<\seq{m}_{d}\seq{u}_d|\widetilde{\seq{m}}_{d}\>\,\,P^{\seq{u}_d} |\Delta_p,\seq{m}_{d}\>,
\ee
where $\widetilde\bfm_d\in\myng{(1)}\otimes\bfm_d$ and $\<\seq{m}_{d}\seq{u}_d|\widetilde{\seq{m}}_{d}\>$ are the vector Clebsch-Gordan coefficients. We can decompose this sum according to $Spin(d-1)$ symmetry of the three-point functions as
\be\label{eq:leftirrdescendants2}
	|P,\Delta_p,\bfm_d;\widetilde{\seq{m}}_{d}\>=&\sum_{\seq{m}_{d}}P^{\myng{(1)},\bullet,\ldots}|\Delta_p,\seq{m}_{d}\>
	\<\seq{m}_{d};\,\myng{(1)},\bullet,\ldots|\widetilde{\seq{m}}_{d}\>
	\nn\\
	&+\sum_{\seq{m}_{d},\seq{u}_d\atop \bfu_{d-1}=\myng{(1)}}\,P^{\seq{u}_d} |\Delta_p,\seq{m}_{d}\>\<\seq{m}_{d}\seq{u}_d|\widetilde{\seq{m}}_{d}\>,\nn\\
	=&
	P^{\myng{(1)},\bullet,\ldots}|\Delta_p,\bfm_d\,\widetilde{\seq{m}}_{d-1}\>\Bigg(
	{\bfm_d\atop \widetilde\bfm_{d-1}}
	{\myng{(1)}\atop\bullet}
	\Bigg\vert
	{\widetilde\bfm_d\atop \widetilde\bfm_{d-1}}
	\Bigg)
	\nn\\
	&+\sum_{\seq{m}_{d},\seq{u}_d\atop \bfu_{d-1}=\myng{(1)}}\,P^{\seq{u}_d} |\Delta_p,\seq{m}_{d}\>
	\Bigg(
	{\bfm_d\atop \bfm_{d-1}}
	{\myng{(1)}\atop\myng{(1)}}
	\Bigg\vert
	{\widetilde\bfm_d\atop \widetilde\bfm_{d-1}}
	\Bigg)
	\<\seq{m}_{d-1}\seq{u}_{d-1}|\widetilde{\seq{m}}_{d-1}\>.
\ee
Here we made use of~\eqref{eq:isoscalar} and of the triviality of CG coefficients involving the trivial representation. Using equations~\eqref{eq:vece1phase},~\eqref{eq:PscalarAction} and~\eqref{eq:PvectorAction} we then find
\be\label{eq:leftuniversal}
	\<0|\phi_4(\infty)\phi_3(1)|P,\Delta_p,\bfm_d;\widetilde{\seq{m}}_{d}\>=&\left[{\widetilde\bfm_d\,\,\bfm_d\atop \widetilde\bfm_{d-1}
	}\right]^{34}_p\<0|\phi_4(\infty)\phi_3(1)|\Delta_p,\bfm_d\,\widetilde{\seq{m}}_{d-1}\>,
\ee
where
\be\label{eq:leftuniversalmatrixpre}
	\left[{\widetilde\bfm_d\,\,\bfm_d\atop \widetilde\bfm_{d-1}
	}\right]^{34}_p=&(-1)^d(\Delta_p+\Delta_3-\Delta_4)\Bigg(
	{\bfm_d\atop \widetilde\bfm_{d-1}}
	{\myng{(1)}\atop\bullet}
	\Bigg\vert
	{\widetilde\bfm_d\atop \widetilde\bfm_{d-1}}
	\Bigg)\nn\\
	&\qquad\quad+\sum_{\bfm_{d-1}}\Bigg(
	{\bfm_{d}\atop\widetilde\bfm_{d-1}}
	\Bigg\vert
	M^{\myng{(1)}}
	\Bigg\vert
	{\bfm_{d}\atop\bfm_{d-1}}
	\Bigg)\Bigg(
	{\bfm_d\atop \bfm_{d-1}}
	{\myng{(1)}\atop\myng{(1)}}
	\Bigg\vert
	{\widetilde\bfm_d\atop \widetilde\bfm_{d-1}}
	\Bigg).
\ee
As we discuss in appendix~\ref{app:formulae:SumRule}, the two terms in the last expression are in fact proportional to each other,
\be
	\sum_{\bfm_{d-1}}\Bigg(
	{\bfm_{d}\atop\widetilde\bfm_{d-1}}
	\Bigg\vert
	M^{\myng{(1)}}
	\Bigg\vert
	{\bfm_{d}\atop\bfm_{d-1}}
	\Bigg)\Bigg(
	{\bfm_d\atop \bfm_{d-1}}
	{\myng{(1)}\atop\myng{(1)}}
	\Bigg\vert
	{\widetilde\bfm_d\atop \widetilde\bfm_{d-1}}
	\Bigg)=(-1)^{d-1}(\bfm_d\,\myng{(1)}|\widetilde{\bfm}_{d})
	\Bigg(
	{\bfm_d\atop \widetilde\bfm_{d-1}}
	{\myng{(1)}\atop\bullet}
	\Bigg\vert
	{\widetilde\bfm_d\atop \widetilde\bfm_{d-1}}
	\Bigg),
\ee
where $(\bfm_d\,\myng{(1)}|\widetilde{\bfm}_{d})$ is given by~\eqref{eq:weirdfactor1}-\eqref{eq:weirdfactor2}. This leads to
\be\label{eq:leftuniversalmatrix}
\left[{\widetilde\bfm_d\,\,\bfm_d\atop \widetilde\bfm_{d-1}
}\right]^{34}_p=&(-1)^d\big(\Delta_p+\Delta_{34}-(\bfm_d\,\myng{(1)}|\widetilde{\bfm}_{d})\big)\Bigg(
{\bfm_d\atop \widetilde\bfm_{d-1}}
{\myng{(1)}\atop\bullet}
\Bigg\vert
{\widetilde\bfm_d\atop \widetilde\bfm_{d-1}}
\Bigg).
\ee
Note that we have not yet actually specialized to the case of scalar operators, except in deriving~\eqref{eq:PvectorAction}.\footnote{For more general operators there will be extra contributions (which we discuss in section~\ref{sec:casimir:SpinningConformalBlocks}) to~\eqref{eq:PvectorAction} and thus also to~\eqref{eq:leftuniversal}. The formula~\eqref{eq:leftuniversalmatrix} for the universal contribution~\eqref{eq:leftuniversal} will remain the same.} Let us do this now.

We start by observing that we necessarily have $\widetilde{\bfm}_{d-1}=\bullet$ in order for both sides of~\eqref{eq:leftuniversal} to be non-trivial -- both sides are proportional to $Spin(d-1)$ CG coefficient $\<\bullet,\ldots;\bullet,\ldots|\widetilde{\seq{m}}_{d-1}\>$ which defines the three-point structures, see equation~\eqref{eq:left3pt}. The selection rule $\widetilde{\bfm}_{d}\in\bfm_d\otimes\myng{(1)}$, combined with the requirement that in the scalar case $\bfm_{d}=\bfj$ and $\widetilde{\bfm}_{d}$ are both traceless-symmetric, leaves only two options, $\widetilde{\bfm}_d=\bfj(\pm 1)$, in notation of appendix~\ref{app:formulae}. We therefore only need to compute
\be
	\left[{\bfj(\pm 1)\,\,\bfj}\atop\bullet\right]^{34}_p.
\ee
According to~\eqref{eq:leftuniversalmatrix} we  have
\be
	&\left[{\bfj(\pm 1)\,\,\bfj}\atop\bullet\right]^{34}_p= (-1)^d\big(\Delta_p+\Delta_{34}-(\bfj\,\myng{(1)}|\bfj(\pm1)\big)\rcgc{\bfj}{\bullet}{\myng{(1)}}{\bullet}{\bfj(\pm 1)}{\bullet}
\ee
By using the explicit expressions from appendix~\ref{app:formulae} we find
\be\label{eq:jminus}
	\left[{\bfj(-1)\,\,\bfj}\atop\bullet\right]^{34}_p&=(-1)^d(\Delta_p+\Delta_{34}-j-d+2)\sqrt\frac{j}{2j+d-2}\\
	\left[{\bfj(+1)\,\,\bfj}\atop\bullet\right]^{34}_p
	&=(-1)^d(\Delta_p+\Delta_{34}+j)\sqrt\frac{j+d-2}{2j+d-2}
	\label{eq:jplus}
\ee
One can already recognize here parts of the recursion coefficients $\Gamma^\pm_p$ in~\eqref{eq:gammacoeffs}. In order to obtain the complete expressions, we need to consider the right three-point structure.

\paragraph{Right three-point structure}
We now consider the right part of~\eqref{eq:casimirrhs},
\be
	\<\Delta_p,\seq{m}_{d}| K_\mu s^{D}e^{\theta M_{12}}\phi_2\phi_1|0\>=s^{\Delta_p+1}\<\Delta_p,\seq{m}_{d}| K_\mu e^{\theta M_{12}}\phi_2\phi_1|0\>.
\ee
Let us denote 
\be
	\<\Delta_p,\seq{m}_{d};K,\seq{u}_d|\equiv \<\Delta_p,\seq{m}_{d}| K_{\seq{u}_d},
\ee
and write
\be\label{eq:rightdecoupling}
	\<\Delta_p,\seq{m}_{d}| K_{\seq{u}_d} e^{\theta M_{12}}\phi_2\phi_1|0\>=&\sum_{\seq{m}'_{d},\seq{u}'_d}\<\seq{m}_{d}\seq{u}_d|e^{\theta M_{12}}|\seq{m}'_{d}\seq{u}'_d\>\times\nn\\
	&\qquad\qquad\times \<\Delta_p,\seq{m}'_{d}| K_{\seq{u}'_d}\phi_2\phi_1|0\>.
\ee
We first compute $\<\Delta_p,\seq{m}'_{d}| K_{\seq{u}'_d}\phi_2\phi_1|0\>$ in the same way as we computed the left three-point function. We can make a shortcut by noting
\be
	\<\Delta_p,\seq{m}'_{d}| K_{\seq{u}'_d}\phi_2\phi_1|0\>=\left(\<0| \phi_2\phi_1 P^{\seq{u}'_d}|\Delta,\seq{m}'_{d}\>\right)^*
\ee
and reusing the results for the left three-point function. This gives us
\be\label{eq:rightuniversal}
	\<K,\Delta_p,\bfm_d;\widetilde{\seq{m}}'_d|\phi_2\phi_1|0\>=\left(\left[{\widetilde\bfm'_d\,\,\bfm_d\atop \widetilde\bfm'_{d-1}
	}\right]^{21}_p\right)^*\<\Delta_p,\bfm_d\,\widetilde{\seq{m}}'_{d-1}|\phi_2\phi_1|0\>,
\ee
where 
\be\label{eq:rightuniversalmatrix}
\left[{\widetilde\bfm'_d\,\,\bfm_d\atop \widetilde\bfm'_{d-1}
}\right]^{21}_p
\ee
is given by an analogue~\eqref{eq:leftuniversalmatrix} with $\Delta_3,\Delta_4$ replaced by $\Delta_2,\Delta_1$, and we defined
\be\label{eq:rightirrdescendants}
	\<K,\Delta_p,\bfm_d;\widetilde{\seq{m}}'_d|=\sum_{\seq{m}'_{d}{\seq{u}}'_d}\<\widetilde{\seq{m}}'_d|\seq{m}'_{d}{\seq{u}}'_d\>\<\Delta_p,\seq{m}'_{d}|K_{\seq{u}'_d}.
\ee

Finally, note that we can rewrite the $Spin(d)$ matrix element in~\eqref{eq:rightdecoupling} as
\be\label{eq:matrixelementCG}
	\<\seq{m}_{d}\seq{u}_d|e^{\theta M_{12}}|\seq{m}'_{d}\seq{u}'_d\>=\sum_{\widetilde\bfm_d=\widetilde\bfm'_d}\sum_{\widetilde{\seq{m}}_d,\widetilde{\seq{m}}'_d} \<\seq{m}_d\seq{u}_d|\widetilde{\seq{m}}_d\>\<\widetilde{\seq{m}}_d|e^{\theta M_{12}}|\widetilde{\seq{m}}'_d\>\<\widetilde{\seq{m}}'_d|\seq{m}'_d\seq{u}'_d\>,
\ee
where the summation is over $\widetilde\bfm_d\in \bfm_d\otimes \myng{(1)}$. Note that the CG coefficients here are the same as in~\eqref{eq:leftirrdescendants} and~\eqref{eq:rightirrdescendants}, explaining the usefulness of these definitions.

\paragraph{Combining the results}

By combining equations~\eqref{eq:leftirrdescendants}, \eqref{eq:leftuniversal}, \eqref{eq:rightdecoupling}, \eqref{eq:rightuniversal}, \eqref{eq:rightirrdescendants} and~\eqref{eq:matrixelementCG} we can rewrite~\eqref{eq:casimirrhs} as
\be\label{eq:casimirrhsrewriting}
	&\sum_{\seq{m}_{d}}\<0|\phi_4\phi_3 P_\mu |\Delta_p,\seq{m}_{d}\>\<\Delta_p,\seq{m}_{d}| K^\mu s^{D}e^{\theta M_{12}}\phi_2\phi_1|0\>=\nn\\
	&=s^{\Delta_p+1}\sum_{\widetilde\bfm_d\in\myng{(1)}\otimes \bfm_d}\sum_{\widetilde{\seq{m}}_d,\widetilde{\seq{m}}'_d}
	\left[{\widetilde\bfm_d\,\,\bfm_d\atop \widetilde\bfm_{d-1}
	}\right]^{34}_p
	\left(\left[{\widetilde\bfm_d\,\,\bfm_d\atop \widetilde\bfm'_{d-1}
	}\right]^{21}_p\right)^*\times\nn\\
	&\hspace{4.5cm}\times\<0|\phi_4(\infty)\phi_3(1)|\Delta_p,\bfm_d\,\widetilde{\seq{m}}_{d-1}\>\times\nn\\
	&\hspace{4.5cm}\times \<\widetilde{\seq{m}}_{d}|e^{\theta M_{12}}|\widetilde{\seq{m}}'_d\>\times\nn\\
	&\hspace{4.5cm}\times\<\Delta_p,\bfm_d\,\widetilde{\seq{m}}'_d|\phi_2(1)\phi_1(0)|0\>.
\ee
Here $\widetilde\bfm'_d=\widetilde{\bfm}_d$. The right hand side of~\eqref{eq:casimirrhsrewriting} now has the same form as the generic contribution~\eqref{eq:bigscaryformula}, except that the state $(\Delta_p,\bfm_d)$ now contributes as a state with dimension $\Delta_{p+1}$ and spin $\widetilde\bfm_d\in \myng{(1)}\otimes \bfm_d$ with a relative coefficient determined by the representation-theoretic data through~\eqref{eq:leftuniversalmatrix}. In the scalar correlator case these contributions have the form determined by~\eqref{eq:scalarcorrelatorcontribution}. It is trivial to account for possible degeneracies and arrive at the following result
\be\label{eq:casimirrhsfinal}
	\sum_{p,\bfm_d,\seq{m}_{d},q}&\<0|\phi_4\phi_3 P_\mu|\Delta_p,\seq{m}_{d},q\>\<\Delta_p,\seq{m}_{d},q|K^\mu s^{D}e^{\theta M_{12}}\phi_2\phi_1|0\>=\nn\\
	&=\sum_{p=0}^{\infty}\sum_{j=0}^\infty (\Gamma^+_{p-1,j-1}\Lambda_{p-1,j-1}+\Gamma^-_{p-1,j+1}\Lambda_{p-1,j+1})s^{\Delta_\cO+p}\frac{C_j^{(\nu)}(\cos\theta)}{C_j^{(\nu)}(1)},
\ee
where 
\be
	\Gamma^\pm_{p,j}=\left[{\bfj(\pm 1)\,\,\bfj\atop \bullet 
	}\right]^{34}_p
	\left(\left[{\bfj(\pm 1)\,\,\bfj\atop \bullet
	}\right]^{21}_p\right)^*.
\ee
Given the definition of~\eqref{eq:rightuniversalmatrix} together with the formulas~\eqref{eq:jminus} and~\eqref{eq:jplus} we immediately recover the result~\eqref{eq:gammacoeffs} of~\cite{Hogervorst:2013sma}. By comparing~\eqref{eq:casimirrhsfinal} with~\eqref{eq:casimirlhsfinal} we also recover the required recursion relation~\eqref{eq:scalarrecursion}.

This derivation may seem much more elaborate than that of~\cite{Hogervorst:2013sma}. However, it has several advantages. The first is that the recursion relation is determined not by some particular identities satisfied by Gegenbauer polynomials,\footnote{Of course, given the representation-theoretic interpretation of Gegenbauer polynomials from~\eqref{eq:gegenbauer}, the identities satisfied by Gegenbauer polynomials can also be understood from representation-theoretic point of view.} but instead by a simple set of representation-theoretic data -- by the reduced matrix elements and isoscalar factors. The second is that it is completely general and only a few modifications are required to find the recursion relations for the most general conformal blocks, as we now discuss.

\subsection{Spinning conformal blocks}
\label{sec:casimir:SpinningConformalBlocks}
\subsubsection{Difference from the scalar case}

Let us now consider the general case of spinning conformal blocks. Looking at the derivation of scalar recursion relation, one can see that the first essential deviation in the spinning case happens in~\eqref{eq:PvectorActionPre}, which needs to be replaced by (recall that $\mu\neq 1$ in this context)
\be
	\<0|\cO^{\seq{m}_d^4}_4(\infty)\ptl_\mu\cO^{\seq{m}_d^3}_3(1)|\Delta_p,\seq{m}_{d}\>=&
	-\<0|\cO^{\seq{m}_d^4}_4(\infty)\cO^{\seq{m}_d^3}_3(1)M_{1 \mu}|\Delta_p,\seq{m}_{d}\>\nn\\
	&-\sum_{\seq{m}^{\prime 3}_d} \<\seq{m}^{\prime 3}_d|M_{1\mu}|\seq{m}^{3}_d\>\<0|\cO^{\seq{m}_d^4}_4(\infty)\cO^{\seq{m}_d^{\prime3}}_3(1)|\Delta_p,\seq{m}_{d}\>\nn\\
	&+\sum_{\seq{m}^{\prime 4}_d} \<\seq{m}^{\prime 4}_d|M_{1\mu}|\seq{m}^{4}_d\>\<0|\cO^{\seq{m}_d^{\prime4}}_4(\infty)\cO^{\seq{m}_d^{3}}_3(1)|\Delta_p,\seq{m}_{d}\>.	
\ee
Analogously to~\eqref{eq:PscalarAction}, the relative sign for action on $\cO_4$ is required because we have placed that operator at infinity. This forces this operator to transform in the reflected representation, which is essentially defined by replacing the generators for $M_{1\mu}$ with $-M_{1\mu}$, hence the relative sign.\footnote{This is most easily understood by considering the radial quantization as the limit of NS quantization~\cite{Rychkov:2016iqz} with poles at the positions of $\cO_1$ and $\cO_4$ as $\cO_4$ is taken to $+\infty$.} Note that this does not affect the $Spin(d-1)$ representations, and so the results of section~\ref{sec:structure} regarding three-point functions still hold.

To proceed, we need to put these new contributions into a form similar to~\eqref{eq:leftuniversal}. Let us focus on the contribution from $\cO_3$ which is proportional to
\be\label{eq:spin3contrib}
\sum_{\seq{m}^{\prime 3}_d} \<\seq{m}^{\prime 3}_d|M^{1\,\seq{u}_{d-1}}|\seq{m}^{3}_d\>\<0|\cO^{\seq{m}_d^4}_4(\infty)\cO^{\seq{m}_d^{\prime3}}_3(1)|\Delta_p,\seq{m}_{d}\>.
\ee
As is already familiar, we start by writing out the matrix element as
\be
	\<\seq{m}^{\prime 3}_d|M^{1\,\seq{u}_{d-1}}|\seq{m}^{3}_d\>=\rmel{\bfm_d^3}{\bfm^{\prime 3}_{d-1}}{M^{\myng{(1)}}}{\bfm_d^3}{\bfm^3_{d-1}}\<\seq{m}^{\prime3}_{d-1}|\seq{m}^3_{d-1}\seq{u}_{d-1}\>.
\ee
We then recall from~\eqref{eq:leftirrdescendants2} that in the end we would like to contract~\eqref{eq:spin3contrib} with $\<\seq{m}_{d-1}\seq{u}_{d-1}|\widetilde{\seq{m}}_{d-1}\>$. We are therefore led to consider the combination (we have temporarily omitted the summation over $\bfm^{\prime 3}_{d-1}$ and $\bfm_{d-1}$)
\be
	\sum_{\seq{u}_{d-1}\seq{m}^{\prime 3}_{d-1}\seq{m}_{d-1}}\<\seq{m}^{\prime3}_{d-1}|\seq{m}^3_{d-1}\seq{u}_{d-1}\>\<\seq{m}_{d-1}\seq{u}_{d-1}|\widetilde{\seq{m}}_{d-1}\>\<0|\cO^{\seq{m}_d^4}_4(\infty)\cO^{\seq{m}_d^{\prime3}}_3(1)|\Delta_p,\seq{m}_{d}\>=
\ee
At this point, we should recall the structure of the three-point functions~\eqref{eq:left3pt}, leading to
\be
	=\sum_{\seq{u}_{d-1}\seq{m}^{\prime 3}_{d-1}\seq{m}_{d-1},t}\<\seq{m}^{\prime3}_{d-1}|\seq{m}^3_{d-1}\seq{u}_{d-1}\>\<\seq{m}_{d-1}\seq{u}_{d-1}|\widetilde{\seq{m}}_{d-1}\>\<0|\seq{m}_{d-1}^4\seq{m}_{d-1}^{\prime3}\seq{m}_{d-1},t\>\bar\lambda^{\bfm_{d-1}^{\prime 3},\bfm_{d-1}^4}_{\bfm_{d-1},t}.
\ee
By separating the sum over $t$, we find the objects
\be
	\sum_{\seq{u}_{d-1}\seq{m}^{\prime 3}_{d-1}\seq{m}_{d-1}}\<\seq{m}^{\prime3}_{d-1}|\seq{m}^3_{d-1}\seq{u}_{d-1}\>\<\seq{m}_{d-1}\seq{u}_{d-1}|\widetilde{\seq{m}}_{d-1}\>\<0|\seq{m}_{d-1}^4\seq{m}_{d-1}^{\prime3}\seq{m}_{d-1},t\>=
\ee
These objects have the same invariance properties as $3j$ symbols, and thus should be expressible in terms of them,
\be
	=\sum_{t'}
	\left\{
		\begin{matrix}
			\bfm_{d-1}^4 & \bfm_{d-1}^3 & \widetilde{\bfm}_{d-1}\\
			\myng{(1)} & \bfm_{d-1} & \bfm^{\prime3}_{d-1}
		\end{matrix}
	\right\}_{tt'}^{(3)}
	\<0|\seq{m}_{d-1}^4\seq{m}_{d-1}^{3}\widetilde{\seq{m}}_{d-1},t'\>.
\ee
The constants
\be\label{eq:sixjoccurence}
\left\{
\begin{matrix}
	\bfm_{d-1}^4 & \bfm_{d-1}^3 & \widetilde{\bfm}_{d-1}\\
	\myng{(1)} & \bfm_{d-1} & \bfm^{\prime3}_{d-1}
\end{matrix}
\right\}_{tt'}^{(3)}
\ee
are known as $6j$-symbols or Racah coefficients of $Spin(d-1)$.\footnote{Up to inessential normalization conventions. We will not make a distinction between the two terms.}${}^{,}$\footnote{Interestingly, a different kind of $6j$ symbols recently played an important role in another approach to conformal blocks~\cite{Karateev:2017jgd}.} We added a label $(3)$ to the notation for the $6j$ symbol to distinguish its definition from the definitions~\eqref{eq:sixj1}-\eqref{eq:sixj4} for the operators $1,2,4$ which will appear later.\footnote{Of course, there is only one type of $6j$ symbols for a given group, and this label is superficial. The $6j$ symbols with different labels can be obtained from the $6j$ symbols of the form~\eqref{eq:sixjoccurence} by certain permutations of columns and introduction of normalization factors. Such relations are, however, convention-dependent, and we therefore avoid using them and instead use the labels such as $(3)$.} 
Note that we can represent this equality schematically as
\be
	\begin{tikzpicture}[anchor=base,baseline]
	\draw [spinning] (0,1) -- (-1,2);
	\draw [spinning] (0,1) -- (0,-1);
	\draw [spinning] (0,-1) -- (-1,-2);
	\draw [spinning] (0,1) -- (2,0);
	\draw [spinning] (0,-1) -- (2,0);
	\draw [spinning] (2,0) -- (3.4,0);
	\node at (-0.5,1.7) [right] {$\bfm_{d-1}^4$};
	\node at (-0.5,-1.7) [right] {$\bfm_{d-1}^3$};
	\node at (0.7,-0.8) [right] {$\myng{(1)}$};
	\node at (0.7,+0.8) [right] {$\bfm_{d-1}$};
	\node at (0,0) [left] {$\bfm^{\prime 3}_{d-1}$};
	\node at (3,0) [below] {$\widetilde\bfm_{d-1}$};
	\end{tikzpicture}
	=
	\Big\{\ldots\Big\}
	\begin{tikzpicture}[anchor=base,baseline]
	\draw [spinning] (2,0) -- (3.4,0);
	\draw [spinning] (2,0) -- (1,1);
	\draw [spinning] (2,0) -- (1,-1);
	\node at (-0.5+2,1.7-1) [right] {$\bfm_{d-1}^4$};
	\node at (-0.5+2,-1.7+1) [right] {$\bfm_{d-1}^3$};
	\node at (3,0) [below] {$\widetilde\bfm_{d-1}$};
	\end{tikzpicture}.
\ee
Restoring the OPE coefficients and the summations over $t,\bfm_{d-1}$ and $\bfm_{d-1}^{\prime3}$, and adding the isoscalar factor from~\eqref{eq:leftirrdescendants2} we find
\be\label{eq:op3action}
	&\sum_{\seq{m}^{\prime 3}_d,\seq{m}_{d},\seq{u}_{d-1}} \<\seq{m}^{\prime 3}_d|M^{1\,\seq{u}_{d-1}}|\seq{m}^{3}_d\>\<0|\cO^{\seq{m}_d^4}_4(\infty)\cO^{\seq{m}_d^{\prime3}}_3(1)|\Delta_p,\seq{m}_{d}\>\<\seq{m}_{d-1}\seq{u}_{d-1}|\widetilde{\seq{m}}_{d-1}\>
	\rcgc{\bfm_d}{\bfm_{d-1}}{\myng{(1)}}{\myng{(1)}}{\widetilde\bfm_d}{\widetilde\bfm_{d-1}}=\nn\\
	&\hspace{5cm} =\<0|\cO^{\seq{m}_d^4}_4(\infty)\cO^{\seq{m}_d^{3}}_3(1)|\Delta_p,\bfm_d\,\widetilde{\seq{m}}_{d-1}\>',
\ee
where prime on the three-point function indicates that the OPE coefficients $\bar\lambda$ have been replaced with $\bar\lambda'$ defined as
\be
	&(\bar\lambda')^{\bfm_{d-1}^3,\bfm_{d-1}^4}_{\widetilde{\bfm}_{d-1},t'}=\nn\\
	&\sum_{\bfm_{d-1},\bfm^{\prime3}_{d-1},t} \rmel{\bfm_d^3}{\bfm^{\prime 3}_{d-1}}{M^{\myng{(1)}}}{\bfm_d^3}{\bfm^3_{d-1}}
	\rcgc{\bfm_d}{\bfm_{d-1}}{\myng{(1)}}{\myng{(1)}}{\widetilde\bfm_d}{\widetilde\bfm_{d-1}}
	\left\{
	\begin{matrix}
		\bfm_{d-1}^4 & \bfm_{d-1}^3 & \widetilde{\bfm}_{d-1}\\
		\myng{(1)} & \bfm_{d-1} & \bfm^{\prime3}_{d-1}
	\end{matrix}
	\right\}_{tt'}\bar\lambda^{\bfm_{d-1}^{\prime3},\bfm_{d-1}^4}_{\bfm_{d-1},t}.
\ee

We can easily perform a similar analysis for the contribution of $\cO_4$ as well as for the operators $\cO_1$ and $\cO_2$ in the right three-point function. Note that the right hand side in~\eqref{eq:op3action} has essentially the same form as the universal contribution~\eqref{eq:leftuniversal}, and thus we can continue to derive the recursion relation in an exact analogy with the scalar case.

\subsubsection{The general form of the recursion relation}

It is now straightforward to finish the derivation of the Casimir recursion relation. The operator version of the Casimir equation is given by the spinning analogue of~\eqref{eq:operatorCasimirEquation},
\be
\label{eq:operatorCasimirEquationSpin}
	\<0|\cO^{\seq{m}_d^4}_4\cO^{\seq{m}_d^{3}}_3 |\cO|(\widetilde\cC-\cC) s^{D}e^{\theta M_{12}}\cO^{\seq{m}_d^2}_2\cO^{\seq{m}_d^{1}}_1|0\>=\<0|\cO^{\seq{m}_d^4}_4\cO^{\seq{m}_d^{3}}_3 P^\mu |\cO| K_\mu s^{D}e^{\theta M_{12}}\cO^{\seq{m}_d^2}_2\cO^{\seq{m}_d^{1}}_1|0\>.
\ee
Completely analogously to the scalar case, the contribution of a single term of~\eqref{eq:conformalprojector} to the left hand side of this equation is given by~\eqref{eq:biggerscaryformula} multiplied by the difference of $\widetilde\cC$ and $\cC$ eigenvalues,
\be
	\sum_{\seq{m}_{d}}&\langle 0|\cO_4^{\seq{m}^4_{d}} \cO_3^{\seq{m}^3_{d}}|\Delta_p,\seq{m}_{d}\rangle\langle\Delta_p,\seq{m}_{d}|(\widetilde\cC-\cC)s^{D}e^{\theta M_{12}}\cO_2^{\seq{m}^2_{d}}\cO_1^{\seq{m}^1_{d}}|0\rangle=\nn\\
	=&
	\sum_{\widetilde{\bfm}^i_{d-1}}\sum_{\bfm_{d-1},t\atop \bfm'_{d-1},t'}\sum_{\bfm_{d-2}}(C(\Delta_p,\bfm_d)-C(\cO))
	\lambda^{\widetilde\bfm^1_{d-1},\widetilde\bfm^2_{d-1}}_{\bfm'_{d-1},t'}
	\bar\lambda^{\widetilde\bfm_{d-1}^3,\widetilde\bfm_{d-1}^4}_{\bfm_{d-1},t}
	s^{\Delta_p} P^{\bfm_d,\bfm_{d-2}}_{\bfm_{d-1},\bfm'_{d-1}}(\theta)\times\nn\\
	&\times\left[
	{\seq{m}^3_{d}\atop\seq{m}^4_{d}}\Big\vert
	{\widetilde{\bfm}^3_{d-1}\atop\widetilde{\bfm}^4_{d-1}}
	\,\bfm_{d-1},t
	\Big\vert
	\bfm_{d-2}
	\Big\vert
	\bfm'_{d-1},t'\,{\widetilde{\bfm}^1_{d-1}\atop\widetilde{\bfm}^2_{d-1}}\Big\vert
	{\seq{m}^1_{d}\atop\seq{m}^2_{d}}
	\right].
\ee
Introducing the shorthand notation~\eqref{eq:3ptrightshorthand} and~\eqref{eq:3ptleftshorthand}, restoring the dependence of $\lambda$ on $p,\bfm_d$ and $q$, and summing over the possible degeneracies $q$ we find
\be\label{eq:spincasimirlhs}
	\sum_{\seq{m}_{d},q}&\langle 0|\cO_4^{\seq{m}^4_{d}} \cO_3^{\seq{m}^3_{d}}|\Delta_p,\seq{m}_{d},q\rangle\langle\Delta_p,\seq{m}_{d},q|(\widetilde\cC-\cC)s^{D}e^{\theta M_{12}}\cO_2^{\seq{m}^2_{d}}\cO_1^{\seq{m}^1_{d}}|0\rangle=\nn\\
=&
\sum_{a,b}\sum_{\bfm_{d-2}}(C(\Delta_p,\bfm_d)-C(\cO))
\,\Lambda^{ba}_{p,\bfm_d}\,
s^{\Delta_p} P^{\bfm_d,\bfm_{d-2}}_{\bfm_{d-1},\bfm'_{d-1}}(\theta)
\left[
{\seq{m}^3_{d}\atop\seq{m}^4_{d}}\Big\vert
b
\Big\vert
\bfm_{d-2}
\Big\vert
a\Big\vert
{\seq{m}^1_{d}\atop\seq{m}^2_{d}}
\right],
\ee
where the OPE matrix $\Lambda$ is defined as
\be
	\Lambda^{ba}_{p,\bfm_d}\equiv \sum_{q}\lambda^a_{p,\bfm_d,q}\bar\lambda^b_{p,\bfm_d,q}.
\ee
Following the discussion of scalar recursion relations in section~\ref{sec:casimir:ReviewScalar} and the modifications mentioned in the beginning of this section, we can find
\be\label{eq:spincasimirrhspre}
	\sum_{\seq{m}_{d},\seq{u}_d}&\langle 0|\cO_4^{\seq{m}^4_{d}} \cO_3^{\seq{m}^3_{d}}P^{\seq{u}_d}|\Delta_p,\seq{m}_{d}\rangle
	\langle\Delta_p,\seq{m}_{d}|K_{\seq{u}_d}s^{D}e^{\theta M_{12}}\cO_2^{\seq{m}^2_{d}}\cO_1^{\seq{m}^1_{d}}|0\rangle=\nn\\
	=&
	\sum_{\widetilde\bfm_d\in\myng{(1)}\otimes\bfm_d}\sum_{a,b}\sum_{\bfm_{d-2}}
	(\bar\gamma_{p,\bfm_d,\widetilde\bfm_d}\bar\lambda)^b
	(\lambda\gamma_{p,\bfm_d,\widetilde\bfm_d})^a
	s^{\Delta_p+1} P^{\widetilde\bfm_d,\bfm_{d-2}}_{\bfm_{d-1},\bfm'_{d-1}}(\theta)\left[
	{\seq{m}^3_{d}\atop\seq{m}^4_{d}}\Big\vert
	b
	\Big\vert
	\bfm_{d-2}
	\Big\vert
	a\Big\vert
	{\seq{m}^1_{d}\atop\seq{m}^2_{d}}
	\right].
\ee
Here the matrix $\gamma$ is defined as
\be\label{eq:gammadefn}
	&(\lambda\gamma_{p,\bfm_d\widetilde\bfm_d})^{\bfm_{d-1}^1\bfm_{d-1}^2}_{\widetilde\bfm'_{d-1},t''}=
	(-1)^d\big(\Delta_p-\Delta_{12}-(\bfm_d\,\myng{(1)}|\widetilde{\bfm}_{d})^*\big)
	\rcgcConj
	{\widetilde\bfm_{d}}{\widetilde\bfm'_{d-1}}
	{\bfm_d}{\widetilde\bfm'_{d-1}}
	{\myng{(1)}}{\bullet}
	\lambda^{\bfm_{d-1}^1\bfm_{d-1}^2}_{\widetilde\bfm'_{d-1},t''}\nn\\
	&\quad+\sum_{\bfm^{\prime2}_{d-1},t',\bfm'_{d-1}}
	\rmel
	{\bfm_d^2}{\bfm_{d-1}^{2}}
	{M^{\myng{(1)}}}
	{\bfm_d^2}{\bfm^{\prime 2}_{d-1}}^*
	\rcgcConj
	{\widetilde\bfm_d}{\widetilde\bfm'_{d-1}}
	{\bfm_d}{\bfm'_{d-1}}
	{\myng{(1)}}{\myng{(1)}}
	\sixj
		{\widetilde\bfm'_{d-1}}{\bfm^2_{d-1}}{\bfm^1_{d-1}}
		{\bfm^{\prime 2}_{d-1}}{\bfm'_{d-1}}{\myng{(1)}}
		_{t't''}^{(2)}
	\lambda^{\bfm_{d-1}^{1},\bfm_{d-1}^{\prime 2}}_{\bfm'_{d-1},t'}\nn\\
	&\quad-\sum_{\bfm^{\prime1}_{d-1},t',\bfm'_{d-1}}
	\rmel
	{\bfm_d^1}{\bfm_{d-1}^{1}}
	{M^{\myng{(1)}}}
	{\bfm_d^1}{\bfm^{\prime 1}_{d-1}}^*
	\rcgcConj
	{\widetilde\bfm_d}{\widetilde\bfm'_{d-1}}
	{\bfm_d}{\bfm'_{d-1}}
	{\myng{(1)}}{\myng{(1)}}
	\sixj
		{\widetilde\bfm'_{d-1}}{\bfm^2_{d-1}}{\bfm^1_{d-1}}
		{\bfm^{\prime 1}_{d-1}}{\myng{(1)}}{\bfm'_{d-1}}
		_{t't''}^{(1)}
	\lambda^{\bfm_{d-1}^{\prime 1},\bfm_{d-1}^{2}}_{\bfm'_{d-1},t'},
\ee
while the matrix $\bar\gamma$ is defined as
\be\label{eq:bargammadefn}
	&(\bar\gamma_{p,\bfm_d\widetilde\bfm_d}\bar\lambda)^{\bfm_{d-1}^3\bfm_{d-1}^4}_{\widetilde\bfm_{d-1},t'}=
	(-1)^d\big(\Delta_p+\Delta_{34}-(\bfm_d\,\myng{(1)}|\widetilde{\bfm}_{d})\big)
	\rcgc
		{\bfm_d}{\widetilde\bfm_{d-1}}
		{\myng{(1)}}{\bullet}
		{\widetilde\bfm_{d}}{\widetilde\bfm_{d-1}}
	\bar\lambda^{\bfm_{d-1}^3\bfm_{d-1}^4}_{\widetilde\bfm_{d-1},t'}\nn\\
	&\quad+\sum_{\bfm^{\prime3}_{d-1},t,\bfm_{d-1}}
	\rmel
		{\bfm_d^3}{\bfm_{d-1}^{\prime3}}
		{M^{\myng{(1)}}}
		{\bfm_d^3}{\bfm^3_{d-1}}
	\rcgc
		{\bfm_d}{\bfm_{d-1}}
		{\myng{(1)}}{\myng{(1)}}
		{\widetilde\bfm_d}{\widetilde\bfm_{d-1}}
	\sixj
		{\bfm^4_{d-1}}{\bfm^3_{d-1}}
		{\widetilde\bfm_{d-1}}{\myng{(1)}}
		{\bfm_{d-1}}{\bfm_{d-1}^{\prime 3}}
		^{(3)}_{tt'}
	\bar\lambda^{\bfm_{d-1}^{\prime3},\bfm_{d-1}^4}_{\bfm_{d-1},t}\nn\\
	&\quad-\sum_{\bfm^{\prime4}_{d-1},t,\bfm_{d-1}}
	\rmel
		{\bfm_d^4}{\bfm_{d-1}^{\prime4}}
		{M^{\myng{(1)}}}
		{\bfm_d^4}{\bfm^4_{d-1}}
	\rcgc
		{\bfm_d}{\bfm_{d-1}}
		{\myng{(1)}}{\myng{(1)}}
		{\widetilde\bfm_d}{\widetilde\bfm_{d-1}}
	\sixj
		{\bfm^4_{d-1}}{\bfm^3_{d-1}}{\widetilde\bfm_{d-1}}
		{\bfm_{d-1}}{\myng{(1)}}{\bfm_{d-1}^{\prime 4}}
		^{(4)}_{tt'}
	\bar\lambda^{\bfm_{d-1}^{3},\bfm_{d-1}^{\prime4}}_{\bfm_{d-1},t}.
\ee
The $6j$ symbols are defined as solutions to the following equations
\be\label{eq:sixj1}
	\sum_{\seq{m}_{d-1}^{\prime 1},\seq{u}'_{d-1},\seq{m}'_{d-1}}
	&\<\seq{m}'_{d-1},t'|\seq{m}^2_{d-1}\seq{m}^{\prime 1}_{d-1}\>\<\seq{m}^{\prime 1}_{d-1}\seq{u}'_{d-1}|\seq{m}^1_{d-1}\>\<\widetilde{\seq{m}}'_{d-1}|\seq{m}'_{d-1}\seq{u}'_{d-1}\>=\nn\\
	&=\sum_{t''}\sixj{\widetilde\bfm'_{d-1}}{\bfm^2_{d-1}}{\bfm^1_{d-1}}{\bfm^{\prime 1}_{d-1}}{\myng{(1)}}{\bfm'_{d-1}}_{t't''}^{(1)}
	\<\widetilde{\seq{m}}'_{d-1},t''|\seq{m}^2_{d-1}\seq{m}^1_{d-1}\>,\\
	\sum_{\seq{m}_{d-1}^{\prime 2},\seq{u}'_{d-1},\seq{m}'_{d-1}}
	&\<\seq{m}'_{d-1},t'|\seq{m}^{\prime 2}_{d-1}\seq{m}^1_{d-1}\>\<\seq{m}^{\prime 2}_{d-1}\seq{u}'_{d-1}|\seq{m}^2_{d-1}\>\<\widetilde{\seq{m}}'_{d-1}|\seq{m}'_{d-1}\seq{u}'_{d-1}\>=\nn\\
	&=\sum_{t''}\sixj{\widetilde\bfm'_{d-1}}{\bfm^2_{d-1}}{\bfm^1_{d-1}}{\bfm^{\prime 2}_{d-1}}{\bfm'_{d-1}}{\myng{(1)}}_{t't''}^{(2)}
	\<\widetilde{\seq{m}}'_{d-1},t''|\seq{m}^2_{d-1}\seq{m}^1_{d-1}\>,\\
	\sum_{\seq{m}_{d-1}^{\prime 3},\seq{u}_{d-1},\seq{m}_{d-1}}
	&\<0|\seq{m}_{d-1}^4\seq{m}_{d-1}^{\prime3}(\seq{m}_{d-1},t)\>\<\seq{m}_{d-1}^{\prime3}|\seq{m}_{d-1}^3\seq{u}_{d-1}\>\<\seq{m}_{d-1}\seq{u}_{d-1}|\widetilde{\seq{m}}_{d-1}\>=\nn\\
	&=\sum_{t'}\sixj{\bfm^4_{d-1}}{\bfm^3_{d-1}}{\widetilde\bfm_{d-1}}{\myng{(1)}}{\bfm_{d-1}}{\bfm_{d-1}^{\prime 3}}^{(3)}_{tt'}\<0|\seq{m}^4_{d-1}\seq{m}^3_{d-1}(\widetilde{\seq{m}}_{d-1},t')\>,\\
	\sum_{\seq{m}_{d-1}^{\prime 4},\seq{u}_{d-1},\seq{m}_{d-1}}
	&\<0|\seq{m}_{d-1}^{\prime4}\seq{m}_{d-1}^{3}(\seq{m}_{d-1},t)\>\<\seq{m}_{d-1}^{\prime4}|\seq{m}_{d-1}^4\seq{u}_{d-1}\>\<\seq{m}_{d-1}\seq{u}_{d-1}|\widetilde{\seq{m}}_{d-1}\>=\nn\\
	&=\sum_{t'}\sixj{\bfm^4_{d-1}}{\bfm^3_{d-1}}{\widetilde\bfm_{d-1}}{\bfm_{d-1}}{\myng{(1)}}{\bfm_{d-1}^{\prime 4}}^{(4)}_{tt'}\<0|\seq{m}^4_{d-1}\seq{m}^3_{d-1}(\widetilde{\seq{m}}_{d-1},t')\>.\label{eq:sixj4}
\ee
Reintroducing the degeneracy index $q$ in~\eqref{eq:spincasimirrhspre} we find
\be\label{eq:spincasimirrhs}
	&\sum_{\seq{m}_{d},\seq{u}_d,q}\langle 0|\cO_4^{\seq{m}^4_{d}} \cO_3^{\seq{m}^3_{d}}P^{\seq{u}_d}|\Delta_p,\seq{m}_{d}\rangle
	\langle\Delta_p,\seq{m}_{d}|K_{\seq{u}_d}s^{D}e^{\theta M_{12}}\cO_2^{\seq{m}^2_{d}}\cO_1^{\seq{m}^1_{d}}|0\rangle=\nn\\
	&=
	\sum_{\widetilde\bfm_d\in\myng{(1)}\otimes\bfm_d}\sum_{a,b}\sum_{\bfm_{d-2}}
	(\bar\gamma_{p,\bfm_d,\widetilde\bfm_d}\Lambda_{p,\bfm_d}\gamma_{p,\bfm_d,\widetilde\bfm_d})^{ba}
	s^{\Delta_p+1} P^{\widetilde\bfm_d,\bfm_{d-2}}_{\bfm_{d-1},\bfm'_{d-1}}(\theta)\left[
	{\seq{m}^3_{d}\atop\seq{m}^4_{d}}\Big\vert
	b
	\Big\vert
	\bfm_{d-2}
	\Big\vert
	a\Big\vert
	{\seq{m}^1_{d}\atop\seq{m}^2_{d}}
	\right].
\ee
Comparing~\eqref{eq:spincasimirlhs} and~\eqref{eq:spincasimirrhs} we arrive at the following recursion relation
\be\label{eq:casimirfinal}
	(C(\Delta_{p+1},\widetilde\bfm_d)-C(\cO))\,\Lambda^{ba}_{p+1,\widetilde\bfm_d}=\sum_{\bfm_d\in\myng{(1)}\otimes\widetilde\bfm_d}
	(\bar\gamma_{p,\bfm_d,\widetilde\bfm_d}\Lambda_{p,\bfm_d}\gamma_{p,\bfm_d,\widetilde\bfm_d})^{ba}.
\ee
Equation~\eqref{eq:casimirfinal} represents the main result of this paper. It gives a recursion relation for the power series coefficients $\Lambda$ of a completely general conformal block. This relation has the same structure as the scalar recursion relation~\eqref{eq:scalarrecursion} and can be solved starting from $p=0$ in a straightforward way. The main difficulty lies in evaluation of the coefficient matrices $\gamma$ and $\bar\gamma$, so let us discuss this in some more detail. 

Suppose that we have chosen a concrete four-point function for which we wish to evaluate the conformal blocks, i.e.\ we made a choice of $\bfm_d^i$. If we look at, say,~\eqref{eq:bargammadefn}, we see that all the sums are finite and the number of terms is independent of $m_{d,1}$ or $\widetilde{m}_{d,1}$, which are the only weights that can be arbitrarily large for the given four-point function. Moreover, each term contributes to a single element of the matrix $\bar\gamma$. Furthermore, we see that $\bfm_d$ and $\widetilde{\bfm}_d$ only enter into the simple quantities (isoscalar factors for vector representation and reduced matrix elements) for which closed-form expressions are known (see appendix~\ref{app:formulae}). Similar remarks apply to~\eqref{eq:gammadefn}. This means that if we compute for the given four-point function a finite number of $6j$ symbols~\eqref{eq:sixj1}-\eqref{eq:sixj4}, we can then express the matrices $\gamma$ and $\bar\gamma$ as closed-form analytic expressions in $\bfm_d$ and $\widetilde\bfm_d$, thus obtaining a closed-form analytic expression for the recursion relation~\eqref{eq:casimirfinal}. If we know all the CG coefficients in~\eqref{eq:sixj1}-\eqref{eq:sixj4}, then the calculation of a finite number of $6j$ symbols is a simple linear algebra problem, so we can assume their knowledge to be equivalent to the knowledge of CG coefficients.

As discussed in section~\ref{sec:reptheory:ClebshGordan}, in several important cases the CG coefficients are known analytically (and so are $6j$ symbols). In these cases we can write closed-form expressions for $\bar\gamma$ and $\gamma$. In the rest of this section we consider two such situations: general blocks in $d=3$ and seed blocks for general $d$.

\subsection{Example: General conformal blocks in 3 dimensions}
\label{sec:casimir:General3d}
As discussed above, the only non-trivial ingredients in the recursion relation~\eqref{eq:casimirfinal} are the $6j$ symbols entering the expressions~\eqref{eq:gammadefn} and~\eqref{eq:bargammadefn}. In $d=3$ these symbols simplify dramatically. However, before computing them, we need to understand a small subtlety which arises in $d=3$. 

In the derivation of the recursion relation, we have encountered isoscalar factors such as
\be\label{eq:isoscalarproblem}
	\rcgc{\bfm_d}{\bfm_{d-1}}{\myng{(1)}}{\myng{(1)}}{\bfm_{d}'}{\bfm_{d-1}'}.
\ee
In $d=3$ this presents a problem since we should instead use the isoscalar factors
\be\label{eq:isoscalardesired}
	\rcgc{\bfm_3}{\bfm_{2}}{\myng{(1)}}{\pm 1}{\bfm_{3}'}{\bfm_{2}'},
\ee
because the vector representation is reducible in 2d. One can still use the formulas of appendix~\ref{app:formulae} to compute the value of~\eqref{eq:isoscalarproblem}, but we need to interpret it in terms of~\eqref{eq:isoscalardesired}. Such an interpretation, together with a analogous discussion for reduced matrix elements is given in~\ref{app:formulae:3dcomment}. Using these, one can check that~\eqref{eq:leftuniversalmatrixpre} still holds in $d=3$ and we can still simplify it using the sum rule from appendix~\ref{app:formulae:SumRule}. The formulas of section~\ref{sec:casimir:SpinningConformalBlocks} can also be seen to remain valid if we interpret the sum over $\seq{u}_2$ in~\eqref{eq:sixj1}-\eqref{eq:sixj4} as a sum over $\bfu_{2}=(+1)$ and $\bfu_{2}=(-1)$.

Consider, for example, the equation~\eqref{eq:sixj1} for the $6j$ symbol related to $\cO_1$,
\be
	\sum_{\seq{m}_{d-1}^{\prime 1},\seq{u}'_{d-1},\seq{m}'_{d-1}}
	&\<\seq{m}'_{d-1},t'|\seq{m}^2_{d-1}\seq{m}^{\prime 1}_{d-1}\>\<\seq{m}^{\prime 1}_{d-1}\seq{u}'_{d-1}|\seq{m}^1_{d-1}\>\<\widetilde{\seq{m}}'_{d-1}|\seq{m}'_{d-1}\seq{u}'_{d-1}\>=\nn\\
	&=\sum_{t''}\sixj{\widetilde\bfm'_{d-1}}{\bfm^2_{d-1}}{\bfm^1_{d-1}}{\bfm^{\prime 1}_{d-1}}{\myng{(1)}}{\bfm'_{d-1}}_{t't''}^{(1)}
	\<\widetilde{\seq{m}}'_{d-1},t''|\seq{m}^2_{d-1}\seq{m}^1_{d-1}\>.\\
\ee
In $d=3$, taking into account the subtlety discussed above, this equation simplifies to
\be
	&
	\sum_{u'=\pm 1}\delta_{m',m_2+m'_{1}}
	\delta_{m'_{1}+u',m_1}
	\delta_{\widetilde{m}',m'+u'}=\sixj{\widetilde m'}{m_2}{m_1}{m'_{1}}{\myng{(1)}}{m'}^{(1)}
	\delta_{\widetilde{m}',m_2+m_1}.
\ee
It is solved by
\be
	\sixj{\widetilde m'}{m_2}{m_1}{m'_{1}}{\myng{(1)}}{m'}^{(1)}=\begin{cases}
		1, & m_1-m'_{1}=\widetilde{m}'-m'=\pm 1\\
		0, & \text{otherwise}
	\end{cases}.
\ee
Similarly, we find
\be
	&\sixj{\widetilde{m}'}{m_2}{m_1}{m'_{2}}{m'}{\myng{(1)}}^{(2)}=\begin{cases}
		1, & m_2-m'_{2}=\widetilde{m}'-m'=\pm 1\\
		0, & \text{otherwise}
	\end{cases},\\
	&\sixj{m_4}{m_3}{\widetilde{m}}{\myng{(1)}}{m}{m'_{3}}^{(3)}=\begin{cases}
		1, & m'_{3}-m_3=\widetilde{m}-m=\pm 1\\
		0, & \text{otherwise}
	\end{cases},\\
	&\sixj{m_4}{m_3}{\widetilde{m}}{m}{\myng{(1)}}{m'_{4}}^{(4)}=\begin{cases}
		1, & m'_{4}-m_4=\widetilde{m}-m=\pm 1\\
		0, & \text{otherwise}
	\end{cases}.
\ee

Recall that the right OPE coefficients in 3d are parametrized as $\lambda^{m_1m_2}$. We then have, according to~\eqref{eq:gammadefn} for $\widetilde j=j\pm 1$,
\be\label{eq:3dgamma}
	&(\lambda\gamma_{p,j,j\pm 1})^{m_1m_2}=\nn\\
	&-(\Delta_p-\Delta_{12}\pm j-\delta_{\pm,-})\sqrt{
		\frac{(j-m_1-m_2+\delta_{\pm,+})(j+m_1+m_2+\delta_{\pm,+})}
		{(2j+1)(j+\delta_{\pm,+})}
	}\lambda^{m_1m_2}\nn\\
	&-\sum_{u=\pm 1} \pm \sqrt{
		\frac{(j_2+um_2)(j_2-um_2+1)(j\pm um_1\pm um_2)(j\pm um_1\pm um_2+1)}
		{4(j+\delta_{\pm,+})(2j+1)}
	}\lambda^{m_1(m_2-u)}\nn\\
	&+\sum_{u=\pm 1} \pm\sqrt{
	\frac{(j_1+um_1)(j_1-um_1+1)(j\pm um_1\pm um_2)(j\pm um_1\pm um_2+1)}
	{4(j+\delta_{\pm,+})(2j+1)}
	}\lambda^{(m_1-u)m_2},
\ee
and for $\widetilde j=j$
\be
	&(\lambda\gamma_{p,j,j})^{m_1m_2}=\nn\\
	&-(\Delta_p-\Delta_{12}-1)
		\frac{m_1+m_2}
		{\sqrt{j(j+1)}}\lambda^{m_1m_2}\nn\\
	&+\sum_{u=\pm 1} u\sqrt{
		\frac{(j_2+um_2)(j_2-um_2+1)(j+um_1+um_2)(j-um_1-um_2+1)}
		{4j(j+1)}
	}\lambda^{m_1(m_2-u)}\nn\\
	&-\sum_{u=\pm 1} u\sqrt{
		\frac{(j_1+um_1)(j_1-um_1+1)(j+um_1+um_2)(j-um_1-um_2+1)}
		{4j(j+1)}
	}\lambda^{(m_1-u)m_2}.
\ee
Similarly, from~\eqref{eq:bargammadefn} we find
\be
	&(\bar \gamma_{p,j,j\pm 1}\bar\lambda)^{m_3m_4}=\nn\\
	&-(\Delta_p+\Delta_{34}\pm j-\delta_{\pm,-})\sqrt{
		\frac{(j-m_3-m_4+\delta_{\pm,+})(j+m_3+m_4+\delta_{\pm,+})}
		{(2j+1)(j+\delta_{\pm,+})}
	}\bar\lambda^{m_3m_4}\nn\\
	&-\sum_{u=\pm 1} \pm \sqrt{
		\frac{(j_3-um_3)(j_3+um_3+1)(j\mp um_3\mp um_4)(j\mp um_3\mp um_4+1)}
		{4(j+\delta_{\pm,+})(2j+1)}
	}\bar\lambda^{(m_3+u)m_4}\nn\\
	&+\sum_{u=\pm 1} \pm\sqrt{
		\frac{(j_4-um_4)(j_4+um_4+1)(j\mp um_3\mp um_4)(j\mp um_3\mp um_4+1)}
		{4(j+\delta_{\pm,+})(2j+1)}
	}\bar\lambda^{m_3(m_4+u)},
\ee
\be
	&(\bar\gamma_{p,j,j}\bar\lambda)^{m_3m_4}=\nn\\
	&\quad(\Delta_p+\Delta_{34}-1)
	\frac{m_3+m_4}
	{\sqrt{j(j+1)}}\bar\lambda^{m_3m_4}\nn\\
	&+\sum_{u=\pm 1} u\sqrt{
		\frac{(j_3-u m_3)(j_3+u m_3+1)(j-um_3-um_3)(j+um_3+um_4+1)}
		{4j(j+1)}
	}\bar\lambda^{(m_3+u)m_4}\nn\\
	&-\sum_{u=\pm 1} u\sqrt{
		\frac{(j_4-um_4)(j_4+um_4+1)(j-um_3-um_4)(j+um_3+um_4+1)}
		{4j(j+1)}
	}\bar\lambda^{m_3(m_4+u)}.\label{eq:3dbargamma}
\ee

\subsubsection{Scalar-fermion block in 3 dimensions}

As a concrete example, consider the scalar-fermion blocks in 3d~\cite{Iliesiu:2015akf,Karateev:2017jgd}. In this case we have $j_1=j_4=\thalf$ and $j_2=j_3=0$. Matrices $\Lambda$ then have the indices
\be\label{eq:lambda3d}
	\Lambda^{m_4,m_1}_{p,j}, \quad m_1,m_4=\pm\thalf.
\ee
In terms of these coefficients the conformal block takes the form, according to~\eqref{eq:3dcontribution},
\be\label{eq:3dfermionexpansion}
	\<0|\psi_4^{m_4}\phi_3|\cO|s^D e^{\theta M_{12}} \phi_2\psi_1^{m_1}|0\>=\sum_{\widetilde m_1,\widetilde m_4}\sum_{p=0}^\infty\sum_{j=0}^\infty \Lambda^{\widetilde m_4,\widetilde m_1}_{p,j} d^j_{-\widetilde m_4,\widetilde m_1}(-\theta)
	\delta_{m_4,\widetilde m_4}
	\delta_{m_1,\widetilde m_1}.
\ee
The intermediate representations are $\bfm_3=(j)$ with half-integral $j\geq \thalf$. The Casimir eigenvalue is given by
\be
	C_{p,j}=\Delta_p(\Delta_p-3)+j(j+1).
\ee
Using equations~\eqref{eq:3dgamma}-\eqref{eq:3dbargamma} we find
\be\label{eq:3dfemiongamma}
	(\lambda\gamma_{p,j,j+1})^{\pm\half}&=
	-(\Delta_p-\Delta_{12}+j)\sqrt{\frac{j+\tfrac{3}{2}}{2(j+1)}}\lambda^{\pm\half}
	+\half\sqrt{\frac{j+\tfrac{3}{2}}{2(j+1)}}\lambda^{\mp\half},\\
	(\lambda\gamma_{p,j,j-1})^{\pm\half}&=
	-(\Delta_p-\Delta_{12}-j-1)\sqrt{\frac{j-\tfrac{1}{2}}{2j}}\lambda^{\pm\half}
	-\half\sqrt{\frac{j-\tfrac{1}{2}}{2j}}\lambda^{\mp\half},\\
	(\lambda\gamma_{p,j,j})^{\pm\half}&=
	\mp(\Delta_p-\Delta_{12}-1)\half\frac{1}{\sqrt{j(j+1)}}\lambda^{\pm\half}
	\mp\half\frac{(j+\thalf)}{\sqrt{j(j+1)}}\lambda^{\mp\half},\\
	(\bar\gamma_{p,j,j+1}\bar\lambda)^{\pm\half}&=
	-(\Delta_p+\Delta_{34}+j)\sqrt{\frac{j+\tfrac{3}{2}}{2(j+1)}}\bar\lambda^{\pm\half}
	+\half\sqrt{\frac{j+\tfrac{3}{2}}{2(j+1)}}\bar\lambda^{\mp\half},\\
	(\bar\gamma_{p,j,j-1}\bar\lambda)^{\pm\half}&=
	-(\Delta_p+\Delta_{34}-j-1)\sqrt{\frac{j-\tfrac{1}{2}}{2j}}\lambda^{\pm\half}
	-\half\sqrt{\frac{j-\tfrac{1}{2}}{2j}}\bar\lambda^{\mp\half},\\
	(\bar\gamma_{p,j,j}\bar\lambda)^{\pm\half}&=
	\pm(\Delta_p+\Delta_{34}-1)\half\frac{1}{\sqrt{j(j+1)}}\lambda^{\pm\half}
	\pm\half\frac{(j+\thalf)}{\sqrt{j(j+1)}}\bar\lambda^{\mp\half}.\label{eq:3dfermionbargamma}
\ee
Using this in~\eqref{eq:casimirfinal} we immediately obtain the recursion relation for coefficients~\eqref{eq:lambda3d}. For example, we have
\be
	(C_{p,j}-C_{0,j_\cO})\Lambda_{p,j}^{+\half,+\half}=&\,(\Delta_{p-1}-\Delta_{12}+j-1)(\Delta_{p-1}+\Delta_{34}+j-1)\frac{j+\thalf}{2j}\Lambda^{+\half,+\half}_{p-1,j-1}\nn\\
	&-(\Delta_{p-1}-\Delta_{12}+j-1)\half\frac{j+\thalf}{2j}\Lambda^{-\half,+\half}_{p-1,j-1}\nn\\
	&-\half(\Delta_{p-1}+\Delta_{34}+j-1)\frac{j+\thalf}{2j}\Lambda^{+\half,-\half}_{p-1,j-1}\nn\\
	&+\frac{1}{4}\frac{j+\thalf}{2j}\Lambda^{-\half,-\half}_{p-1,j-1}+\ldots,
\ee
where $\ldots$ represent contributions from $\Lambda_{p-1,j}$ and $\Lambda_{p-1,j+1}$. We compare the conformal block generated by this recursion relation with the known results~\cite{Iliesiu:2015akf,Karateev:2017jgd} in appendix~\ref{app:ScalarFermionChecks}, finding a perfect agreement.

\subsection{Example: Seed conformal blocks in general dimensions}
\label{sec:casimir:SeedBlocks}

We have already considered the seed blocks in section~\ref{sec:structure:SeedConformalBlocks}. Here, as in previous subsections, we start by computing the $6j$ symbols~\eqref{eq:sixj1}-\eqref{eq:sixj4}. Since in the seed block case the operators $\cO_1$ and $\cO_3$ are scalars, we do not need the $6j$ symbols for them.

For $\cO_2$ the equation for the $6j$ symbol specializes to
\be
	\sum_{\seq{m}_{d-1}^{\prime 2},\seq{u}'_{d-1},\seq{m}'_{d-1}}
	&\<\seq{m}'_{d-1}|\seq{m}^{\prime 2}_{d-1}\>\<\seq{m}^{\prime 2}_{d-1}\seq{u}'_{d-1}|\seq{m}^2_{d-1}\>\<\widetilde{\seq{m}}'_{d-1}|\seq{m}'_{d-1}\seq{u}'_{d-1}\>=\nn\\
	&=\sixj{\bfm^2_{d-1}}{\bfm^2_{d-1}}{\bullet}{\bfm^{\prime2}_{d-1}}{\bfm^{\prime2}_{d-1}}{\myng{(1)}}^{(2)}
	\<\widetilde{\seq{m}}'_{d-1}|\seq{m}^2_{d-1}\>,
\ee
and we can simplify the left-hand side to
\be
	\<\widetilde{\seq{m}}'_{d-1}|\seq{m}^2_{d-1}\>
\ee
which implies that simply
\be
	\sixj{\bfm^2_{d-1}}{\bfm^2_{d-1}}{\bullet}{\bfm^{\prime2}_{d-1}}{\bfm^{\prime2}_{d-1}}{\myng{(1)}}^{(2)}=1,
\ee
whenever the selection rules are satisfied.
Similarly, equation~\eqref{eq:sixj4} specializes to
\be
	\sum_{\seq{m}_{d-1}^{\prime 4},\seq{u}_{d-1},\seq{m}_{d-1}}
	&\<0|\seq{m}_{d-1}^{\prime4}\seq{m}_{d-1}\>\<\seq{m}_{d-1}^{\prime4}|\seq{m}_{d-1}^4\seq{u}_{d-1}\>\<\seq{m}_{d-1}\seq{u}_{d-1}|\widetilde{\seq{m}}_{d-1}\>=\nn\\
	&=\sixj{\bfm^4_{d-1}}{\bullet}{\overline{\bfm^4_{d-1}}}{\overline{\bfm^{\prime 4}_{d-1}}}{\myng{(1)}}{\bfm_{d-1}^{\prime 4}}^{(4)}\<0|\seq{m}^4_{d-1}\widetilde{\seq{m}}_{d-1}\>,
\ee
and the left hand side can be reduced to
\be
	\pm \<0|\seq{m}^4_{d-1}\widetilde{\seq{m}}_{d-1}\>,
\ee
where the sign is equal to\footnote{Here, as before, $(-1)^{\bfm_d}$ is defined as $1$ unless $d=4k+2$ in which case it is equal to $(-1)^{m_{4k+2,2k+1}}$.} $(-1)^{\bfm_{d-1}^4-{\bfm^{\prime 4}_{d-1}}}$ unless $\bfm_{d-1}^4={\bfm^{\prime 4}_{d-1}}$ and $d=4k$ in which case it is equal to $-1$. To see this, one can use the identity
\be
	\sum_{\seq{m}_{d-1}^{\prime 4}}
	\<0|\seq{m}_{d-1}^{\prime4}\seq{m}_{d-1}\>\<\seq{m}_{d-1}^{\prime4}|\seq{m}_{d-1}^4\seq{u}_{d-1}\>=\pm
		\sum_{\seq{m}'_{d-1}}
	\<0|\seq{m}^4_{d-1}\seq{m}'_{d-1}\>\<\seq{m}'_{d-1}|\seq{m}_{d-1}\seq{u}_{d-1}\>,
\ee
where the sign is as above. Up to normalization, it has to be true because both sides have the same $Spin(d-1)$ invariance properties. Up to a phase, the normalization can be determined by fully contracting each side with its Hermitian conjugate. The sign can then be found by setting $\seq{u}_{d-1}=(\myng{(1)},\bullet,\ldots)$ and examining the phase on both sides using~\eqref{eq:2jfinal} and the formulas in section~\ref{app:formulae:Isoscalar}.

This implies that
\be
	\sixj{\bfm^4_{d-1}}{\bullet}{\overline{\bfm^4_{d-1}}}{\overline{\bfm^{\prime 4}_{d-1}}}{\myng{(1)}}{\bfm_{d-1}^{\prime 4}}^{(4)}=\begin{cases}
		-1 & \bfm_{d-1}^4=\bfm_{d-1}^{\prime 4}\text{ and } d=4k\\
		(-1)^{\bfm_{d-1}^4-\bfm_{d-1}^{\prime 4}} & \text{otherwise}
	\end{cases}.
\ee
It is now straightforward to substitute these $6j$ symbols into the expressions~\eqref{eq:gammadefn} and~\eqref{eq:bargammadefn} for the matrices $\gamma$ and $\bar\gamma$ to obtain closed-form analytic expressions for them. The final general expression is not particularly illuminating, so we do not write it out explicitly. Instead, let us again consider a specific example, the scalar-fermion seed blocks in $d=2n$ dimensions.

\subsubsection{Scalar-fermion blocks in $d=2n$ dimensions}

We have considered the structure of these blocks in section~\ref{sec:structure:SeedConformalBlocks}. The OPE matrices $\Lambda$ are $1\times 1$ and there are two types of exchanged representations, $\bfj^\pm \equiv (j,\thalf,\ldots,\thalf,\pm\thalf)$. Thus, we can label the OPE matrices as
\be
	\Lambda_{p,j,\pm}.
\ee
We can arrange them into a vector as in section~\ref{sec:structure:Pfunctions:matromForm},
\be
	\mathbf{\Lambda}_{p,j}=\begin{pmatrix}
		\Lambda_{p,j,+}\\
		\Lambda_{p,j,-}
	\end{pmatrix}.
\ee
We furthermore have
\be
	\myng{(1)}\otimes \bfj^\pm = (\bfj+1)^\pm \oplus (\bfj-1)^\pm \oplus \bfj^\mp.
\ee
Equation~\eqref{eq:gammadefn} reduces to
\be
	\lambda\gamma_{p,\bfj^\pm,(\bfj+ 1)^\pm}=&\,(\Delta_p-\Delta_{12}+j)\sqrt{\frac{1}{2}\frac{j+2n-\tfrac{3}{2}}{j+n-\tfrac{1}{2}}}\lambda\nn\\
	&-\left(\frac{i}{2}\sqrt{2n-1}\right)\left(
	\frac{\pm i}{\sqrt{2n-1}}
	\sqrt{\frac{1}{2}\frac{j+2n-\tfrac{3}{2}}{j+n-\tfrac{1}{2}}}
	\right)\lambda\nn\\
	=&\,(\Delta_p-\Delta_{12}+j\pm\thalf)\sqrt{\frac{1}{2}\frac{j+2n-\tfrac{3}{2}}{j+n-\tfrac{1}{2}}}\lambda,\\
	\lambda\gamma_{p,\bfj^\pm,(\bfj-1)^\pm}=&\,
	(\Delta_p-\Delta_{12}-j-d+2)\sqrt{
		\frac{1}{2}
		\frac{j-\thalf}
		{j+n-\tfrac{3}{2}}
	}\lambda\nn\\
	&-\left(\frac{i}{2}\sqrt{2n-1}\right)\left(
	\frac{\mp i}{\sqrt{2n-1}}
	\sqrt{
		\frac{1}{2}
		\frac{j-\thalf}
		{j+n-\tfrac{3}{2}}
	}
	\right)\lambda\nn\\
	=&\,(\Delta_p-\Delta_{12}-j-d+2\mp \tfrac{1}{2})\sqrt{
		\frac{1}{2}
		\frac{j-\thalf}
		{j+n-\tfrac{3}{2}}
	}\lambda,\\
	\lambda\gamma_{p,\bfj^\pm,\bfj^\mp}=&\,
		(\Delta_p-\Delta_{12}-n+\thalf)\sqrt{
			\frac{2n-2}
			{(2j+2n-3)(2j+2n-1)}
		}\lambda\nn\\
	&-\left(\frac{i}{2}\sqrt{2n-1}\right)\left(
	\frac{\mp 2i (j+n-1)}{\sqrt{2n-1}}
	\sqrt{
		\frac{2n-2}
		{(2j+2n-3)(2j+2n-1)}
	}
	\right)\lambda\nn\\
	=&\,(\Delta_p-\Delta_{12}-n+\thalf\mp(j+n-1))\sqrt{
		\half\frac{n-1}
		{(j+n-\tfrac{3}{2})(j+n-\thalf)}
	}\lambda.
\ee
Similarly, we find from~\eqref{eq:bargammadefn}
\be
\bar\gamma_{p,\bfj^\pm,(\bfj+ 1)^\pm}\lambda=
	&\,(\Delta_p+\Delta_{34}+j\pm(-1)^{n-1}\thalf)\sqrt{\frac{1}{2}\frac{j+2n-\tfrac{3}{2}}{j+n-\tfrac{1}{2}}}\lambda,\\
\bar\gamma_{p,\bfj^\pm,(\bfj-1)^\pm}\lambda=
	&(\Delta_p+\Delta_{34}-j-d+2\mp(-1)^{n-1}\tfrac{1}{2})\sqrt{
		\frac{1}{2}
		\frac{j-\thalf}
		{j+n-\tfrac{3}{2}}
	}\lambda,\\
\bar\gamma_{p,\bfj^\pm,\bfj^\mp}\lambda=
	&(\Delta_p+\Delta_{34}-n+\thalf\mp(-1)^{n-1}(j+n-1))\sqrt{
	\half\frac{n-1}
	{(j+n-\tfrac{3}{2})(j+n-\thalf)}
	}\lambda.
\ee
Finally, the Casimir eigenvalue is given, according to~\eqref{eq:spindcasimir} and~\eqref{eq:confcasimir},
\be
	C_{p,j}=\Delta_p(\Delta_p-2n)+j(j+2n-2)+\frac{(2n-2)(2n-3)}{8}.
\ee

The recursion relation~\eqref{eq:casimirfinal} can then be put into the form
\be\label{eq:recursionScalarFermion2n}
	(C_{p,j}-C_{0,j_\cO})\mathbf\Lambda_{p,j}=\Gamma^+_{p-1,j-1}\mathbf\Lambda_{p-1,j-1}+\Gamma^-_{p-1,j-1}\mathbf\Lambda_{p-1,j+1}+\Gamma^0_{p-1,j-1}\mathbf\Lambda_{p-1,j},
\ee
where 
\be
	\Gamma^+_{p,j}&=
	(\Delta_p-\Delta_{12}+j+\thalf)
	(\Delta_p+\Delta_{34}+j+(-1)^{n-1}\thalf)\frac{j+2n-\tfrac{3}{2}}{2j+2n-1}
	\begin{pmatrix}
		1 & 0 \\
		0 & 0
	\end{pmatrix}+\nn\\
	&\quad +
	(\Delta_p-\Delta_{12}+j-\thalf)
	(\Delta_p+\Delta_{34}+j-(-1)^{n-1}\thalf)\frac{j+2n-\tfrac{3}{2}}{2j+2n-1}
	\begin{pmatrix}
		0 & 0 \\
		0 & 1
	\end{pmatrix},\\
	\Gamma^-_{p,j}&=
	(\Delta_p+\Delta_{12}-j-2n+2-\thalf)
	(\Delta_p+\Delta_{34}-j-2n+2-(-1)^{n-1}\thalf)\frac{j-\thalf}{2j+2n-3}
	\begin{pmatrix}
		1 & 0 \\
		0 & 0
	\end{pmatrix}+\nn\\
	&\quad +
	(\Delta_p+\Delta_{12}-j-2n+2+\thalf)
	(\Delta_p+\Delta_{34}-j-2n+2+(-1)^{n-1}\thalf)\frac{j-\thalf}{2j+2n-3}
	\begin{pmatrix}
		0 & 0 \\
		0 & 1
	\end{pmatrix},\\
	\Gamma^0_{p,j}&=
	(\Delta_p-\Delta_{12}-j-2n+\tfrac{3}{2})
	(\Delta_p+\Delta_{34}-n+\thalf-(-1)^{n-1}(j+n-1))\times\nn\\
	&\hspace{6.5cm}\times\frac{2n-2}{(2j+2n-3)(2j+2n-1)}
	\begin{pmatrix}
		0 & 0 \\
		1 & 0
	\end{pmatrix}+\nn\\
	&\quad +
	(\Delta_p-\Delta_{12}+j-\thalf)
	(\Delta_p+\Delta_{34}-n+\thalf+(-1)^{n-1}(j+n-1))\times\nn\\
	&\hspace{6.5cm}\times\frac{2n-2}{(2j+2n-3)(2j+2n-1)}
	\begin{pmatrix}
		0 & 1 \\
		0 & 0
	\end{pmatrix}.
\ee

The full conformal block can then be expanded by using a generalization of~\eqref{eq:matromForm},
\be\label{eq:scalarFermionExpansion}
	\<0|\overline\psi_4\psi_3|\cO| s^\Delta e^{\theta M_{12}}\psi_2\phi_1|0\>=
	\sum_{p=0}^{\infty}\sum_{j=0}^\infty s^{\Delta+p} \mathbf{\Lambda}_{p,j}\cdot \bfP^j(\theta)\cdot \bfT,
\ee
where $\bfT=(t_+,t_-)$ and the matrom $\bfP^j$ is given by~\eqref{eq:fermionmatrom}. In appendix~\ref{app:ScalarFermionChecks} we compare the conformal blocks obtained from this recursion relation with the known expressions in 2d ($n=1$) and 4d ($n=2$), finding a perfect agreement.

\subsection{An efficient implementation?}
\label{sec:implementationremarks}

We have derived the Casimir recursion relation for general conformal blocks. Our derivation relies on the knowledge of a number of $6j$-symbols of $Spin(d-1)$. As we have discussed, there are important cases, such as general blocks in 3d and 4d or seed blocks in general dimensions, where these symbols are readily available. In other cases, they can be computed as soon as the relevant Clebsch-Gordan coefficients are known. These Clebsch-Gordan coefficients are needed anyway for the three-point functions (and can be derived from them), so it is reasonable to assume that the $6j$ symbols are computable in all cases of interest. 

If the relevant $6j$ symbols are known, then our results provide a closed-form expression for the recursion relation~\eqref{eq:casimirfinal}. This is a quite general result, so it is interesting to discuss the possibility of employing it for an efficient computation of spinning conformal blocks. Assume that we have fixed numerical values for scaling dimensions and spins of the external operators and the spin of the intermediate primary and would like to compute the conformal block and its derivatives as a function of the intermediate dimension $\Delta_\cO$. The simplest approach is to naively iterate the recursion relation and find the coefficients of the power series expansion in $z$-coordinates.

This approach has several obvious disadvantages. Firstly, the $z$-coordinate expansion converges much slower than the $\rho$-coordinate expansion~\cite{Hogervorst:2013sma}. Secondly, the coefficients of the expansion are going to be some complicated rational functions of $\Delta_\cO$, manipulations with which are costly. Moreover, the difference of Casimir eigenvalues in~\eqref{eq:casimirfinal},
\be\label{eq:casimirdifference}
	C(\Delta_\cO+n,\widetilde\bfm_d)-C(\cO)=2n\Delta_\cO+n^2-nd+C(\widetilde\bfm_d)-C(\bfm_d^\cO),
\ee
produces a lot of apparent poles at various rational values of~$\Delta_\cO$. We however know that the conformal blocks can only have poles at (half-)integral values of $\Delta_\cO$~\cite{Penedones:2015aga}. This implies that there must be a lot of cancellations, which make the direct analytic even less optimal. Let us discuss some possible solutions to these problems.

The first problem can be in principle avoided by converting the $z$-coordinate expansion into a $\rho$-coordinate expansion. It is possible because we have the relation $z=4\rho+O(\rho^2)$, so if we know the expansion of $f(z)$ to order $z^N$, we can compute expansion of $f(z(\rho))$ to the same order $\rho^N$. If the coefficients in expansion of $f(z)$ are numbers, and we aim to evaluate $f(\thalf)$, then this conversion can be done efficiently by defining $z_N^k$ to be equal to the $\rho$-series of $z^k$, truncated at order $\rho^N$ and with $\rho$ set to $\rho=3-2\sqrt{2}$ (the value corresponding to $z=\thalf$). Then the number $f(1/2)$ can be computed by simply replacing $z^k$ in its $z$-expansion by the numbers $z^k_N$. These numbers can be precomputed once for any given $N$.

However, as we noted above, in our case the coefficients of $z$-expansion are complicated rational functions and thus this conversion would have to be performed using symbolic algebra. To solve this problem, it is convenient to recall that for any conformal block $G(\Delta_\cO)$ (for simplicity of notation we keep the dependence only on $\Delta_\cO$ explicit) the function $H(\Delta_\cO)=|\rho|^{-\Delta_\cO} G(\Delta_\cO)$ is a meromorphic function of $\Delta_\cO$ with either single or double\footnote{We are not aware of a direct proof that at most second-order poles appear in even $d$ (see e.g.~\cite{Penedones:2015aga,Yamazaki:2016vqi} for a discussion). However, since the scalar blocks have at most second-order poles, the results of~\cite{Karateev:2017jgd} imply that there are at most finitely many higher-order poles in any given conformal block. Also, standard arguments from complex analysis show that at most double poles can appear from collision of two single poles, which can possibly be used to show that at least the blocks which can be analytically continued in dimension $d$ have at most second-order poles.} poles and a finite limit at infinity\footnote{At least for $\Delta_\cO$-independent choice of three-point functions.}~\cite{Zamolodchikov:1987,Kos:2013tga,Penedones:2015aga}. In odd dimensions this function only has single poles, so let us consider this case for simplicity.\footnote{The same approach should work in even dimensions, with minor modifications.} We then can write
\be\label{eq:poleexpansion}
	H(\Delta_\cO)=H(\infty)+\sum_i \frac{R_i}{\Delta_\cO-\Delta_i},
\ee
where $\Delta_i$ are the locations of the poles and $R_i$ are some coefficients.\footnote{$R_i$ are known to be proportional to other conformal blocks. We do not use this fact here.} The function $H(\infty)$ can be computed in closed form for a general conformal block by a suitable choice of the basis of four-point structures. Expansion~\eqref{eq:poleexpansion} is often used to derive rational approximations to conformal blocks, required for numerical analysis using~\texttt{SDPB}~\cite{Kos:2013tga,Simmons-Duffin:2015qma}. For this, note that different terms in this expansion are suppressed by powers $\rho^{n_i}$ for some positive $n_i$. Thus, one can keep only the finite number of terms with $n_i\leq M$ for some sufficiently large $M$. Since the derivatives of $G$ are determined by derivatives of $H$, it is sufficient to compute the derivatives of $R_i$ and $H(\infty)$ numerically in order to obtain the rational approximations required for numerical bootstrap applications.

Our recursion relation can be used to determine $R_i$ and their derivatives numerically. Indeed, on each step of the recursion relation we explicitly divide by a linear function of $\Delta_\cO$~\eqref{eq:casimirdifference}. Thus, we know exactly when we produce poles and we can compute their residues and how they change on each step of the recursion. If we select a subset of $\Delta_i$, we only need to track the derivatives of the residues at these poles, which are simply numbers. We can avoid dealing with the apparent poles at rational $\Delta_\cO$ by tracking only the $\Delta_i$ allowed by representation theory~\cite{Penedones:2015aga}. This is similar in spirit to multiplication of polynomials in Fourier space (as in FFT polynomial multiplication), except we are working with rational functions. This approach should allow us to efficiently compute the numerical $z$-series of derivatives of $R_i$. We can then use the aforementioned procedure to resum it into $\rho$-series at $z=\half$.

Note that in this scheme it is most convenient to take the derivatives in $z$-coordinate. These derivatives do not necessarily have the fastest rate of convergence among other simple choices.\footnote{Choice of the coordinate matters: the derivative $df(z)/df(z)$ converges much faster than the derivative of $df(z)/dz$.} A related problem is that it is not obvious what is the best basis of four-point tensor structures in terms of convergence.\footnote{The choice of basis matters as well, because the bases can differ by $z$-dependent factors: even if $f(z)$ converges quickly, $f(z)/(1-z)^{100}$ may converge much slower.} The approach based on~\eqref{eq:poleexpansion} somewhat solves this ambiguity -- it is a well-defined procedure to keep a finite number of poles in~\eqref{eq:poleexpansion}, and we can then compute $R_i$ to an order $N$ higher than $M$, eliminating the possible discrepancies between various choices. Indeed, if we keep the number of poles that we track fixed, then the complexity of computing each new order grows only because the range of allowed values for $m_{d,1}$ expands.

In order for the above program to succeed, we need to be able to efficiently compute derivatives of these $P$-functions. It appears that this problem is largely solved by the recursion relation~\eqref{eq:matromrecursion} which can be easily implemented numerically for any choice of representations given the availability of closed-form formulas for vector isoscalar factors. We still need an initial condition for the recursion relation. As we discussed previously, it can be obtained by direct exponentiation of $M_{12}$. However, in numerical applications we do not even need this. We only need a first few derivatives of $P$-functions at $\theta=0$, which are given by matrix elements of powers of $M_{12}$, making the computation even easier.

\section{Conclusions}
\label{sec:conclusions}
The two major results of this paper are 
\begin{enumerate}
	\item The general form~\eqref{eq:biggerscaryformula} of a $\bR\times Spin(d)$-multiplet contribution to a general four-point function of operators with spins.
	\item The Casimir recursion relation~\eqref{eq:casimirfinal} (and the formulas~\eqref{eq:gammadefn} and~\eqref{eq:bargammadefn} for the relevant coefficients) for the amplitudes $\Lambda_{p,\bfm_d}$ of these contributions to a general spinning conformal block.
\end{enumerate}

The first result is expressed in terms of certain special functions $P$~\eqref{eq:Pdefinition}, which we have studied in detail in section~\ref{sec:structure:Pfunctions}. We have described the basic properties of these functions (including orthogonality relations) as well as a practical approach to their calculation. In appendix~\ref{app:tensors} we have furthermore related these functions to the irreducible projectors of~\cite{Costa:2016hju}.\footnote{We believe that this is not the most optimal way for computation of explicit examples of functions $P$, and one instead should use the methods described in~\ref{sec:structure:Pfunctions}. Nevertheless, this relation does provide expressions which may be useful in analytical applications.} We have studied how~\eqref{eq:biggerscaryformula} simplifies in some special cases, namely for $d=3,4$ and for seed blocks in general $d$. We have also proven the folklore theorem which states that the number of four-point tensor structures is the same as the number of classes of conformal blocks.

Our second result paves a way to an algorithmic computation of general conformal blocks. The expressions~\eqref{eq:casimirfinal}, \eqref{eq:gammadefn} and~\eqref{eq:bargammadefn} give a closed-form recursion relation for the coefficients of the $z$-coordinate expansion of a general conformal block, if the relevant $6j$ symbols of $Spin(d-1)$ are known. There is a finite number of such $6j$ symbols for any given conformal block, and they can be straightforwardly computed if the corresponding Clebsch-Gordan coefficients are known. The required CG coefficients are indeed known in many important cases. In particular, we have explicitly worked out the case of general conformal blocks in 3 dimensions and the seed blocks in general dimensions. To illustrate the recursion relation in explicit examples, we have studied the scalar-fermion seed blocks in $d=3$ and $d=2n$, comparing to the known results when possible. Finally, in section~\ref{sec:implementationremarks} we have briefly discussed a strategy for an efficient numerical implementation of the recursion relation~\eqref{eq:casimirfinal}.

Many extensions of these results are possible. For example, the scalar-fermion seed blocks can also be straightforwardly obtained for $d=2n+1$, we have omitted this case only to keep the size of the paper reasonable. For the same reason we have not written down the explicit formulas for the case of general blocks in $d=4$, even though these can be obtained (in terms of $SU(2)$ $6j$-symbols) mechanically from the general expressions. Extension to $d=5$ is also possible, due to $Spin(5-1)\simeq SU(2)\times SU(2)$. An interesting problem is to develop a numerical algorithm for computation of general $Spin(d-1)$ CG coefficients and $6j$ symbols. Combined with the recursion relation~\eqref{eq:casimirfinal} this would constitute the first completely general algorithm for computation of conformal blocks.\footnote{Here by an ``algorithm'' we mean an actual complete algorithm which can be straightforwardly translated into a computer program. Techniques (not algorithms) for computing completely general spinning conformal blocks are already known~\cite{SimmonsDuffin:2012uy,Penedones:2015aga,Costa:2016xah,Karateev:2017jgd}.} It is also interesting to implement this recursion relation efficiently, perhaps along the lines of section~\ref{sec:implementationremarks}. Finally, there is always the question whether these results can be extended to superconformal case. We hope to address some of these questions in future work.

\section*{Acknowledgments}
I would like to thank Denis Karateev, Jo\~ao Penedones, Fernando Rej\'on-Barerra, Slava Rychkov, David Simmons-Duffin, Emilio Trevisani, and the participants of Simons Bootstrap Collaboration workshop on numerical bootstrap for valuable discussions. Special thanks to David Simmons-Duffin for comments on the draft. I am grateful to the authors of~\cite{Costa:2016hju} for making their~\texttt{Mathematica} code publicly available. I also thank the Institute for Advanced Study, where part of this work was completed, for hospitality. This work was supported by DOE grant DE-SC0011632.

\appendix
\section{Conformal algebra and its representation on local operators}
\label{app:conformalalgebra}
Here we describe our conventions for the conformal algebra. The commutation relations are as follows,
\be
	[D,P_\mu]&=P_\mu,\quad [D,K_\mu]=-K_\mu,\\
	[K_\mu,P_\nu]&=2\delta_{\mu\nu}D+2M_{\mu\nu},\\
	[M_{\mu\nu},P_\rho]&=\delta_{\mu\rho}P_\nu-\delta_{\nu\rho}P_\mu,\quad [M_{\mu\nu},K_\rho]=\delta_{\mu\rho}K_\nu-\delta_{\nu\rho}K_\mu,\\
	[M_{\mu\nu},M_{\rho\sigma}]&=\delta_{\mu\rho}M_{\nu\sigma}-\delta_{\nu\rho}M_{\mu\sigma}+\delta_{\mu\sigma}M_{\rho\nu}-\delta_{\nu\sigma}M_{\rho\mu}.
\ee
The generators obey the following conjugation properties,
\be
	D^\dagger=D,\quad P^\dagger=K, \quad M^\dagger_{\mu\nu}=- M_{\mu\nu}.
\ee
The generators act on primary operators as follows,
\be
	[D,\cO(x)]&=x\cdot\ptl\cO(x)+\Delta\cO(x),\\
	[P_\mu,\cO(x)]&=\ptl_\mu\cO(x),\\
	[M_{\mu\nu},\cO(x)]&=(x_\mu\ptl_\nu-x_\nu\ptl_\mu)\cO(x)+S_{\mu\nu}\cO(x),\\
	[K_\mu,\cO(x)]&=(2x_\mu x^\sigma-x^2\delta^\sigma_\mu)\ptl_\sigma\cO(x)+2x^\sigma (\Delta\delta_{\mu\sigma}+S_{\mu\sigma})\cO(x).
\ee
Here $S_{\mu\nu}$ are the generators which act on the spin indices of $\cO(x)$ and satisfy the commutation relations opposite to $M_{\mu\nu}$. Our convention for $M_{\mu\nu}$ differs by a minus sign from that of~\cite{Simmons-Duffin:2016gjk}. $M_{\mu\nu}$ in our case has the interpretation of rotating $e_\mu$ towards $e_\nu$.

\section{Reduced matrix elements and vector isoscalar factors}
\label{app:formulae}

In order to write down the formulas for isoscalar factors and reduced matrix elements, we need to take some preliminary steps. First, let us consider the decomposition of the tensor product $\bfm_d\otimes \myng{(1)}$. Generically, we have in even dimensions, according to Brauer's formula,
\be\label{eq:evendtensor}
	\bfm_d\otimes \myng{(1)}\simeq \bigoplus_{i=1}^n \bfm_d(+i)\oplus \bfm_d(-i),\qquad d=2n,
\ee
where $\bfm_d(\pm i)$ is the same as $\bfm_d$ but with the component $m_{d,i}$ shifted by $\pm 1$. Similarly, in odd dimensions we have, generically,
\be\label{eq:odddtensor}
	\bfm_d\otimes\myng{(1)}\simeq \bfm_d\oplus \bigoplus_{i=1}^n \bfm_d(+i)\oplus \bfm_d(-i),\qquad d=2n+1.
\ee
These formulas are valid for generic $\bfm_d$, i.e.\ those with all components non-zero and sufficiently large. For concrete representations, some of the direct summands may disappear if there are non-dominant weights in the right hand side. By applying Brauer's formula, we see that to find the final tensor product rule we just need to remove all non-dominant weights and, if $d=2n+1$ and $m_{d,n}=0$, also remove $\bfm_d$.\footnote{This can be seen by analyzing the situations in which $\bfm_d(\pm i)$ may fail to be dominant. It turns out that in most cases there is an affine Weyl reflection which stabilizes the non-dominant $\bfm_d(\pm i)$ and thus such weights simply have to be removed. The exception is the case $m_{2n+1,n}=0$: $\bfm_{2n+1}(-n)$ can be turned into $\bfm_{2n+1}$ with one affine Weyl reflection, and thus cancels it.}

We now define the following new parameters,
\be
	x_{2n+1,j}&=m_{2n+1,j}+n-j,\\
	x_{2n,j}&=m_{2n,j}+n-j.
\ee
Note that regardless of the dimension, $m_{d,j}$ is a non-increasing function of $j$. Since we add to it a strictly decreasing function of $j$, we find that $x_{d,j}$ is a strictly decreasing function of $j$. In particular $x_{d,j}\neq x_{d,i}$ for $i\neq j$. Furthermore, $x_{d,j}>0$ except possibly for $j=n$ when it can be zero (for $d=2n+1$) or negative (for $d=2n$). We can also easily check that $|x_{d,j}|$ is strictly decreasing and thus in fact $x_{d,j}\neq \pm x_{d,i}$ for $i\neq j$.

In terms of these parameters the dimensions of the representations $\bfm_d$ have the following expressions
\be
	\dim \bfm_{2n} &= \prod_{1\leq i < j\leq n}\frac{(x_{2n,i}+x_{2n,j})(x_{2n,i}-x_{2n,j})}{(y_{2n,i}+y_{2n,j})(y_{2n,i}-y_{2n,j})},\\
	\dim \bfm_{2n+1} &= \prod_{i=1}^n \frac{x_{2n+1,i}+\thalf}{y_{2n+1,i}+\thalf}\prod_{1\leq i<j \leq n}\frac{(x_{2n+1,i}+x_{2n+1,j}+1)(x_{2n+1,i}-x_{2n+1,j})}{(y_{2n+1,i}+y_{2n+1,j}+1)(y_{2n+1,i}-y_{2n+1,j})},
\ee
where $y_{d,k}=n-k$ is $x_{d,k}$ for the trivial representation (so that $\dim \bullet=1$).

\subsection{Reduced matrix elements}
\label{app:formulae:ReducedMatrixElements}
We are now ready to write the formulas for the reduced matrix elements~\eqref{eq:reducedmatelement}. We will give formulas for
\be
\rmel{\bfm_d}{\bfm_{d-1}}{M_{12}}{\bfm_{d}}{\bfm'_{d-1}}\equiv 
(-1)^{d-1}\rmel{\bfm_d}{\bfm_{d-1}}{M^{\myng{(1)}}}{\bfm_{d}}{\bfm'_{d-1}},
\ee
which is more natural from the point of view of~\eqref{eq:M12reduced}. We have in odd dimensions
\be
\rmel{\bfm_{2n+1}}{\bfm_{2n}}{M_{12}}{\bfm_{2n+1}}{\bfm_{2n}(\pm j)}=
	\pm\sqrt{
	\frac{\prod_{k=1}^n(x_{2n+1,k}\mp x_{2n,j})(x_{2n+1,k}\pm x_{2n,j}+1)}
	{2\prod_{k=1\atop k\neq j}^{n}(x_{2n,k}-x_{2n,j})(x_{2n,k}+x_{2n,j})}
}.
\ee
According to~\eqref{eq:evendtensor} this gives all possible reduced matrix elements in even dimensions. Note that according to the discussion above, the factors in the denominator are never zero (assuming that all weights are dominant).

In even dimensions we have
\be\label{eq:evendrmel1}
&\qquad\rmel{\bfm_{2n}}{\bfm_{2n-1}}{M_{12}}{\bfm_{2n}}{\bfm_{2n-1}}=
\frac{
	-i\prod_{k=1}^n x_{2n,k}
}
{
	\sqrt{\prod_{k=1}^{n-1}x_{2n-1,k}(x_{2n-1,k}+1)}
},\\ \label{eq:evendrmel2}
&\rmel{\bfm_{2n}}{\bfm_{2n-1}}{M_{12}}{\bfm_{2n}}{\bfm_{2n-1}(\pm i)}=\nn\\
&=\pm\sqrt{
	-\frac{
		\prod_{k=1}^n(x_{2n,k}-x_{2n-1,i}-\delta_{\pm,+})(x_{2n,k}+x_{2n-1,i}+\delta_{\pm,+})
	}{
		(x_{2n-1,i}+\delta_{\pm,+})(2x_{2n-1,i}+1)\prod_{k=1\atop k\neq i}^{n-1}(x_{2n-1,k}-x_{2n-1,i})(x_{2n-1,k}+x_{2n-1,i}+1)
	}
},
\ee
where $\delta_{\pm,+}$ is equal to $1$ for $+$ sign and to $0$ for $-$ sign. According to~\eqref{eq:odddtensor}, this account for all reduced matrix elements in even dimensions. The only potential zero in the denominator of~\eqref{eq:evendrmel1} is from $x_{2n-1,n-1}$. However, if $x_{2n-1,n-1}=m_{2n-1,n-1}=0$, then $\bfm_{2n-1}$ does not appear in $\bfm_{2n-1}\otimes\myng{(1)}$, and this reduced matrix element has to be set to $0$. Similarly, the only potential zero in the denominator of~\eqref{eq:evendrmel2} appears for $(-)$ sign and $i=n-1$, when we have a factor of $x_{2n-1,n-1}$. Again, it is only a problem if $x_{2n-1,n-1}=m_{2n-1,n-1}=0$, in which case $\bfm_{2n-1}(-n+1)$ does not appear in $\bfm_{2n-1}\otimes\myng{(1)}$ so we need to set this matrix element to~$0$.

\subsection{Isoscalar factors}
\label{app:formulae:Isoscalar}
The isoscalar factors are given by formulas of a very similar form. In odd dimensions we have
\be
&\qquad\rcgc{\bfm_{2n+1}}{\bfm_{2n}}{\myng{(1)}}{\bullet}{\bfm_{2n+1}}{\bfm_{2n}}=
\frac{
	\prod_{k=1}^n x_{2n,k}
}{
	\sqrt{\prod_{k=1}^n x_{2n+1,k}(x_{2n+1,k}+1)}
},\\
&\rcgc{\bfm_{2n+1}}{\bfm_{2n}}{\myng{(1)}}{\bullet}{\bfm_{2n+1}(\pm i)}{\bfm_{2n}}=\nn\\&=
\sqrt\frac{
	\prod_{k=1}^n(x_{2n+1,i}-x_{2n,k}+\delta_{\pm,+})(x_{2n+1,i}+x_{2n,k}+\delta_{\pm,+})
}{
	(x_{2n+1,i}+\delta_{\pm,+})(2x_{2n+1,i}+1)\prod_{k=1\atop k\neq i}^n (x_{2n+1,i}-x_{2n+1,k})(x_{2n+1,i}+x_{2n+1,k}+1)
},
\ee
and the same comments as for the reduced matrix elements apply about the possible zeros in denominators. In even dimensions the isoscalar factors are given by
\be\label{eq:isoscalarevend}
&\rcgc{\bfm_{2n}}{\bfm_{2n-1}}{\myng{(1)}}{\bullet}{\bfm_{2n}(\pm i)}{\bfm_{2n-1}}=\sqrt{
	\frac{
		\prod_{k=1}^{n-1}(x_{2n,i}\mp x_{2n-1,k})(x_{2n,i}\pm x_{2n-1,k}\pm 1)
	}
	{
		2\prod_{k=1\atop k\neq i}^{n}(x_{2n,i}-x_{2n,k})(x_{2n,i}+x_{2n,k})
	}
}.
\ee
To derive the isoscalar factor for $(\myng{(1)},\myng{(1)})$ pattern in vector representation, we consider the following expression,
\be
	\<\seq{m}_d;\myng{(1)},\bullet,\ldots|M_{12}|\seq{m}'_d\>.
\ee
Acting with $M_{12}$ on the left, we find
\be
	&\<\seq{m}_d;\myng{(1)},\myng{(1)},\bullet,\ldots|\seq{m}'_d\>
	+\sum_{\widetilde{\seq{m}}_d}\<\seq{m}_d|M_{12}|\widetilde{\seq{m}}_d\>\<\widetilde{\seq{m}}_d;\myng{(1)},\bullet,\ldots|\seq{m}'_d\>=\nn\\
	&=\rcgc{\bfm_d}{\bfm_{d-1}}{\myng{(1)}}{\myng{(1)}}{\bfm'_d}{\bfm'_{d-1}}
	\rcgc{\bfm_{d-1}}{\bfm_{d-2}}{\myng{(1)}}{\bullet}{\bfm'_{d-1}}{\bfm'_{d-2}}\delta_{\seq{m}_{d-2},\seq{m}'_{d-2}}-\nn\\
	&\qquad 
		-\sum_{\widetilde{\seq{m}}_d}\rmel{\bfm_d}{\widetilde\bfm_{d-1}}{M_{12}}{\bfm_d}{\bfm_{d-1}}^*
		\rcgc{\bfm_{d-1}}{\bfm_{d-2}}{\myng{(1)}}{\bullet}{\widetilde\bfm_{d-1}}{\widetilde\bfm_{d-2}}
		\rcgc{\bfm_{d}}{\widetilde\bfm_{d-1}}{\myng{(1)}}{\bullet}{\bfm'_{d}}{\bfm'_{d-1}}
		\delta_{\seq{m}_{d-2},\widetilde{\seq{m}}_{d-2}}\delta_{\seq{m}'_{d-1},\widetilde{\seq{m}}_{d-1}}\nn\\
	&=\rcgc{\bfm_d}{\bfm_{d-1}}{\myng{(1)}}{\myng{(1)}}{\bfm'_d}{\bfm'_{d-1}}
	\rcgc{\bfm_{d-1}}{\bfm_{d-2}}{\myng{(1)}}{\bullet}{\bfm'_{d-1}}{\bfm'_{d-2}}\delta_{\seq{m}_{d-2},\seq{m}'_{d-2}}-\nn\\
	&\qquad 
		-\rmel{\bfm_d}{\bfm'_{d-1}}{M_{12}}{\bfm_d}{\bfm_{d-1}}^*
		\rcgc{\bfm_{d-1}}{\bfm_{d-2}}{\myng{(1)}}{\bullet}{\bfm'_{d-1}}{\bfm'_{d-2}}
		\rcgc{\bfm_{d}}{\bfm'_{d-1}}{\myng{(1)}}{\bullet}{\bfm'_{d}}{\bfm'_{d-1}}
		\delta_{\seq{m}_{d-2},\seq{m}'_{d-2}}.
\ee
Action on the right gives, on the other hand,
\be
	&\sum_{\widetilde{\seq{m}}_d}\<\seq{m}_d;\myng{(1)},\bullet,\ldots|\widetilde{\seq{m}}_d\>\<\widetilde{\seq{m}}_d|M_{12}|\seq{m}'_d\>=\nn\\
	&=-\sum_{\widetilde{\seq{m}}_d}
		\rcgc{\bfm_d}{\bfm_{d-1}}{\myng{(1)}}{\bullet}{\widetilde\bfm_d}{\widetilde\bfm_{d-1}}
		\rmel{\bfm_d'}{\bfm'_{d-1}}{M_{12}}{\bfm'_{d}}{\widetilde\bfm_{d-1}}^*
		\rcgc{\widetilde\bfm_{d-1}}{\widetilde\bfm_{d-2}}{\myng{(1)}}{\bullet}{\bfm'_{d-1}}{\bfm'_{d-2}}
		\delta_{\seq{m}_{d-1},\widetilde{\seq{m}}_{d-1}}\delta_{\seq{m}'_{d-2},\widetilde{\seq{m}}_{d-2}}\nn\\
	&=-
		\rcgc{\bfm_d}{\bfm_{d-1}}{\myng{(1)}}{\bullet}{\bfm'_d}{\bfm_{d-1}}
		\rmel{\bfm_d'}{\bfm'_{d-1}}{M_{12}}{\bfm'_{d}}{\bfm_{d-1}}^*
		\rcgc{\bfm_{d-1}}{\bfm'_{d-2}}{\myng{(1)}}{\bullet}{\bfm'_{d-1}}{\bfm'_{d-2}}
		\delta_{{\seq{m}}_{d-2},\seq{m}'_{d-2}}.
\ee
By comparing these expressions and choosing $\bfm'_{d-2}$ such that 
\be
		\rcgc{\bfm_{d-1}}{\bfm'_{d-2}}{\myng{(1)}}{\bullet}{\bfm'_{d-1}}{\bfm'_{d-2}}
\ee
is non-vanishing, we conclude
\be
	\rcgc{\bfm_d}{\bfm_{d-1}}{\myng{(1)}}{\myng{(1)}}{\bfm'_d}{\bfm'_{d-1}}=&
	-\rcgc{\bfm_d}{\bfm_{d-1}}{\myng{(1)}}{\bullet}{\bfm'_d}{\bfm_{d-1}}
		\rmel{\bfm_d'}{\bfm'_{d-1}}{M_{12}}{\bfm'_{d}}{\bfm_{d-1}}^*\nn\\
	&+\rcgc{\bfm_{d}}{\bfm'_{d-1}}{\myng{(1)}}{\bullet}{\bfm'_{d}}{\bfm'_{d-1}}
		\rmel{\bfm_d}{\bfm'_{d-1}}{M_{12}}{\bfm_d}{\bfm_{d-1}}^*.
\ee

\subsection{Comments on $d=3$}
\label{app:formulae:3dcomment}
A few modifications to the above formulas are required in the case $d=3$. This is because the $d-1=2$ and vector representation in $d=2$ is not irreducible.

The formulas for the reduced matrix elements of remain valid if they are used together with~\eqref{eq:isoscalarevend}. Indeed, we can compute
\be
	\<j,m\pm 1|M_{12}|j,m\>=
	\rmel{j}{m\pm 1}{M_{12}}{j}{m}
	\rcgcConj{m\pm 1}{\bullet}{\myng{(1)}}{\bullet}{m}{\bullet}=\mp \frac{1}{2}\sqrt{(j\mp m)(j\pm m+1)},
\ee
which coincides with the standard expression for $M_{12}$ which follows from
\be
	M_{12}=-iJ_{\hat 2}=-\frac{J_+}{2}+\frac{J_-}{2},
\ee
as discussed in section~\ref{sec:reptheory:GelfandTsetlin:3d}. Alternatively, the formula for the reduced matrix can be interpreted as
\be
	\rmel{j}{m\pm 1}{M_{12}}{j}{m}=
	 \rmel{j}{m\pm 1}{M^{1,+1}}{j}{m}
	-\rmel{j}{m\pm 1}{M^{1,-1}}{j}{m},
\ee
where $M^{1,\pm}=-\frac{J_\pm}{\sqrt{2}}$ are defined according to~\eqref{eq:veceplus} and~\eqref{eq:veceminus} (treating the second index of $M$ as a vector index). The matrix elements in the right hand side should be used with the CG coefficients of $Spin(2)$, $\<m\pm 1|\pm1,m\>=1$.

The isoscalar factors can interpreted as 
\be
	\rcgc{j}{m}{\myng{(1)}}{\myng{(1)}}{j'}{m'}
	=
	\rcgc{j}{m}{\myng{(1)}}{+1}{j'}{m'}
	-\rcgc{j}{m}{\myng{(1)}}{-1}{j'}{m'}.
\ee
The isoscalar factors in the right hand side are to be combined with the CG coefficients of $Spin(2)$, $\<m\pm 1|\pm1,m\>=1$. This can be checked against the known formulas for $Spin(3)$ CG coefficients.\footnote{Recall that the $m$-independent phase of CG coefficients is convention-dependent. The formulas given here agree with the conventions of~\cite{edmonds1996angular} (the conventions used in~\texttt{Mathematica} as of version \texttt{11.0}) for $j'=j,j+1$ and differ by a sign for $j'=j-1$.}

\subsection{A sum rule for reduced matrix elements and isoscalar factors}
\label{app:formulae:SumRule}
As discussed in the main text, the following identity holds,
\be
	\sum_{\bfm_{d-1}}\Bigg(
{\bfm_{d}\atop\widetilde\bfm_{d-1}}
\Bigg\vert
M^{\myng{(1)}}
\Bigg\vert
{\bfm_{d}\atop\bfm_{d-1}}
\Bigg)\Bigg(
{\bfm_d\atop \bfm_{d-1}}
{\myng{(1)}\atop\myng{(1)}}
\Bigg\vert
{\widetilde\bfm_d\atop \widetilde\bfm_{d-1}}
\Bigg)=(-1)^{d-1}(\bfm_d\,\myng{(1)}|\widetilde{\bfm}_{d})
\Bigg(
{\bfm_d\atop \widetilde\bfm_{d-1}}
{\myng{(1)}\atop\bullet}
\Bigg\vert
{\widetilde\bfm_d\atop \widetilde\bfm_{d-1}}
\Bigg).
\ee
We are not aware of a simple derivation of this fact or of the coefficients $(\bfm_d\,\myng{(1)}|\widetilde{\bfm}_{d})$. We note, however, that this identity is required for existence of certain weight-shifting operators in vector representation. The coefficients $(\bfm_d\,\myng{(1)}|\widetilde{\bfm}_{d})$ are given by the following formulas
\be
	(\bfm_{2n}\,\myng{(1)}|\bfm_{2n}(\pm i))&=n\mp x_{2n,i}-1,\\
	(\bfm_{2n+1}\,\myng{(1)}|\bfm_{2n+1}(\pm i))&=n\mp x_{2n,i}-\delta_{\pm+},\\
	(\bfm_{2n+1}\,\myng{(1)}|\bfm_{2n+1})&=n.
\ee
We found these formulas by considering a few low-dimensional cases and guessing the general result, which was then verified on a large set of representations in various dimensions. In terms of $m_{d,k}$ these coefficients can be rewritten as
\be\label{eq:weirdfactor1}
	(\bfm_d\,\myng{(1)}|\bfm_d(+i))&=-m_{d,i}+i-1,\\
	(\bfm_d\,\myng{(1)}|\bfm_d(-i))&=m_{d,i}+d-i-1,\\\label{eq:weirdfactor2}
	(\bfm_{2n+1}\,\myng{(1)}|\bfm_{2n+1})&=n.
\ee

\section{Scalar-fermion blocks in various dimensions}
\label{app:ScalarFermionChecks}

\subsection{Comparison in 2 dimensions}
Interestingly, the formulas for scalar-fermion seed blocks in section~\ref{sec:casimir:SeedBlocks} also work in the case $n=1$, i.e.\ $d=2$. We have the following identity,
\be
	s^D e^{\theta M_{12}} = (s e^{i\theta})^{\frac{D-i M_{12}}{2}}(s e^{-i\theta})^{\frac{D+i M_{12}}{2}},
\ee
and so if we define
\be\label{eq:sl2holo}
	L_0&=\frac{D-i M_{12}}{2},\quad L_{-1}=\frac{P_1-i P_2}{2},\quad L_{+1}=\frac{K_1+i K_2}{2},\\
	\bar L_0&=\frac{D+i M_{12}}{2},\quad \bar L_{-1}=\frac{P_1+i P_2}{2},\quad \bar L_{+1}=\frac{K_1-i K_2}{2},\label{eq:sl2anti}
\ee
we find that the conformal block in the form~\eqref{eq:fourptconvenient} is given by
\be
	\<0|\cO_4^{m_4}\cO_3^{m_3}|\cO|z^{L_0}\bar z^{\bar L_0}\cO_2^{m_2}\cO_1^{m_1}|0\>.
\ee
The algebras~\eqref{eq:sl2holo} and~\eqref{eq:sl2anti} satisfy the usual commutation relations
\be
	[L_m,L_n]&=(m-n)L_{m+n},\\
	[\bar L_m,\bar L_n]&=(m-n)\bar L_{m+n}.
\ee

The configuration considered in section~\ref{sec:casimir:SeedBlocks} is $m_2=-m_4=\thalf$ and $m_1=m_3=0$. This corresponds to holomorphic and anti-holomorphic dimensions
\be
	h_1&=\thalf\Delta_1,\quad h_2=\thalf\Delta_2-\tfrac{1}{4},\quad h_3=\thalf\Delta_3,\quad h_4=\thalf\Delta_4+\tfrac{1}{4},\\
	\bar h_1&=\thalf\Delta_1,\quad \bar h_2=\thalf\Delta_2+\tfrac{1}{4},\quad \bar h_3=\thalf\Delta_3,\quad \bar h_4=\thalf\Delta_4-\tfrac{1}{4},
\ee
while the intermediate representation $\bfj^\pm$ corresponds to $h_\cO=\thalf\Delta_\cO\mp \thalf j$ and $\bar h_\cO=\thalf\Delta_\cO\pm \thalf j$. The conformal block for exchange of $\bfj^\pm$ is equal to the usual expression
\be
	z^{h_\cO}{}_2F_1(h_\cO-h_{12},h_\cO+h_{34};2h_\cO;z)\times \bar z^{\bar h_\cO}{}_2F_1(\bar h_\cO-\bar h_{12},\bar h_\cO+\bar h_{34};2\bar h_\cO;\bar z).
\ee
It is straightforward to expand this expression in power series in $s$ and check that it is consistent with the recursion relation~\eqref{eq:recursionScalarFermion2n}.

\subsection{Comparison in 3 dimensions}
To perform the comparison with the known 3d results, we first need to relate the GT basis to the standard basis for 3d fermions. There is a unique fermionic representation in 3d, $\bfm_3=(\thalf)$, with the allowed GT patterns
\be
	\seq{m}_{3,\pm}=(\thalf),(\pm\thalf),
\ee
consistently with the representation being two-dimensional. For 3d spinors we use the conventions as in~\cite{Iliesiu:2015akf,Kravchuk:2016qvl,Karateev:2017jgd}, and we will be comparing with the scalar-fermion blocks in the form of~\cite{Karateev:2017jgd}. These papers use Lorentz signature and thus we need to perform Wick rotation by defining 
\be
	M^{\mu\nu}=-i^{\delta_{\mu,0}+\delta_{\nu,0}}M^{\mu\nu}_L,
\ee
where $M^{\mu\nu}_L$ are the Lorentz generators from~\cite{Karateev:2017jgd}. We also added a $(-)$ due to the difference in conventions for conformal algebra. Furthermore, we need to relabel the indices by defining
\be
1_\text{here}=2_\text{there},\quad 2_\text{here}=0_\text{there},\quad
3_\text{here}=1_\text{there}.
\ee
This is required because of the way the conformal frame is defined in~\cite{Karateev:2017jgd}. Using the explicit expression for the Lorentz generators and the correspondence above, we can identify
\be
	\cO^1=\cO^{\seq{m}_{3,-}},\quad \cO^2=i\cO^{\seq{m}_{3,+}}.
\ee
Contracting the structures~\eqref{eq:fourpt3d} with polarization vectors $s_\alpha$ as in~\cite{Karateev:2017jgd}, we find
\be
	\left[
	{0,0\atop \thalf,m_4}\Big\vert{0\atop +\thalf}
	{+\thalf\atop 0}\Big\vert
	{\thalf,m_1\atop 0,0}
	\right]\rightarrow&-\bar\xi_4\bar\xi_1=-[-\thalf,0,0,-\thalf],\\
	\left[
	{0,0\atop \thalf,m_4}\Big\vert{0\atop +\thalf}
	{-\thalf\atop 0}\Big\vert
	{\thalf,m_1\atop 0,0}
	\right]\rightarrow&i\bar\xi_4\xi_1=i[\thalf,0,0,-\thalf],\\
	\left[
	{0,0\atop \thalf,m_4}\Big\vert{0\atop -\thalf}
	{+\thalf\atop 0}\Big\vert
	{\thalf,m_1\atop 0,0}
	\right]\rightarrow&i\bar\xi_1\xi_4=i[-\thalf,0,0,\thalf],\\
	\left[
	{0,0\atop \thalf,m_4}\Big\vert{0\atop -\thalf}
	{-\thalf\atop 0}\Big\vert
	{\thalf,m_1\atop 0,0}
	\right]\rightarrow&\xi_4\xi_1=[\thalf,0,0,\thalf],
\ee
where the right hand side is in the notation of~\cite{Karateev:2017jgd}. The results (for parity-even components) of~\cite{Karateev:2017jgd} are given in the form
\be
	\<0|\psi_4(\infty,s_4)\phi_3(1)|\cO|\phi_2(z,\bar z)\psi_1(0,s_1)|0\>=&
	\thalf g_1(z,\bar z)[-\thalf,0,0,-\thalf]+
	\thalf g_2(z,\bar z)[\thalf,0,0,-\thalf]+\nn\\
	&+\thalf g_2(z,\bar z)[-\thalf,0,0,\thalf]+
	\thalf g_1(z,\bar z)[\thalf,0,0,\thalf].
\ee
This implies that 
\be
	s^{-\Delta_1-\Delta_2}\<0|\psi_4\phi_3|\cO|s^D e^{\theta M_{12}}\phi_2\psi_1|0\>=&
	\thalf \left(\cos\tfrac{\theta}{2}\,g_1(z,\bar z)+i\sin\tfrac{\theta}{2}\,g_2(z,\bar z)\right)[-\thalf,0,0,-\thalf]+\nn\\
	&+\thalf \left(\cos\tfrac{\theta}{2}\,g_2(z,\bar z)+i\sin\tfrac{\theta}{2}\,g_1(z,\bar z)\right)[\thalf,0,0,-\thalf]+\nn\\
	&+\thalf \left(\cos\tfrac{\theta}{2}\,g_2(z,\bar z)+i\sin\tfrac{\theta}{2}\,g_1(z,\bar z)\right)[-\thalf,0,0,\thalf]+\nn\\
	&+\thalf \left(\cos\tfrac{\theta}{2}\,g_1(z,\bar z)+i\sin\tfrac{\theta}{2}\,g_2(z,\bar z)\right)[\thalf,0,0,\thalf].
\ee

Using this result, we can compute the expansion~\eqref{eq:3dfermionexpansion} in terms of functions $g_1$ and $g_2$. These functions are conveniently computed by acting with the differential operators of~\cite{Karateev:2017jgd} on the scalar conformal block obtained from the recursion relation~\eqref{eq:scalarrecursion}.\footnote{Alternatively, one can use Zamolodchikov-type recursion relations of~\cite{Iliesiu:2015akf}.} We have checked that the resulting expansion is consistent with the recursion relation which follows from~\eqref{eq:3dfemiongamma}-\eqref{eq:3dfermionbargamma} at the first few levels for various choices of $j_\cO$.

\subsection{Comparison in 4 dimensions}

To perform the comparison with the known 4d results, we first need to relate the GT basis to the standard basis for Weyl fermions. We are considering the two fermionic representations $\bfm_4^\pm=(\thalf,\pm\thalf)$. The allowed GT patterns are
\be
	\seq{m}^\pm_{4,+}&=(\thalf,\pm\thalf),(\thalf),(+\thalf),\\
	\seq{m}^\pm_{4,-}&=(\thalf,\pm\thalf),(\thalf),(-\thalf),
\ee
consistently with the representations being two-dimensional. For 4d Weyl spinors, we will use the conventions of~\cite{Cuomo:2017wme}. We need to make a few adaptations from conventions there to the present conventions. First, we need to perform Wick rotation by defining
\be
	M^{\mu\nu}=-i^{\delta_{\mu,0}+\delta_{\nu,0}}M^{\mu\nu}_L,
\ee
where $M_L$ are the Lorentz generators of~\cite{Cuomo:2017wme}. We also added a $(-)$ due to the difference in conventions for conformal algebra. Furthermore, we need to relabel the indices by defining
\be
	1_\text{here}=3_\text{there},\quad 2_\text{here}=0_\text{there},\quad
	3_\text{here}=1_\text{there},\quad
	4_\text{here}=2_\text{there}.
\ee
This is required because of the way the conformal frame is defined in~\cite{Cuomo:2017wme}.

Comparing the transformation properties of $|\seq{m}^\pm_{4,\pm}\>$ and the operators $\cO^{\dot\alpha}$ and $\cO_{\alpha}$, we find that we can set
\be
	\cO_1 &= \cO^{\seq{m}^+_{4,-}},\quad
	\cO_2 = -i\cO^{\seq{m}^+_{4,+}},\\
	\cO^{\dot 1} &= \cO^{\seq{m}^-_{4,-}},\quad
	\cO^{\dot 2} = +i\cO^{\seq{m}^-_{4,+}}.
\ee
According to~\eqref{eq:seed4pt} we find the following non-zero components of tensor structures~\eqref{eq:ferm4pt}
\be
\left[
{\bullet\atop\seq{m}^-_{4,-}}\Big\vert
{\bullet\atop(\thalf)}
\,(\thalf)
\Big\vert
(+\thalf)
\Big\vert
(\thalf)
\,{\bullet\atop(\thalf)}\Big\vert
{\bullet\atop\seq{m}^+_{4,+}}
\right]&=\frac{-i}{\sqrt{2}},\\
\left[
{\bullet\atop\seq{m}^-_{4,+}}\Big\vert
{\bullet\atop(\thalf)}
\,(\thalf)
\Big\vert
(-\thalf)
\Big\vert
(\thalf)
\,{\bullet\atop(\thalf)}\Big\vert
{\bullet\atop\seq{m}^+_{4,-}}
\right]&=\frac{i}{\sqrt{2}}.
\ee
Contracting with polarization vectors as in~\cite{Cuomo:2017wme}, we find 
\be
	t_+=+\frac{\xi_2\bar\xi_4}{\sqrt{2}}=
	+\frac{1}{\sqrt{2}}\struct
		{0}{-\thalf}{0}{0}
		{0}{0}{0}{-\thalf}
	,\\
	t_-=-\frac{\eta_2\bar\eta_2}{\sqrt{2}}=
	-\frac{1}{\sqrt{2}}\struct
		{0}{+\thalf}{0}{0}
		{0}{0}{0}{+\thalf}.
\ee
Using this correspondence, we can find that the primal conformal block has the form
\be\label{eq:ferm4pt4d}
	\<0|\overline\psi_4\phi_3|\cO|s^De^{\theta M_{12}}\psi_2\phi_1|0\>=&
	-\sqrt{2}\left(2\sqrt{z}H^0_1(z,\bar z)+\frac{1}{\sqrt{\bar z}}H^1_1(z,\bar z)
	\right)t_+\nn\\
	&-\sqrt{2}\left(2\sqrt{\bar z}H^0_1(z,\bar z)+\frac{1}{\sqrt{z}}H^1_1(z,\bar z)
	\right)t_-.
\ee
In our terminology it corresponds to exchange of a primary in representation $(\ell+\thalf,\thalf)$ with $\ell$ as in~\cite{Cuomo:2017wme}. Using explicit expressions for functions $H$~\cite{Echeverri:2016dun} in normalization of~\cite{Cuomo:2017wme}, we can check that the leading term in $s=|z|$ coincides with~\eqref{eq:scalarFermionExpansion} and~\eqref{eq:fermionmatrom} with 
\be
	\mathbf{\Lambda}_{0,j_\cO}=\begin{pmatrix}
		-i\frac{(\ell+2)(-1)^\ell }{\sqrt{2}}\\
		0
	\end{pmatrix}.
\ee
We can then use the recursion relation~\eqref{eq:recursionScalarFermion2n} to compute higher order coefficients and plug them into the expansion~\eqref{eq:scalarFermionExpansion}. We can compute the same expansion by plugging the explicit expressions for functions $H$ into~\eqref{eq:ferm4pt4d} using~\texttt{CFTs4D} package from~\cite{Cuomo:2017wme}. We checked that both expansion coincide at the first few levels.

\section{Gelfand-Tsetlin bases for tensor representations}
\label{app:tensors}
To gain some familiarity with GT bases in general dimensions, let us consider how it is related to the usual Cartesian bases for tensor representations. For simplicity of discussion, we avoid dealing with self-duality constraints. This restricts us to the representations $\bfm_d$ with $m_{d,k}=0$ for $k\geq d/2$, i.e.\ to Young diagrams with less than $d/2$ rows. In particular, we will only consider the GT patters in which all representations are of this kind.\footnote{The same general approach works even without these assumptions, and the details are not hard to recover.}

Our goal is for a given GT pattern $\seq{m}_d$ to find the explicit tensor $T^{\mu_1\ldots\mu_{|\bfm_d|}}_{\seq{m}_d}$ which gives the corresponding basis element $|\seq{m}_d\>$, up to a multiplicative factor. We do this recursively, by explicitly constructing the \textit{dimensional induction map}
\be
I_{\bfm_{d-1}}^{\bfm_d}:V_{\bfm_{d-1}}\to V_{\bfm_d},\quad \bfm_{d-1}\in \bfm_{d}
\ee
which is defined, up to normalization, by the requirement that it is $Spin(d-1)$-equivariant and non-trivial. By irreducibility of $\bfm_{d-1}$ it follows that $I$ establishes an isomorphism between $V_{\bfm_{d-1}}$ and the subspace in $V_{\bfm_{d}}$ which transforms according to $\bfm_{d-1}$ under $Spin(d-1)$. Since dimensional reduction is multiplicity-free, this subspace is uniquely determined. 

It then immediately follows from the definition of GT basis that the following relationship between GT basis vectors holds,
\be
|\bfm_{d},\bfm_{d-1},\bfm_{d-2},\ldots\>\propto I_{\bfm_{d-1}}^{\bfm_d}|\bfm_{d-1},\bfm_{d-2},\ldots\>.
\ee
In particular, if $\bfm_{d-k}=\bullet$ is the trivial representation, we find
\be\label{eq:GTrecursion}
|\bfm_{d},\bfm_{d-1},\bfm_{d-2},\ldots\>\propto I_{\bfm_{d-1}}^{\bfm_d}I_{\bfm_{d-2}}^{\bfm_{d-1}}I_{\bfm_{d-3}}^{\bfm_{d-2}}\ldots I_{\bfm_{d-k}}^{\bfm_{d-k+1}}1.
\ee

\begin{figure}[t]
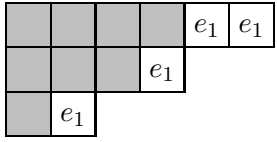

	\centering
	\begin{ytableau}
		*(lightgray)\hfill & *(lightgray) & *(lightgray) & *(lightgray) & e_1 & e_1\\
		*(lightgray) & *(lightgray) & *(lightgray) & e_1 \\
		*(lightgray) & e_1
	\end{ytableau}
	\caption{The relationship between Young diagrams of $\bfm_d$ and $\bfm_{d-1}$. The boxes which belong to $\bfm_{d-1}$ are shaded.}
	\label{fig:ytabrelation}
\end{figure}

To construct $I^{\bfm_d}_{\bfm_{d-1}}$ explicitly, start with a general $U^{\mu_1\ldots \mu_{|\bfm_{d-1}|}}\in V_{\bfm_{d-1}}$. For convenience we assume that the indices of $U$ run from $2$ to $d$.\footnote{Recall that by our choice of $Spin(d-1)\subset Spin(d)$, $Spin(d-1)$ stabilizes $e_1$.} We first extend the definition of $U$ to allow its indices to assume the value $1$ by setting $U^{\mu_1\ldots \mu_{|\bfm_{d-1}|}}=0$ whenever at least one of $\mu_i=1$. We then define
\be\label{eq:Tprimedefn}
T'^{\mu_1\ldots \mu_{|\bfm_{d}|}}=U^{\mu_1\ldots \mu_{|\bfm_{d-1}|}}e^{\mu_{|\bfm_{d-1}|+1}}_1\cdots e^{\mu_{|\bfm_{d}|}}_1-\mathrm{traces}.
\ee
A generic relationship between the Young diagrams $\bfm_{d}$ and $\bfm_{d-1}$ is shown in figure~\ref{fig:ytabrelation}. We can associate the indices of $e_1$ in~\eqref{eq:Tprimedefn} to the unshaded boxes in figure~\ref{fig:ytabrelation} and apply to $T'$ the Young symmetrizer $Y_{\bfm_{d}}$ corresponding to $\bfm_{d}$ to define
\be
I_{\bfm_{d-1}}^{\bfm_d}U\equiv Y_{\bfm_d}T'.
\ee
Note that it is guaranteed by the dimensional reduction rules from section~\ref{sec:reptheory:DimensionalReduction} that no two indices of $e_1$ land in the same column of $\bfm_d$.

As explained above, this map allows us to reconstruct Gelfand-Tsetlin basis vectors up to a phase. Let us look at some examples. First, consider the GT basis vector
\be
|\jyng,\bullet,\ldots\>.
\ee
From~\eqref{eq:GTrecursion} we find
\be\label{eq:GTSTTexample1}
|\jyng,\bullet,\ldots\>\propto I^{\jyng}_{\bullet}1=e^{\mu_{1}}_1\cdots e^{\mu_{j}}_1-\mathrm{traces}.
\ee
This reproduces the result of section~\ref{sec:structure:Pfunctions:scalarmatrom}.

As a more complicated example, consider
\be
|\jyng,\myng{(1)},\bullet,\ldots\>.
\ee
From~\eqref{eq:GTrecursion} we find
\be\label{eq:GTSTTexample2}
|\jyng,\myng{(1)},\bullet,\ldots\>&\propto I^{\jyng}_{\myng{(1)}}|\myng{(1)},\bullet,\ldots\>\nn\\
&\propto I^{\jyng}_{\myng{(1)}} e^{\mu_1}_2=e^{(\mu_1}_2 e_1^{\mu_2}\cdots e^{\mu_j)}_1-\mathrm{traces}.
\ee
Similarly, we can find that
\be
|\jyng,\myng{(1)},\myng{(1)},\bullet,\ldots\>\propto e_3^{(\mu_1}e^{\mu_{2}}_1\cdots e^{\mu_{j})}_1-\mathrm{traces},
\ee
and so on. In the case $j=1$ this reproduces the results of section~\ref{sec:reptheory:GelfandTsetlin:vector} for vector representation. 

Consider now the simplest non-STT example,
\be
|\myng{(2,1)}\raisebox{3.5pt}{$\cdots\!\myng{(1)}$},\myng{(1)},\bullet,\ldots\>.
\ee
Note that $\myng{(1)}$ is the simplest representation to which $\myng{(2,1)}\raisebox{3.5pt}{$\cdots\!\myng{(1)}$}$ can reduce. This differs from~\eqref{eq:GTSTTexample2} only in the Young symmetrizer,
\be\label{eq:GTnonsttexample}
|\myng{(2,1)}\raisebox{3.5pt}{$\cdots\!\myng{(1)}$},\myng{(1)},\bullet,\ldots\>&\propto Y_{\myng{(2,1)}\raisebox{2.8pt}{$\cdots\!\myng{(1)}$}}(e_2^{\nu}e^{\mu_{1}}_1\cdots e^{\mu_{j}}_1-\mathrm{traces})=\nn\\
&=\thalf e_2^{\nu}e_1^{(\mu_1}e_1^{\mu_2}\cdots e_1^{\mu_j)}-\thalf e_1^{\nu}e_2^{(\mu_1}e_1^{\mu_2}\cdots e_1^{\mu_j)}-\mathrm{traces}.
\ee
Similarly,
\be
|\myng{(2,1)}\raisebox{3.5pt}{$\cdots\!\myng{(1)}$},\myng{(1)},\myng{(1)},\bullet,\ldots\>\propto \thalf e_3^{\nu}e_1^{(\mu_1}e_1^{\mu_2}\cdots e_1^{\mu_j)}-\thalf e_1^{\nu}e_3^{(\mu_1}e_1^{\mu_2}\cdots e_1^{\mu_j)}-\mathrm{traces},
\ee
and so on.

It is important that we perform trace subtraction and Young symmetrization in all steps of dimensional induction. Consider for example the state
\be
|\myng{(3)},\myng{(2)},\bullet,\ldots\>.
\ee
We have first in $d-1$ dimensions
\be
|\myng{(2)},\bullet,\ldots\>\propto e_2^{\mu_1} e_2^{\mu_2}-\frac{1}{d-1}\delta^{\mu_1\mu_2},
\ee
and when we lift it to $d$ dimensions, we have agreed to set the new entries of this tensor to $0$, which in this case amounts to replacing 
\be
\delta^{\mu_1\mu_2}\to\widetilde{\delta}^{\mu_1\nu_1}\equiv \delta^{\mu_1\mu_2}-e_1^{\mu_1}e_1^{\mu_2},
\ee
so that indeed $\widetilde{\delta}^{1\mu_2}=0$. We thus have\footnote{This object is automatically traceless in $d$ dimensions so we don't have to subtract $d$-dimensional traces.}
\be\label{eq:exampletracecorrect}
|\myng{(3)},\myng{(2)},\bullet,\ldots\>\propto Y_{\myng{(3)}}\left(e_2^{\mu_1} e_2^{\mu_2}-\frac{1}{d-1}\left(\delta^{\mu_1\mu_2}-e_1^{\mu_1}e_1^{\mu_2}\right)\right)e_1^{\mu_3}.
\ee
Clearly, if we didn't take care with $\widetilde\delta$, or had postponed trace subtraction to $d$ dimensions, we would have never obtained a term $e_1^{\mu_1} e_1^{\mu_2} e_1^{\mu_3}$. These choices would be wrong since for $\mu_1=\mu_2=\mu_3=1$ they would reduce to a non-zero constant and thus their dimensional reduction has a component along the trivial representation of $Spin(d-1)$. On the other hand,~\eqref{eq:exampletracecorrect} is non-zero iff only one of $\mu_i$ is set to $1$, in which case it reduces to $\myng{(2)}$, as required.

Similarly, care should be taken with compositions of Young symmetrizers between dimensions. Consider the state
\be
|\myng{(2,1)},\myng{(2)},\myng{(1)},\bullet,\ldots\>.
\ee
We have successively 
\be
|\myng{(1)},\bullet,\ldots\>&\propto e_3^{\mu_1},\\
|\myng{(2)},\myng{(1)},\bullet,\ldots\>&\propto e_2^{(\mu_1}e_3^{\mu_2)},\label{eq:YsymmExampledm1}\\
|\myng{(2,1)},\myng{(2)},\myng{(1)},\bullet,\ldots\>&\propto \half e_1^{\nu}e_2^{(\mu_1}e_3^{\mu_2)}-\frac{1}{4}\left(e_1^{\mu_1}e_2^{(\nu}e_3^{\mu_2)}+e_1^{\mu_2}e_2^{(\nu}e_3^{\mu_1)}\right).\label{eq:YsymmExampled}
\ee
Here we have applied Young symmetrizer both in~\eqref{eq:YsymmExampledm1} and~\eqref{eq:YsymmExampled}. Had we only applied the $d$-dimensional symmetrizer, we would find
\be
|\myng{(2,1)},\myng{(2)},\myng{(1)},\bullet,\ldots\>&\propto \half e_1^{\nu}e_2^{(\mu_1}e_3^{\mu_2)}-\frac{1}{4}\left(e_1^{\mu_1}e_2^{\nu}e_3^{\mu_2}+e_1^{\mu_2}e_2^{\nu}e_3^{\mu_1}\right).\label{eq:YsymmExampleWrong}
\ee
It is easy to see that~\eqref{eq:YsymmExampleWrong} is wrong: setting $\mu_2=1$ we obtain $-\frac{1}{2}e_2^\nu e_3^{\mu_1}$, which is a tensor with no definite symmetry. On contrary, setting $\mu_2=1$ in~\eqref{eq:YsymmExampled}, we find $-\frac{1}{2}e_2^{(\nu} e_3^{\mu_1)}$ which belongs to $\myng{(2)}$ as required. We thus see that the symmetrizers from different dimensions interact non-trivially to ensure that the dimensional reductions are irreducible.

We have so far avoided the question of normalization of the tensors $T_{\seq{m}_d}$. Up to a phase it is determined by the requirement that GT vectors have unit length. This is straightforward to implement on the tensor side. Sometimes we would like to know the normalization factor as a function of the length of the first row $j$ -- this is perhaps most easily implemented using the irreducible projectors as we explain below. The phases can be chosen based on convenience,\footnote{Of course, for every GT pattern this choice should be made once and for all in order to have consistent expressions.} unless one wants to make contact with the GT formulas in appendix~\ref{app:formulae}. We have not attempted to find the general prescription which would match the phase conventions of these formulas.

\subsection{$P$-functions}

In this section we relate $P^{\bfm_d,\bfm_{d-2}}_{\bfm_{d-1},\bfm'_{d-1}}(\theta)$ in tensor representations to the irreducible projectors studied in~\cite{Costa:2016hju}. 

We start by utilizing the tensor representation of GT basis vectors in the definition of $P$,
\be\label{eq:tensorPstart}
P^{\bfm_d,\bfm_{d-2}}_{\bfm_{d-1},\bfm'_{d-1}}(\theta)&\equiv \<\bfm_d,\bfm_{d-1},\bfm_{d-2},\ldots|e^{\theta M_{12}}|\bfm_d,\bfm'_{d-1},\bfm_{d-2},\ldots\>\nn\\
&=T^{\mu_1\ldots \mu_{|\bfm_d|}}_{\seq{m}_{d}}(e^{\theta M_{12}})_{\mu_1\ldots \mu_{|\bfm_d|},\nu_1\ldots \nu_{|\bfm_d|}}T^{\nu_1\ldots \nu_{|\bfm_d|}}_{\seq{m}'_{d}}\nn\\
&=T^{\mu_1\ldots \mu_{|\bfm_d|}}_{\seq{m}_{d}}T^{\mu_1\ldots \mu_{|\bfm_d|}}_{\seq{m}'_{d}}(\theta),
\ee
where $T_{\seq{m}'_{d}}(\theta)$ is equal to $T_{\seq{m}'_{d}}$ in which all occurrences of $e_1$ and $e_2$ have been replaced with
\be
e_1(\theta)&=e^{\theta M_{12}}e_1=\cos\theta e_1+\sin\theta e_2,\\
e_2(\theta)&=e^{\theta M_{12}}e_2=-\sin\theta e_1+\cos\theta e_2.
\ee
Note that in the first line of~\eqref{eq:tensorPstart} $\ldots$ represent the same sequence in both vectors, which can be chosen arbitrarily. For example, if $\bfm_{d-2}$ is STT, we can choose all representations in $\ldots$ to be trivial. We have also assumed that we had chosen the tensors $T_{\seq{m}_{d}}$ to be real for all relevant $\seq{m}_{d}$.\footnote{This might not be possible it the GT patterns do not satisfy the assumptions discussed in the beginning of this appendix. In that case one needs to add some complex conjugations in the formulas.}

We can further trivially rewrite the last line of~\eqref{eq:tensorPstart} as
\be
T^{\mu_1\ldots \mu_{|\bfm_d|}}_{\seq{m}_{d}}T^{\mu_1\ldots \mu_{|\bfm_d|}}_{\seq{m}'_{d}}(\theta)
=T^{\mu_1\ldots \mu_{|\bfm_d|}}_{\seq{m}_{d}}
{}_{\mu_1\ldots \mu_{|\bfm_d|}}\pi_{\nu_1\ldots \nu_{|\bfm_d|}}
T^{\nu_1\ldots \nu_{|\bfm_d|}}_{\seq{m}'_{d}}(\theta)=T_{\seq{m}_{d}}\cdot\pi\cdot T_{\seq{m}'_{d}},
\ee
where ${}_{\mu_1\ldots \mu_{|\bfm_d|}}\pi_{\nu_1\ldots \nu_{|\bfm_d|}}$ is the projector onto the irreducible representation $\bfm_d$. From our construction of tensors $T_{\seq{m}_{d}}$ we know that we can write $T_{\seq{m}_{d}}$ in terms of the basis vectors $e_i$ and Kronecker deltas $\delta_{\mu_i\mu_j}$. We can thus write
\be
T_{\seq{m}_{d}}&=T_{\seq{m}_{d}}^{(e)}+\text{terms containing }\delta^{\mu_i\mu_j},\\
T_{\seq{m}'_{d}}&=T_{\seq{m}'_{d}}^{(e)}+\text{terms containing }\delta^{\mu_i\mu_j}.
\ee
We then conclude
\be\label{eq:Pfrompi}
P^{\bfm_d,\bfm_{d-2}}_{\bfm_{d-1},\bfm'_{d-1}}(\theta)=T_{\seq{m}_{d}}^{(e)}\cdot\pi\cdot  T_{\seq{m}'_{d}}^{(e)}(\theta).
\ee
Note that it is easy to compute $T^{(e)}_{\seq{m}_{d}}$ for generic $m_{d,1}$, because we do not need to explicitly remove traces in the last step of dimensional induction, while the number of indices in the preceding steps is independent from $m_{d,1}$. 

Furthermore, the right hand side of~\eqref{eq:Pfrompi} contains the irreducible projector $\pi$ contracted with a bunch of vectors (basis vectors $e_i$ or $e_1(\theta),e_2(\theta)$) on both sides. These are precisely the contractions studied recently in~\cite{Costa:2016hju}. Given their results, we then obtain a simple algorithm for computation of $P$-functions. It is best illustrated in examples.

\paragraph{Matrix element $P^{\bfm_d,\bullet}_{\bullet,\bullet}(\theta)$} 
\hfill\\
We start with the simplest example,
\be
P^{\bfm_d,\bullet}_{\bullet,\bullet}(\theta).
\ee
Since in this case $\bfm_{d-1}=\bullet$, we necessarily have $\bfm_d=\bfj$ is a traceless-symmetric tensor representation. Recall from~\eqref{eq:GTSTTexample1} that 
\be
T_{\bfj,\bullet,\ldots}^{\mu_1\ldots \mu_j}=N_j\left(e_1^{\mu_1}\ldots e_1^{\mu_j}-\text{traces}\right),
\ee
where we also introduced the normalization factor $N_j$. We thus conclude
\be
T_{\bfj,\bullet,\ldots}^{(e),\mu_1\ldots \mu_j}&=N_j e_1^{\mu_1}\ldots e_1^{\mu_j},\\
T_{\bfj,\bullet,\ldots}^{(e),\mu_1\ldots \mu_j}(\theta)&=N_j e_1^{\mu_1}(\theta)\ldots e_1^{\mu_j}(\theta).
\ee
The results of~\cite{Costa:2016hju} are formulated in the following way. They define the function
\be
\pi_j(z_1,\bar z_1)=z_1^{\mu_1}\ldots z_1^{\mu_j}{}_{\mu_1\ldots\mu_j}\pi_{\nu_1\ldots \nu_j}\bar{z}_1^{\mu_1}\ldots \bar {z}_1^{\mu_j},
\ee
where $\pi$ is the projector on traceless-symmetric spin-$j$ representation. This function completely encodes the projector since the components can be recovered by taking repeated derivatives in $z_1$ and $\bar z_1$.\footnote{Note that we do not require $z_1\cdot z_1=0$.} It is then can be shown that
\be
\pi_j(z_1,\bar z_1)=\frac{j!}{2^j(\nu)_j}| z_1|^j|\bar z_1|^j C^{(\nu)}_j\left(\frac{z_1\cdot \bar z_1}{|z_1||\bar z_1|}\right),
\ee
where $\nu=\frac{d-2}{2}$. We then immediately find that
\be
P^{\bfj,\bullet}_{\bullet,\bullet}(\theta)&=T_{\bfj,\bullet,\ldots}^{(e)}\cdot\pi\cdot T_{\bfj,\bullet,\ldots}^{(e)}(\theta)\nn\\
&=N_j^2\pi(z_1,\bar z_1)\Big\vert_{z_1=e_1,\,\bar z_1=e_1(\theta)}\nn\\
&=N_j^2\frac{j!}{2^j(\nu)_j}|e_1|^j|e_1(\theta)|^j C^{(\nu)}_j\left(\frac{e_1\cdot e_1(\theta)}{|e_1||e_1(\theta)|}\right)\nn\\
&=\frac{N_j^2 j!}{2^j(\nu)_j}C_j^{(\nu)}(\cos\theta).
\ee
Note that the normalization condition for $|\bfj,\bullet,\ldots\>$ is equivalent to $P^{\bfj,\bullet}_{\bullet,\bullet}(0)=1$, and thus using
\be\label{eq:Gegenbauer1}
C_j^{(\nu)}(1)=\frac{(2\nu)_j}{j!}.
\ee
we find
\be
1=\frac{N_j^2 j!}{2^j(\nu)_j}C_j^{(\nu)}(1)=N_j^2 \frac{(2\nu)_j}{2^j(\nu)_j},
\ee
from where we conclude that\footnote{Here we essentially make a choice of phase for $|\bfj,\bullet,\ldots\>$.}
\be
N_j=\sqrt{\frac{2^j(\nu)_j}{(2\nu)_j}},
\ee
while
\be\label{eq:gegenbauermatel}
P^{\bfj,\bullet}_{\bullet,\bullet}(\theta)=\frac{j!}{(2\nu)_j}C_j^{(\nu)}(\cos\theta).
\ee

\paragraph{Matrix element} $\!\!\!\!P^{\bfj,\bullet}_{\myng{(1)},\myng{(1)}}(\theta)$\hfill\\
We now consider the matrix elements
\be\label{eq:doubleexample}
P^{\bfm_d,\bullet}_{\myng{(1)},\myng{(1)}}(\theta).
\ee
Note that now both $\bfm_{d-1}$ and $\bfm_{d-1}'$ are equal to $\myng{(1)}$ and thus $\bfm_d$ can be either a traceless-symmetric tensor $\bfj$ or a hook diagram $(\bfj,\myng{(1)})$. 

We start from the traceless-symmetric case and will return to the hook exchange later. From~\eqref{eq:GTSTTexample2} we find
\be
T_{\bfj,\myng{(1)},\bullet,\ldots}^{(e),\mu_1,\ldots,\mu_j}&=N_{j,\myng{(1)}}e_2^{(\mu_1}e_1^{\mu_2}\cdots e_1^{\mu_j)},\\
T_{\bfj,\myng{(1)},\bullet,\ldots}^{(e),\mu_1,\ldots,\mu_j}(\theta)&=N_{j,\myng{(1)}}e_2^{(\mu_1}(\theta)e_1^{\mu_2}(\theta)\cdots e_1^{\mu_j)}(\theta).
\ee
We then find
\be
P^{\bfj,\bullet}_{\myng{(1)},\myng{(1)}}(\theta)&=T_{\bfj,\myng{(1)},\bullet,\ldots}^{(e)}\cdot\pi\cdot T_{\bfj,\myng{(1)},\bullet,\ldots}^{(e)}(\theta)\nn\\
&=\frac{1}{j^2}N_{j,\myng{(1)}}^2(e_2\cdot\partial_{z_1})(e_2(\theta)\cdot\partial_{\bar z_1})\pi(z_1,\bar z_1)\Big\vert_{z_1=e_1,\,\bar z_1=e_1(\theta)}\nn\\
&=\frac{N_{j,\myng{(1)}}^2 j!}{j^2 2^j(\nu)_j}\left(\cos\theta\partial C_j^{(\nu)}(\cos\theta)-\sin^2\theta \partial^2C_j^{(\nu)}(\cos\theta)\right)\nn\\
&=-\frac{N_{j,\myng{(1)}}^2 j!}{j^2 2^j(\nu)_j}\partial^2_\theta C_j^{(\nu)}(\cos\theta).
\ee
Again, we have the normalization condition $P^{\bfj,\bullet}_{\myng{(1)},\myng{(1)}}(1)=1$. To solve for $N_{j,\myng{(1)}}$, we need to know $\partial C_j^{(\nu)}(1)$, which can be computed using the identity
\be
\partial_x C^{(\nu)}_j(x)=2\nu\, C^{(\nu+1)}_{j-1}(x).
\ee
We thus find 
\be
\frac{N_{j,\myng{(1)}}^2 j!}{j^2 2^j(\nu)_j}\frac{2\nu (2\nu+2)_{j-1}}{(j-1)!}=1,
\ee
and therefore (adding a phase for future convenience)
\be
N_{j,\myng{(1)}}&=-\sqrt{\frac{2^j j (\nu)_j}{2\nu(2\nu+2)_{j-1}}},\\
P^{\bfj,\bullet}_{\myng{(1)},\myng{(1)}}(\theta)&=-\frac{(j-1)!}{2\nu (2\nu+2)_{j-1}}\partial^2_\theta C_j^{(\nu)}(\cos\theta).
\ee

\paragraph{Matrix elements} $\!\!\!\!P^{\bfm_d,\bullet}_{\myng{(1)},\bullet}(\theta)$ \textbf{and} $ P^{\bfm_d,\bullet}_{\bullet,\myng{(1)}}(\theta)$\\
Having determined the normalization factors $N_j$ and $N_{j,\myng{(1)}}$, we can now address the matrix elements
\be
P^{\bfm_d,\bullet}_{\myng{(1)},\bullet}(\theta),\quad P^{\bfm_d,\bullet}_{\bullet,\myng{(1)}}(\theta),
\ee
which are not subject to a simple normalization condition at $\theta=0$. In particular, their phases are convention-dependent. We have
\be
P^{\bfm_d,\bullet}_{\myng{(1)},\bullet}(\theta)&=T_{\bfj,\myng{(1)},\bullet,\ldots}^{(e)}\cdot\pi\cdot T_{\bfj,\bullet,\ldots}^{(e)}(\theta)\nn\\
&=N_jN_{j,\myng{(1)}}j^{-1}(e_2\cdot \ptl_{z_1})\pi_j(z_1,\bar z_1)\Big\vert_{z_1=e_1,\bar z_1=e_1(\theta)}\nn\\
&=-\frac{ 2\nu j!}{(2\nu)_j}\sqrt{\frac{2\nu+1}{j(2\nu+j)}} \sin\theta\, C^{(\nu+1)}_{j-1}(\cos\theta).
\ee
An analogous calculation shows that 
\be
P^{\bfm_d,\bullet}_{\bullet,\myng{(1)}}(\theta)=\frac{ 2\nu j!}{(2\nu)_j}\sqrt{\frac{2\nu+1}{j(2\nu+j)}} \sin\theta\, C^{(\nu+1)}_{j-1}(\cos\theta)=\left(P^{\bfm_d,\bullet}_{\myng{(1)},\bullet}(-\theta)\right)^*,
\ee
consistently with~\eqref{eq:hermitianconj}. One can check in explicit examples that these results coincide with the direct exponentiation of $M_{12}$, providing a non-trivial check of the formalism and normalization factors.

\paragraph{Matrix element}$\!\!\!\!P^{(\bfj,\myng{(1)}),\bullet}_{\myng{(1)},\myng{(1)}}(\theta)$\hfill\\
Consider now the case of the hook exchange $\bfm_d=(\bfj,\myng{(1)})$ in~\eqref{eq:doubleexample}. We are now dealing with a new type of representations. Correspondingly, in~\cite{Costa:2016hju} the following function is defined
\be
\pi_{(j,1)}(z_1,z_2,\bar z_1,\bar z_2)=z_2^\nu z_1^{\mu_1}\cdots z_1^{\mu_j}{}_{\nu,\mu_1\ldots\mu_j}\pi_{\bar\nu,\bar\mu_1\ldots\bar\mu_j}\bar z_2^{\bar\nu}\bar z_1^{\bar \mu_1}\cdots\bar z_1^{\bar \mu_j}.
\ee
The expression for the full projector is somewhat complicated, so we do not reproduce it here. In practice, we used the Mathematica code supplied with~\cite{Costa:2016hju} to perform the calculations with these projectors.

From equation~\eqref{eq:GTnonsttexample} we find
\be
T_{(\bfj,\myng{(1)}),\myng{(1)},\bullet,\ldots}^{(e),\nu,\mu_1,\ldots,\mu_j}&=N_{(\bfj,\myng{(1)}),\myng{(1)}}\left[ e_1^{\nu}e_2^{(\mu_1}e_1^{\mu_2}\cdots e_1^{\mu_j)}-e_2^{\nu}e_1^{(\mu_1}e_1^{\mu_2}\cdots e_1^{\mu_j)}\right],\\
T_{(\bfj,\myng{(1)}),\myng{(1)},\bullet,\ldots}^{(e),\nu,\mu_1,\ldots,\mu_j}(\theta)&=N_{(\bfj,\myng{(1)}),\myng{(1)}} \left[e_1^{\nu}(\theta)e_2^{(\mu_1}(\theta)e_1^{\mu_2}(\theta)\cdots e_1^{\mu_j)}(\theta)-e_2^{\nu}(\theta)e_1^{(\mu_1}(\theta)e_1^{\mu_2}(\theta)\cdots e_1^{\mu_j)}(\theta)\right].
\ee
This implies
\be
N_{(\bfj,\myng{(1)}),\myng{(1)}}^{-2}T_{(\bfj,\myng{(1)}),\myng{(1)},\bullet,\ldots}^{(e)}\cdot\pi\cdot T_{(\bfj,\myng{(1)}),\myng{(1)},\bullet,\ldots}^{(e)}(\theta)=
&j^{-2}(e_2\cdot\ptl_{z_1})(e_2(\theta)\cdot\ptl_{\bar z_1})\pi_{(j,1)}(e_1,e_1,e_1(\theta),e_1(\theta))\nn\\
&-j^{-1}(e_2\cdot\ptl_{z_1})\pi_{(j,1)}(e_1,e_1,e_1(\theta),e_2(\theta))\nn\\
&-j^{-1}(e_2(\theta)\cdot\ptl_{\bar z_1})\pi_{(j,1)}(e_1,e_2,e_1(\theta),e_1(\theta))\nn\\
&+\pi_{(j,1)}(e_1,e_2,e_1(\theta),e_2(\theta)),
\ee
where the values of the arguments of $\pi_{(j,1)}(z_1,z_2,\bar z_1,\bar z_2)$ should be substituted after taking the derivatives. Using the explicit form of the projector~\cite{Costa:2016hju}, and using the normalization condition $P^{(\bfj,\myng{(1)}),\bullet}_{\myng{(1)},\myng{(1)}}(0)=1$, we find
\be
N_{(\bfj,\myng{(1)}),\myng{(1)}}&=\sqrt{\frac{2^j j (d+j-3)(\nu)_j}{2\nu^2 (j+1)^2(2\nu+2)_{j-1}}},\\
P^{(\bfj,\myng{(1)}),\bullet}_{\myng{(1)},\myng{(1)}}(\theta)&=\frac{(j-1)!}{(2\nu+2)_{j-1}}C^{(\nu+1)}_{j-1}(\cos\theta).
\ee

\bibliographystyle{JHEP}
\bibliography{refs}

\end{document}